\documentclass[%
 reprint,
nofootinbib,
 amsmath,amssymb,
 aps,
]{revtex4-2}

\usepackage{aasmacros}
\usepackage{graphicx}
\usepackage{dcolumn}
\usepackage{multirow}
\usepackage{bm}
\usepackage{newtxtext,newtxmath}
\usepackage{subcaption}
\usepackage[T1]{fontenc}
\usepackage{makecell}
\usepackage{animate}
\usepackage{mathtools}
\usepackage{url}
\usepackage[hidelinks,colorlinks]{hyperref}
\usepackage{xcolor}
\hypersetup{
    colorlinks=true,
    linkcolor={blue},
    citecolor={blue},
    urlcolor={blue}
}
\DeclarePairedDelimiter{\evdel}{\langle}{\rangle}
\newcommand{\ev}{\evdel}
\def\hlinewd#1{%
\noalign{\ifnum0=`}\fi\hrule \@height #1 %
\futurelet\reserved@a\@xhline} 
\usepackage{graphicx}	
\usepackage{amsmath}	

\begin{document}

\preprint{APS/123-QED}

\title{Diversity and universality: Evolution of dwarf galaxies with self-interacting dark matter}

\author{Zhichao Carton Zeng$^{1,2,3}$}\email{E-mail: zczeng@tamu.edu}
\author{Annika H. G. Peter$^{1,2,4}$}
\author{Xiaolong Du$^{5,6}$}
\author{Andrew Benson$^{5}$}
\author{Jiaxuan Li$^{7}$}
\author{Charlie Mace$^{1,2}$}
\author{Shengqi Yang$^{5,8}$}

\affiliation{$^{1}$Department of Physics, The Ohio State University, 191 W. Woodruff Avenue, Columbus, Ohio 43210, USA \\
$^{2}$Center for Cosmology and Astroparticle Physics, The Ohio State University, 191 W. Woodruff Avenue, Columbus, Ohio 43210, USA\\
$^{3}$Department of Physics and Astronomy, Mitchell Institute for Fundamental Physics and
Astronomy, Texas A\&M University, College Station, Texas 77843, USA \\
$^{4}$Department of Astronomy, The Ohio State University, 140 W. 18th Avenue, Columbus, Ohio 43210, USA\\
$^{5}$ Carnegie Observatories, 813 Santa Barbara Street, Pasadena, California 91101, USA\\
$^{6}$ Department of Physics and Astronomy, University of California, Los Angeles, California 90095, USA\\
$^{7}$ Department of Astrophysical Sciences, Princeton University, 4 Ivy Lane, Princeton, New Jersey 08544, USA\\
$^{8}$ Los Alamos National Laboratory, Los Alamos, New Mexico 87545, USA
}

\date{\today}

\begin{abstract}
Dark matter halos with self-interacting dark matter (SIDM) experience a unique evolutionary phenomenon, in that their central regions eventually collapse to high density  through the runaway gravothermal process after initially forming a large and low-density core.  When coupled with orbital evolution, this is expected to naturally produce a large diversity in dark matter halos' inner mass distribution, potentially explaining the diversity problem of dwarf galaxies. However, it remains unknown how the diversity in SIDM halos propagates to the more easily observed luminous matter at the center of the halo, especially the stellar component.  In this work, we use idealized N-body simulations with two species of particles (dark matter and stars) to study the response of the stellar properties of field and satellite dwarf galaxies to SIDM evolution and orbital effects on their halos. Galaxies' stellar components, including galaxy size, mass-to-light ratio, and stellar velocity dispersion, display a much larger scatter in SIDM than the standard cold dark matter model. Importantly, we find  signs of universality in the evolution pathways, or ``tidal tracks'', of SIDM dwarf satellites, which are physically interpretable and potentially parametrizable.  This type of tidal-track model can be layered onto larger-scale, cosmological simulations to reconstruct the evolution of populations of SIDM dwarfs in cases where high-resolution simulations of galaxies are otherwise prohibitively expensive. 
\end{abstract}

                              
\maketitle


\section{Introduction}

Self-interacting dark matter (SIDM) has recently raised interest as an alternative candidate to the standard cold dark matter (CDM) paradigm. 
Particle interactions that are stronger than gravity cause efficient heat transfer within dark matter halos.
This leads to the unique two-phase evolution of SIDM halos: core formation (on shorter time scales) followed by core-collapse (on longer time scales) \cite{spergel00, balberg02, balberg02b, essig19}. SIDM scattering first causes thermalization in the halo center, leading to a uniform low-density core. The core then serves as a heat bath, slowly but endlessly transferring heat to larger radii of the halo, where the ``temperature'' (proxy of velocity dispersion, defined as $k T = \sigma_v^2$ where $\sigma_v$ is the velocity dispersion in analogy with an ideal gas) remains always lower. Due to the negative heat capacity of such self-gravitating systems, the central density core becomes counterintuitively hotter and smaller via this heat outflow, resulting in dark matter falling towards the center, which in turn further strengthens the temperature gradient. This process is thus self-accelerating, and eventually becomes a runaway process known as core-collapse (also called gravothermal collapse/catastrophe) \cite{lyndenbell68, lyndenbell80, inagaki83, balberg02, balberg02b, Colin02, essig19}. 

The density of the central core of SIDM halos can therefore naturally span an enormous diversity, when observed at different life stages during the core-formation to core-collapse evolution \cite{sameie20, correa21, zzc22, zzc23, roberts24}. 
The core-collapse mechanism of SIDM halos is thus considered to have to potential to explain recent observational anomalies on small scales, which appears to be challenging to describe in the CDM paradigm \cite{bullock17, buckley17, tulin17}. These include the ``diversity problem'' of dwarf galaxies \cite{oman15, santos18, revaz18, li20, santos20, hayashi20}, enhanced galaxy-galaxy lensing cross sections \cite{Meneghetti20, meneghetti23, tokayer24, dutra24} and ultracompact substructure lenses \cite{minor21b, nadler23, gzhang23, despali24}. 

However, observations on larger scales set constraints on SIDM cross sections that are too small to cause halo core-collapse within a Hubble time.
On the scales of galaxy clusters, the SIDM cross section is constrained to $\lesssim \mathcal{O}(1)\ \rm cm^2/g$ by central densities \cite{rocha13, elbert18, andrade22, Ragagnin24}, cluster ellipticity \cite{Peter13, robertson19, McDaniel21, shen22, despali22,  robertson23}, and galaxy offsets during cluster mergers \cite{randall08, kim17, robertson17, cross23, fischer23}. Consequently, additional degrees of freedom, such as velocity-dependent SIDM cross sections \cite{mv12, Sagunski20, nadler20, nadler23, correa21, sq22, sq23, zzc23} and energy dissipation during particle scattering events \cite{Choquette18, essig19, huo19, hyxiao21, shen22, roy24}, have been introduced to produce core-collapse in the low-mass halos that are relevant for the small-scale problems while evading cluster-scale constraints. 

Importantly, the speed of core-collapse depends on environment. 
Core-collapse in SIDM subhalos is driven by the heat outflow from subhalo center and can thus be accelerated by tidal stripping, which strengthens the subhalo's temperature gradient by removing dark matter from the outskirts \cite{Kahlhoefer19, nishikawa20, sameie20, correa21, zzc22, neev23}. On the other hand, it can be slowed down (or even disrupted) by tidal heating and evaporation (also known as SIDM ram-pressure, which refers to the dark matter (DM) self-interaction between the subhalo and the host \cite{kummer18,  zzc22, slone21}). These effects inject heat into the subhalo center, causing the central core to puff up, thus countering the core-collapse process \cite{zzc22, zzc23}. 
Considering the complicated interplay between various orbital effects and subhalos' internal heat transfer, \cite{zzc23} has simulated realistic merger trees of host-subhalos systems and found that for certain velocity-dependent SIDM models, core-collapse does happen in a significant fraction ($\gtrsim 10\%$) of subhalos within the cosmic time, while also satisfying the constraints of $\sigma/m\lesssim \mathcal{O}(1)\ \rm cm^2/g$ for cluster-scale velocities. \cite{zzc23} also confirmed that such significant core-collapse does greatly raise the diversity in subhalos' inner density/mass as well as the inner density slope.

The next step is to understand how this predicted diversity in dark matter structure in (sub)halo centers 
affects the visible baryons in the system.  How do baryons  (specifically stars) respond to the SIDM halo evolution, including also the complicated environment effects? Exploring the evolution of stellar components in the context of SIDM will help to see if SIDM can provide a plausible answer to the diversity in dwarf galaxies, including the recent discovery of various extremes: ultradiffuse \cite{vDk15, koda15, sales22, jxli23, forbes23}, ultracompact \cite{drinkwater03, czliu15, Saifollahi21, kxwang23}, DM-free \cite{df2, df4}, and DM-dominant \cite{smith24, errani24} galaxies.
However, most of the previous works on SIDM core-collapse are in the dark matter only context, or at most including a central baryonic potential to account for its contracting effect on dark matter \cite{fz19, wxfeng21, fz22, ymzhong23, sq23}. This is because explicitly including baryons/stars in simulations with possible SIDM core-collapse presents a tremendous challenge to computational power. Recent convergence tests of dark matter only simulations show that $\gtrsim 10^5$ particles per halo are required to ensure converged evolution in the halo's inner density at late stages of core-collapse \cite{charlie24, Palubski24, Fischer24}.  When the task is to resolve stellar components in low-mass (sub)halos $\lesssim \mathcal{O}(10^{10})M_\odot$, for which the stellar mass fraction is low $\lesssim \mathcal{O}(10^{-3})$ \cite{behroozi19}, it becomes more demanding in particle resolution. As such, it is difficult to simulate SIDM dwarf galaxies in a cosmological context with full hydrodynamics, especially in the context of core0collapse (Cruz \textit{et al}. in prep.). 

An alternative ``tidal track'' approach has been developed for CDM in previous work to trace the evolution of stellar components at low computational expense\cite{Penarrubia07, Penarrubia10, errani15}. This approach carefully traces the evolution of single satellite galaxies using high-resolution controlled simulations, in which the evolution of stellar properties in a tidal field can be more accurately tracked than in cosmological simulations. The results are then used to build parametrized, empirical relations between the evolution of stellar properties such as stellar size and the mass-to-light ratio, which demands high particle resolution, and the evolution of the total (DM+stellar) inner mass, which is less demanding in resolution.  Such tidal tracks can then be applied in larger-scale simulations with lower resolution to reconstruct the stellar evolution \cite{sales19, carleton19}. In this paper, we explore the possibility of building such tidal evolution tracks for SIDM cases, following a similar spirit.  


In this work, we examine the dynamical evolution of an SIDM dwarf galaxy, both in isolation and as a satellite in a host halo, testing multiple SIDM models.  This work is organized as follows: In Sec. \ref{sec:method} we illustrate the setup of our controlled simulations, each consisting of a single dwarf galaxy-(sub)halo system, composed of two species of particles, DM and stars. We also include a detailed convergence test regarding particle mass resolution and particle mass ratio between the two species, which are the two key parameters that can cause spurious effects in the simulation. 
In Sec. \ref{sec:result}, we show the evolution of dwarf (satellite) galaxies in different SIDM models and orbits, including the evolution of their stellar properties in time and a tidal track study that correlates with the change in the (sub)halos' inner mass. We summarize and discuss limitations and possible future works in Sec. \ref{sec:summary}. 

\section{Setup of simulations}\label{sec:method}

\subsection{Overview}\label{sec:method-overview}
In this section, we describe how we set up idealized simulations of an orbiting subhalo with a galaxy embedded in it, using two species of particles to trace DM and stars.  Importantly, we perform a number of convergence tests for our scheme. Although particular choices are specific to the goals of this study, we identify a number of issues that are general and which we highlight for the benefit of the simulation community. In Sec. \ref{sec:res}, we show the results of our convergence tests, which have additional implications for choices simulators make for cosmological simulations. 

We aim to trace the dynamical evolution of the stellar component within a typical dwarf (satellite) galaxy in different SIDM models. We use the N-body initial condition generator \texttt{SpherIC} \cite{gk2013} to set up initial conditions for the ``DM+star'' simulation, where the initial density profiles of DM and stellar components are given as input and the initial velocities are calculated using Eddington inversion \cite{eddington16, lacroix18}. The contribution from both DM and stars to the gravitational potential is included, such that the system is in equilibrium at the beginning. 

Because our initial scientific motivation for this study was the formation of ultradiffuse galaxies (UDGs) in group-scale environments \cite{merritt16, df2, df4, zlshen24}, we choose a relatively massive dwarf galaxy and its halo as the starting point of our investigations.  In this context, we use initial conditions that we had in hand from \cite{zzc23}.  Specifically, we pick one subhalo from a halo merger tree of a galaxy-group-sized system, which is generated by the galaxy evolution code \texttt{Galacticus} and was one of the four realizations used in a recent study of ensemble evolution of SIDM subhalos \cite{zzc23}. We select a $M_{200c}=3.01\times10^{10} M_\odot$ subhalo that forms (defined by the time when its mass reaches half its infall mass in \texttt{Galacticus}) at $z\sim4$ (equivalently $t=$1.5 Gyr since the big bang). The subhalo is chosen such that its mass is low enough to be simulated in an analytic host ($\lesssim 0.3\%$ host mass, see \cite{zzc22}), with negligible dynamical friction 
For the reasons we discuss below, lower-mass subhalos, which typically have lower stellar mass fraction \cite{behroozi19, behroozi19b, Erfanianfar19}, demand much higher resolution for our simulations. We additionally pick this subhalo because of its early formation time to make sure that it has a long evolution time inside the host halo, so that the study of this single satellite can exhaust the evolution pathways of satellites with short to long in-orbit times.   For all DM models, the DM component of the subhalo is initialized as a Navarro-Frenk-White (NFW) profile, with characteristic parameters $\rho_s=1.45\times10^7 M_\odot/ {\rm kpc^3}$, $r_s=5.12$ kpc, $M_{200c}=3.01\times10^{10} M_\odot$ and $r_{200c}=37.32$ kpc (with the critical density corresponding to the beginning of the simulation $t=$1.5 Gyr or $z\sim4$). 

For the satellite galaxy embedded in the subhalo, we set the initial stellar mass as $2.48\times10^7 M_\odot$ according to the stellar-mass-halo-mass relation in \cite{behroozi19}, with an initial size (half-light radius)\footnote{Note that the ``half-light'' radius in this work is more precisely ``half-stellar-mass'' radius, since we only have identical star particles in the simulation, without differentiating color and luminosity.} as 0.37 kpc, according to the stellar-size-halo-size relation in \cite{fz19}. To account for the diversity in satellites' initial density and to partially compensate for the missing baryonic feedback in our DM+star simulations \cite{tkchan15, lazar20, sales22}, we test two scenarios of the initial density profiles for stars, a Hernquist \cite{hernquist90} profile to represent a cuspy stellar distribution (characteristic of massive dwarfs), and a Plummer \cite{plummer} profile to represent a cored distribution (representing the more commonly observed properties of classical dwarf galaxies). The corresponding scale radii for the Hernquist and Plummer galaxies are then $r_{\rm hern} = $0.15 kpc and $r_p=$0.28 kpc.

To trace the evolution of the subhalo and its embedded satellite galaxy orbiting in the host halo (the combined object consisting of the DM subhalo and its dwarf galaxy will be denoted as ``satellite'' hereafter, for simplicity), we use the hybrid N-body + semianalytical method introduced in \cite{zzc22}, which is built upon the hydrodynamical code \texttt{Arepo} \cite{Springel10} with an SIDM module \cite{mv12, mv14, mv19}. The host halo is modeled with an analytic mass and particle velocity distribution. The analytic mass profile is needed for the calculation of the gravitational force on the satellite. To account for the self-interaction effects between the host and the subhalo (the `evaporation'), we need both the mass distribution and the velocity dispersion of dark matter particles within the host.  To build such an analytic distribution of the host given different SIDM models, we initialize the host with an NFW profile, evolve it in a separate simulation with DM self-interaction for 4 Gyr to reach equilibrium, and then measure the resulting density and radial velocity dispersion profiles. The NFW initial condition of the host has parameters $\rho_s=3.92\times10^6 M_\odot/ {\rm kpc^3}$, $r_s=53.76$ kpc, $M_{200c}=1.01\times10^{13} M_\odot$ and $r_{200c}=436.5$ kpc, corresponding to the merger tree at $t=9$ Gyr ($z\approx0.5$).  For simplicity and better control of the idealized simulations in this work, the analytic host is static, not growing with time, as in the treatment of \cite{zzc22}. 

\begin{figure}
    \includegraphics[width=\columnwidth]{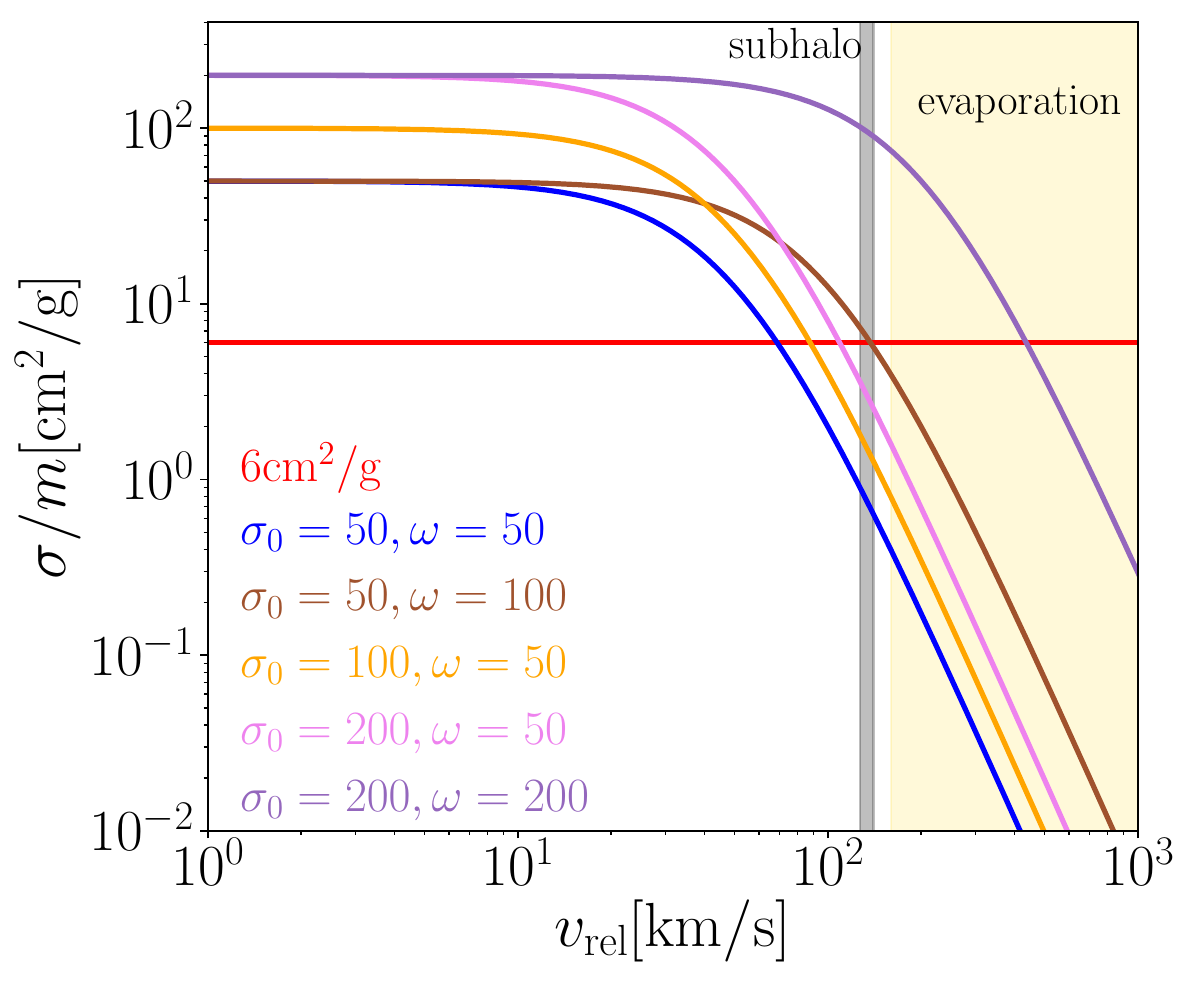}
    \caption{The $\sigma/m - v_{\rm rel}$ relation for the SIDM models we use in this work. The vertical shaded region in yellow indicates characteristic orbital velocities of the host, and this is the scale relevant for the evaporation of subhalo particles by the host. The gray shaded region represents the velocity scale relevant for particle scatterings within the subhalo, using an empirical relation $\ev{v_{\rm rel}} \approx 3.8 \sigma_v$ from \protect{\cite{zzc23}}. }
    \label{fig:sig-v}
\end{figure}

The SIDM models we study in this work are mostly the same as in \cite{zzc23}, including a constant model $\sigma/m=6\ \rm cm^2/g$ and a group of velocity-dependent models, which follow a generic form characterized by two parameters $\sigma_0$ and $\omega$:
\begin{equation}
    \frac{\sigma}{m} = \frac{\sigma_0}{[1+v_{\rm rel}^2/\omega^2]^2}.
\end{equation}
We plot the $\sigma/m - v_{\rm rel}$ relation for these models in Fig. \ref{fig:sig-v}. As we will see in later sections, the combination of these $\{\sigma_0, \omega\}$ models and multiple orbits covers the main evolutionary pathways of subhalos of SIDM subhalos: cored with and without evaporation, and core-collapsed with and without evaporation.

In this work, we follow \cite{zzc22, zzc23} to use the (sub)halo's central density as the main tracer of the SIDM core-collapse process. At every time step, the local SIDM density of each SIDM particle (where `local' is defined by a sphere enclosing its 32 nearest SIDM neighbors, also known as the SIDM softening length) is measured for evaluation of the scattering probability. A (sub)halo's central density is then estimated as the average of the highest 50 values of these local densities, denoted as $\rho_{\rm cen50}$. When $\rho_{\rm cen50}$ reaches five times its initial value, we mark this (sub)halo as core collapsed and terminate the simulation. 
This choice is a compromise given two considerations.  On one hand, the computational cost becomes prohibitively high for the gravity solver and scattering algorithm to resolve the physics in the denser and denser region of the runaway collapse. Moreover, as the scattering mean-free-path decreases and the core-collapse region becomes more fluidlike, new techniques beyond N-body simulation are required to accurately track the physics. On the other hand, as found in \cite{zzc22}, orbital effects such as tidal heating and host-subhalo evaporation can disrupt the early-stage core-collapse process. If the termination density threshold is set too low, it may misclassify a galaxy as core-collapsed when its evolution is instead disrupted by evaporation. In the real physical scenario, however, the core-collapse process continues beyond this ``5 times'' cutoff threshold (see \cite{sophia23} for more discussion). We thus note that the results of core-collapsed (sub)halos and their stellar properties that we present in this work should be viewed as a lower limit of the dense state produced by the core-collapse process.

\subsection{Resolution and mass ratio tests}\label{sec:res}

In this subsection, we describe the numerical recipe for setting up an idealized N-body simulation of a stellar system embedded in a dark matter (sub)halo, with two types of particles: DM and stars. Our goal is to determine how to set up the simulations so that the evolution of the half-light radius (and thus the stellar properties therein) is accurately resolved. We vary two numerical parameters to test the convergence: the DM particle resolution, and the mass ratio between dark matter and star particles. While it is a conventional practice to use a dark matter to baryon (gas/star) particle mass ratio of approximately 5:1 following the cosmological energy density partition \cite{tng, mtng, firebox} (see \href{https://www.tng-project.org/data/landscape/}{here} for an overview), it has been argued that using unequal masses for two species of particles can exacerbate the two-body scattering and cause spurious diffusion in the less massive particles \cite{binney02, ludlow19}.  This leads to spuriously larger sizes of galaxies. To explore the size of the effect for our simulations, we compare the simulation results produced by $5:1$ and $1:1$ mass ratios between DM and star particles.

\begin{figure}
    \includegraphics[width=0.9\columnwidth]{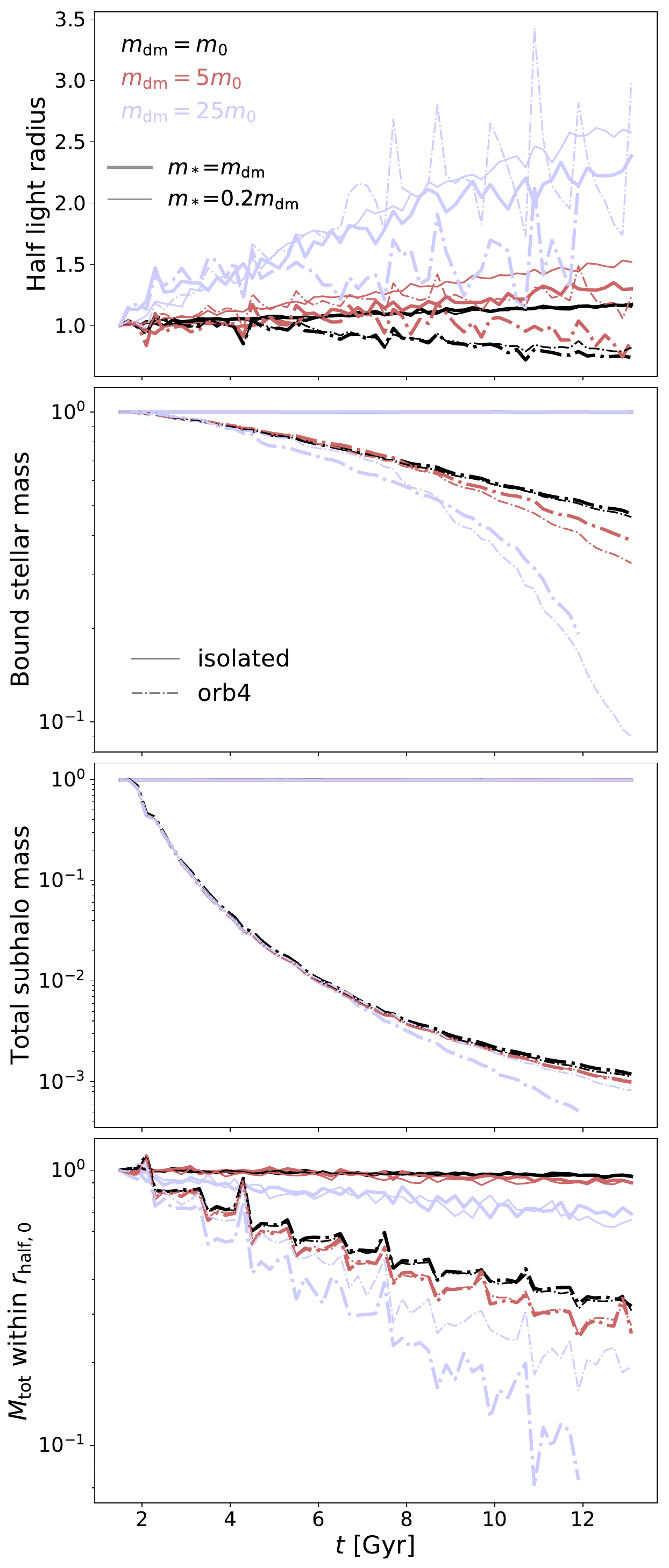}
    \caption{Convergence test for resolution and star-DM particle mass ratio with CDM (sub)halos. From top to bottom are the time evolution of the $r_{\rm half}$ of bound star particles, the bound stellar mass, total mass within $r_{200c}$ of the (sub)halo, and the total mass within the initial half-light radius $r_{\rm half, 0}$. Black lines ($m_{\rm dm}=m_0$) are simulations with our highest particle resolution, with red and purple lines having $5\times$ and $25\times$ the DM particle mass (thus lower resolution). The thick lines have equal mass ratio between star and DM particles, and the thin lines have a mass ratio of $1:5$. All the combinations of resolution and mass-ratios are listed in Table \protect\ref{table:reso}. The solid lines represent galaxy-halo systems in isolation, and the dash-dotted lines are for orb4 subhalos (the most eccentric orbit we test in this work, see Table \protect\ref{table:orb}). }
    \label{fig:mp-cdm}
\end{figure}

We impose two physically motivated conditions to check that the aforementioned spurious effects are minimal.  First, we consider a galaxy in an isolated CDM halo.  The galaxy size should remain stable over time, since this galaxy-halo system is generated in equilibrium and should therefore not evolve.  Second, we consider systems such as CDM subhalos and both isolated and satellite SIDM halos, which are intrinsically expected to evolve over time due to orbital effects, SIDM scattering, or both.  The lower resolution simulations should asympotically converge to the higher resolution ones. We focus on two metrics to quantify the evolution of the galaxy: its size, as defined by the half-light radius, and its mass, as defined by the stellar mass bound to the (sub)halo. In this section, we test the combination of three dark matter particle masses and two star-to-DM particle mass ratios, as detailed in Table \ref{table:reso}. 

\begin{table*} 
	\centering
	\begin{tabular}{lcccccr} 
		\hline
		Label & $m_{\rm dm}[M_\odot]$ & $N_{\rm dm}$ & $\epsilon_{\rm dm}$[kpc] & $m_{*}:m_{\rm dm}$ & $m_*[M_\odot]$ & $N_*$ \\
		\hline
        \multirow{2}{*}{$m_{\rm dm}=m_0$}  & \multirow{2}{*}{$5\times10^3$} & \multirow{2}{*}{$8.5\times10^6$}  & \multirow{2}{*}{0.026} & 1:5 & $10^3$ & 24820 \\ \cline{5-7} & & &  & 1:1 & $5\times10^3$ & 4964
    	\\	\hline
        \multirow{2}{*}{$m_{\rm dm}=5m_0$}  & \multirow{2}{*}{$2.5\times10^4$} & \multirow{2}{*}{$1.7\times10^6$} & \multirow{2}{*}{0.044}  & 1:5 & $5\times10^3$ & 4964  \\ \cline{5-7} & & & & 1:1 & $2.5\times10^4$ & 993
    	\\	\hline
        \multirow{2}{*}{$m_{\rm dm}=25m_0$}  & \multirow{2}{*}{$1.25\times10^5$} & \multirow{2}{*}{$3.4\times10^5$} & \multirow{2}{*}{0.075}  & 1:5 & $2.5\times10^4$ & 993  \\ \cline{5-7} & & &  & 1:1 & $1.25\times10^5$ & 199
    	\\	\hline
	\end{tabular}
	\caption{ The particle masses and number of particles for different resolution and star-DM particle mass ratio. $m_0$ is the DM particle mass from our highest-resolution simulations and $\epsilon_{\rm dm}$ is the gravitational softening length for DM particles. }
 \label{table:reso}
\end{table*}

\begin{table*} 
	\centering
	\begin{tabular}{lccccc} 
		\hline
		Orbits & $R_{\rm apo}/R_{\rm vir}$(host) & $R_{\rm apo}:R_{\rm peri}$ & $R_{\rm apo}$ [kpc]  &  $R_{\rm peri}$ [kpc] & $v_0$(sub) [km/s] \\
		\hline
        orb1 & 1 & 5:1 & 436.5 & 87.3 & 127.8  \\   \hline
        orb2 & 0.7 & 10:1 & 305.5 & 30.6 & 75.0  \\   \hline
        orb3 & 0.4 & 10:1 & 174.6 & 17.5 & 71.7  \\   \hline        
        orb4 & 0.4 & 20:1 & 174.6 & 8.7 & 38.2  \\   \hline  
	\end{tabular}
	\caption{Parameters of orbits we sample in this work. From orb1 to orb4, the host effects get stronger. }
    \label{table:orb}
\end{table*}

\begin{table*} 
	\centering
	\begin{tabular}{lcc || cc} 
        \hline
        & \multicolumn{2}{c}{Hernquist I.C. for stars}& \multicolumn{2}{c}{Plummer I.C. for stars} \\ 
        \hline
		\hline
		Orbits & $\sigma_0=200, \omega=50$ & $\sigma_0=200, \omega=200$ & $\sigma_0=200, \omega=50$ & $\sigma_0=200, \omega=200$  \\
		\hline
        isolated & - & $t_{\rm cc}=8.7$ Gyr  & - &  $t_{\rm cc}=8.7$ Gyr \\   \hline
        orb1 & - & $t_{\rm cc}=7.7$ Gyr & - & $t_{\rm cc}=7.7$ Gyr  \\   \hline
        orb2 & $t_{\rm cc}=12.9$ Gyr & -   
 & - & - \\   \hline
        orb3 & $t_{\rm cc}=10.3$ Gyr & - & $t_{\rm cc}=13.1$ Gyr & - \\   \hline        
        orb4 & - & -  & - & - \\   \hline  
	\end{tabular}
	\caption{ The core-collapse time, $t_{\rm cc}$, of the test (sub)halo, combining different SIDM models and orbits. Only the cases that core-collapse before the end of the simulation at $t=13.1$ Gyr have a $t_{\rm cc}$. Other (sub)halos in this table may eventually core-collapse given a longer time, such as the isolated ones, or may never collapse due to the interactions with the host. Apart from the two SIDM models in this table, other SIDM models we test in this work have no core-collapse happening in the test subhalo. }
    \label{table:ccp-orb}
\end{table*}


\begin{figure}
    \includegraphics[width=\columnwidth]{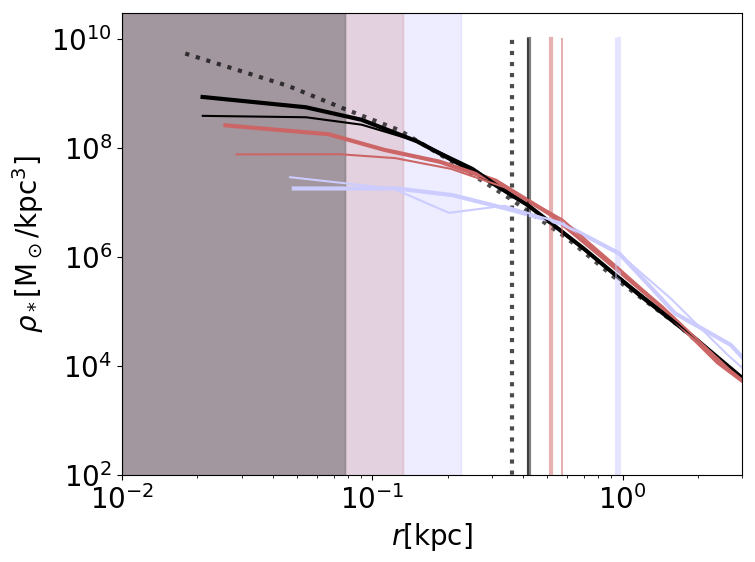}
    \caption{Stellar density profiles of the isolated CDM halo under different resolution and particle mass ratios, at the end of the simulation. The resolution and particle mass ratios are presented in the same manner as in Fig. \protect\ref{fig:mp-cdm}. The black dashed line shows the stellar density profile of the initial condition that is the same for all three resolutions. The shaded regions mark the radius of three times the gravitational softening length of each resolution run. The vertical lines mark the $r_{\rm half}$ of corresponding resolution cases at the end of the simulation. }
    \label{fig:mp-cdm-pro}
\end{figure}

We consider the collisionless cases first; both isolated and orbiting subhalos in CDM.  Our results are shown in the time evolution plot Fig. \ref{fig:mp-cdm}. We show the results with highest resolution in black lines, which is also the default resolution we use for the production runs in the results section. We then increase the DM particle mass (thus lower resolution) by 5 and 25 times respectively, as shown in red and purple lines. We assign star particle mass as either equal to DM, in thick lines, or 20\% of that of DM, in thin lines. The test halos in isolation are shown by solid lines. For the orbiting case, for the purpose of testing extremes, we put the test halos in the orb4 orbit (see Table \ref{table:orb}), as shown by dash-dotted lines, which is the most plunging orbit in our production runs.  From top to bottom, we show the galaxy half-light radius, stellar mass that remains bound to the system, the total (star+DM) bound (sub)halo mass, and the enclosed total mass within the initial half-light radius from the (sub)halo center. Throughout this work, the stellar half-light radius $r_{\rm half}$ is determined by first identifying the stellar particles that remain bound to the (sub)halo with the \texttt{AHF} \cite{ahf} halo finder, then finding the galaxy center by iteratively shrinking the sphere that encloses these bound stars, and setting $r_{\rm half}$ to the median value of the star particles' distance to the galaxy center. 

We describe the results of the isolated CDM simulations first.  In the top panel for $r_{\rm half}$, galaxies in isolated halos (solid lines) increase in size over time, as a result of two-body relaxations among particles near the center \cite{binney02, power03, bt08, ludlow19a}. Our highest resolution runs (black solid lines) have the lowest amount of this spurious diffusion, $<10\%$ increase in half-light radius, while the galaxies in the lowest resolution runs (light-purple solid lines) puff up by more than a factor of 2. Our mid-resolution runs (red solid lines) show a $\sim 20\%$ increase in galaxy size. 

When comparing different particle mass ratios we find that, at each resolution, assigning an equal mass ratio among the two particle species (stars and DM) produces a more stable system with less spurious diffusion than the 1:5 mass ratio among stars and DM which is common in cosmological simulations (see thin vs thick solid lines, all colors/linestyles). As a reference, we plot the stellar density profiles of isolated CDM systems for each mass resolution and particle mass ratio at the end of the simulation in Fig. \ref{fig:mp-cdm-pro}. Here we also include the initial stellar profiles (black dotted line), which ideally the end-state simulations should still follow if free from numerical artifacts. The corresponding half-light radii are shown in the vertical lines for each simulation. We can see that nearly all simulations experience spurious diffusion in the stellar distribution to some extent.  The high-resolution run with $8.5\times10^6$ particles and equal particle mass suffers the least. The regions shaded in different colors in Fig. \ref{fig:mp-cdm-pro} correspond to where the radii are less than three times the gravitational softening lengths\footnote{We determine the gravitational softening length of DM particles $\epsilon_{\rm dm}$ according to Eqn. (20) and (21) of \cite{vdb18}, for the high-resolution run. We then scale $\epsilon_{\rm dm}$ by a factor of $5^{1/3}$ when setting $\epsilon_*$ for stars with 1:5 mass ratio, or when scaling down DM to the mid-resolution ($25^{1/3}$ for low-resolution). The corresponding $\epsilon_{\rm dm}$ are listed in Table \ref{table:reso}. } of corresponding resolutions, which is conventionally denoted as the ``untrustworthy'' scale for N-body simulation \cite{power03}. The high-resolution stellar density profile evolves to be shallower within the three-times-softening region, but is robust above this scale. The mid- and low-resolution runs show a similar behavior, only with slightly worse robustness above their three-times-softening scales.  
This again highlights that the required particle resolution should be high enough such that the corresponding three-times-softening radius is well below the half-light radius of the galaxy. 

For the orbiting systems with CDM, there is now tidal mass loss in stellar components, as we show with the remaining bound fraction of stellar mass in the second panel of Fig. \ref{fig:mp-cdm}. We can see that lower resolution runs have faster stellar mass loss, because the stellar particles diffuse out to larger radii as we have discussed in the isolated cases, and become less resistant to tidal fields. Similar to the galaxy size in isolation, our low-resolution results deviate substantially from the high-resolution runs, while the mid-resolution runs indicate convergence with the high-resolution runs. Comparing the choice of 1:1 vs. 1:5 particle mass ratios, we again find better converged results with the equal mass ratio. For another metric, the galaxy size of orbiting systems in the top panel of Fig. \ref{fig:mp-cdm}, the convergence tests show similar results. Our mid- and high-resolution runs are converging to a similar decrease in $r_{\rm half}$, while the low-resolution run still has a comparable $r_{\rm half}$ to its initial condition, despite the significant tidal mass loss.  


The total bound mass (DM+stars) is much less sensitive to the resolution than the stellar properties.  In the third panel of Fig. \ref{fig:mp-cdm}, we show the total bound mass of the (sub)halo. We can see that the mid- and high-resolution results are indistinguishable, even when the mass loss is $99.9\%$, and the low-resolution run also shows good convergence until the stripped mass reaches more than $99\%$. This, in contrast with the not-as-good convergence we find for stellar properties, highlights that parameter choices previously determined to be suitable for subhalo-wide properties are not automatically guaranteed to be sufficient when the satellite galaxy is to be resolved. 

Given the requirement of high resolution to accurately track the evolution of stellar properties as we have just shown, an alternative ``tidal track'' approach has been proposed as a substitute for direct simulations with star particles \cite{Penarrubia07, errani15}. The idea is to use a less demanding quantity (one that is more robustly measured when simulated with lower resolution) as a proxy to anchor the evolution of stellar properties \cite{sales19}. Such an ``anchor'' quantity is chosen to be the total enclosed mass within the initial half-light radius $r_{\rm half,0}$ in \cite{errani15}. In this work, we follow this prescription and explore the tidal tracks for SIDM cases. Here we examine the convergence results of this $M(<r_{\rm half,0})$ in the bottom panel of Fig. \ref{fig:mp-cdm}. We can see that for both the isolated and orbiting CDM satellites, the low-resolution runs (purple lines) still significantly diverge from the high-resolution runs, even though $M(<r_{\rm half,0})$ is expected to be a more robust tracer. The mid-resolution runs (red lines), however, do achieve better convergence with the high-resolution runs (black lines) in the $M(<r_{\rm half,0})$ panel than in the $r_{\rm half}$ panel. For example, the mid-resolution runs have a maximum discrepancy of $\sim10\%$ (isolated) and $\sim20\%$ (orb4) in the top $r_{\rm half}$ panel when compared against the high-resolution runs (thick red solid vs. thick black solid, and thick red dash-dotted vs. thick black dash-dotted lines), while the corresponding contrast in the bottom $M(<r_{\rm half,0})$ panel is only $\sim5\%$ and $\sim10\%$ respectively. Considering the tidal track scaling relation in \cite{errani15} is $r_{\rm half} \sim M(<r_{\rm half,0}) ^\beta$ with $\beta \sim 1/3$, the reconstructed stellar properties would yield even smaller errors. Thus we can claim that the proposed tracer $M(<r_{\rm half,0})$ is more robust when the particle resolution is reduced, unless the resolution is overly low (as in our low-resolution examples).

\begin{figure}
    \includegraphics[width=0.97\columnwidth]{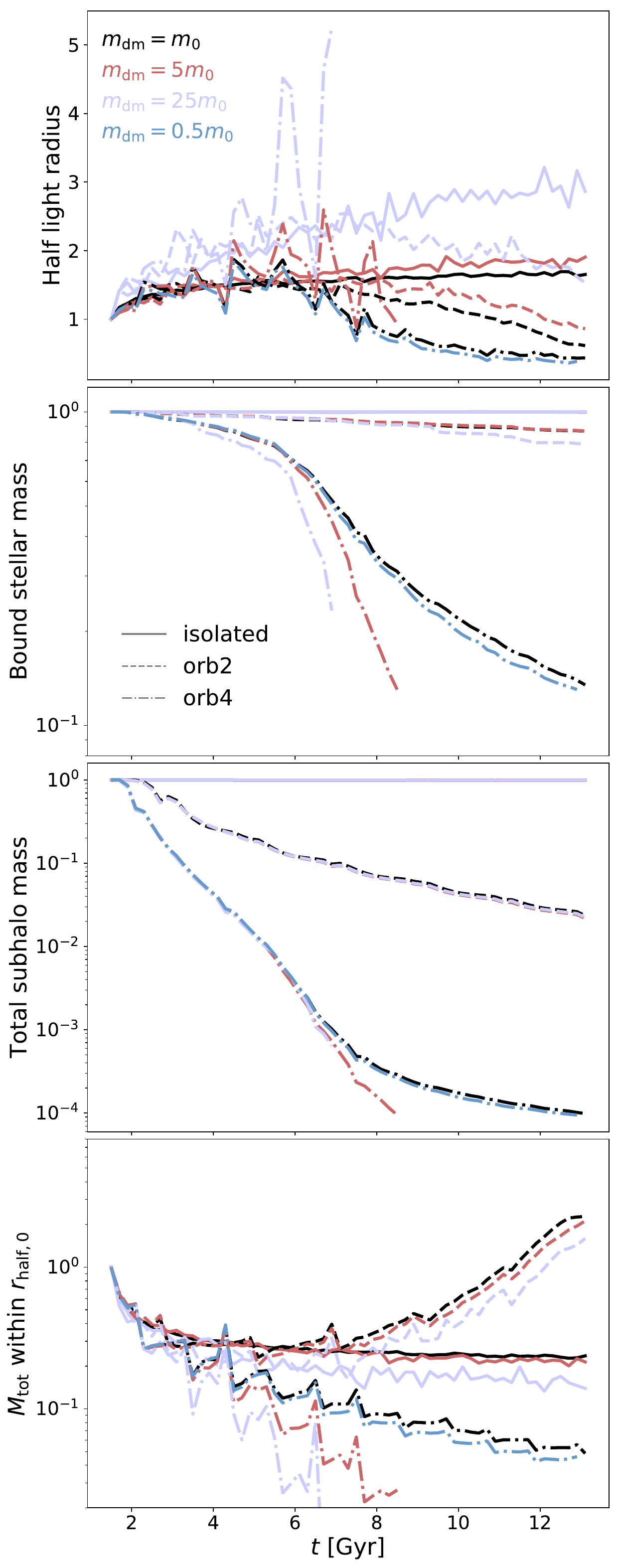}
    \vspace{-0.3cm}
    \caption{Convergence test for resolution with $\{\sigma_0=200, \omega=50\}$ (sub)halos and their galaxies. The panel series, colors for different DM resolutions, and line types for different orbits are identical to the resolution test figure of CDM in Fig. \protect\ref{fig:mp-cdm}. Note that since we have conducted the particle mass ratio test for CDM and found equal mass ratio to be a better choice, we only test the resolution convergence for SIDM here, without further testing the particle mass ratios. }
    \label{fig:mp-sidm}
\end{figure}

We pivot to the SIDM cases in Fig. \ref{fig:mp-sidm}, to determine if scattering changes the resolution criteria. Since we have concluded from our previous discussion of CDM cases that using an equal mass ratio between DM and star particles is preferred to avoid spurious mass segregation effects, here we only test different resolutions, fixing an equal mass ratio for all cases. We test three orbits for the SIDM model $\{\sigma_0=200, \omega=50\}$: a) isolated, in order to test SIDM halos in the cored regime; b) orb2, where the subhalo with this SIDM model core-collapses; c) orb4, where the subhalo suffers from extreme mass loss.  We find that the main results from resolution tests for CDM are still valid for SIDM: a) the highest resolution is needed to trace the evolution of stellar properties in all orbital cases--the low-resolution case is catastrophically wrong, and the mid-resolution simulation shows convergence with the high-resolution; b) the comparison between the mid- to high-resolution simulations indicates convergence in $r_{\rm half}$ for both the cored SIDM halo in isolation and the core-collapsing one on orb2, suggesting that the current high-resolution we use for tracing stellar properties should be sufficient for the core-collapse regime at least up to our termination of the simulation; c) the inner mass $M(<r_{\rm half,0})$ as the proposed tracer/proxy for tidal tracks has better convergence at lower resolutions compared to the stellar size $r_{\rm half}$ (e.g., red dashed vs. black dashed lines in bottom and top panels); d) when the subhalo's mass loss is extreme, e.g. $\gtrsim 99.9\%$ in the orb4 case in Fig. \ref{fig:mp-sidm}, even the inner mass $M(<r_{\rm half,0})$ can fail to remain robust in the mid-resolution run, and high-resolution is definitely needed. This is relevant for the formation of DM-free galaxies that we show in Sec. \ref{sec:time-evo-mlratio} and \ref{sec:track}. Note that given the non-converging behavior between our mid- and high-resolution runs in the extreme orbital mass-loss scenario, for this orb4 case with $\{\sigma_0=200, \omega=50\}$, we specifically prepare an ultra-high resolution simulation, reducing the particle mass further by a factor of two, as shown in the blue lines in Fig. \ref{fig:mp-sidm}. The results between our default high-resolution and this ultra-high resolution simulations now show good convergence, indicating that our default choice of $m_{\rm dm}=m_0$ in Table \ref{table:reso} is sufficient for the scope of this study. This result, that SIDM simulations do not require a higher resolution than CDM, is because the self-interaction timescale, $t_{\rm SI} \approx \frac{\lambda}{\sigma_v}=\frac{1}{\rho (\sigma/m)\sigma_v}$, does not explicitly depend on the particle resolution, unlike the gravitational relaxation timescale $t_{\rm relax}\approx\frac{N_p}{8\ln{(r_{\rm half}}/{\epsilon_{\rm soft}})} \frac{r_{\rm half}}{\sigma_v}$ \cite{bt08}.  

Based on the discussion above, we draw the following conclusions from our convergence tests. For dwarf galaxies in both CDM and SIDM (cored or core-collapsing) (sub)halos, our highest particle resolution ``$m_{\rm dm}=m_0$'' ($\sim 10^7$ particles, see Table \ref{table:reso}) is required to trace their evolution accurately. 
Our mid-resolution ``$m_{\rm dm}=5m_0$'' ($\sim 2\times 10^6$ particles) shows borderline-satisfactory convergence with $\sim 10\%$ error from the high-resolution results for stellar size. An unequal particle mass ratio of 1:5 between DM and star particles brings a second-order spurious effect on top of insufficient resolution. The proposed proxy for tidal track studies \cite{errani15}, the inner mass $M(<r_{\rm half,0})$ does display increased robustness compared to the stellar size $r_{\rm half}$ when simulated with (slightly) lower resolutions. Therefore, for the rest of this work, we choose the highest resolution we tested with equal mass ratio as the default choice for all our following production simulations.

\section{Results}\label{sec:result}

In this section, we present the results of satellite galaxy-subhalo co-evolution, varying DM models and orbits. We first show the time evolution of stellar properties such as the half-light radius $r_{\rm half}$ and the mass-to-light ratio (which we define as the inverse of stellar mass fraction) within $r_{\rm half}$ in Sec. \ref{sec:time-evo-rhalf} and Sec. \ref{sec:time-evo-mlratio}. Next, we present the tidal evolution tracks of these properties, as a function of the total mass enclosed in the initial half-light radius $M(<r_{\rm half,0})$ in Sec. \ref{sec:track}. As we showed in the previous section, this measure of satellite evolution is less sensitive to numerical resolution effects. Note that the satellites we analyze in Sec. \ref{sec:time-evo-rhalf} to \ref{sec:track} are generated with the Hernquist profile initial conditions. Comparison of the evolution tracks between Hernquist (cuspy)  and Plummer (cored)  stellar initial conditions is included in Sec. \ref{sec:hevspl}. We also test the robustness of these tidal track results by varying the initial stellar mass and size, as shown in section \ref{sec:trackvar}.

\subsection{Time evolution: half-light radius}\label{sec:time-evo-rhalf}

\begin{figure*}
    \centering
    \begin{subfigure}[t]{0.48\textwidth}
        \centering
        \includegraphics[width=\textwidth, clip,trim=0.2cm 0cm 0.2cm 0cm]{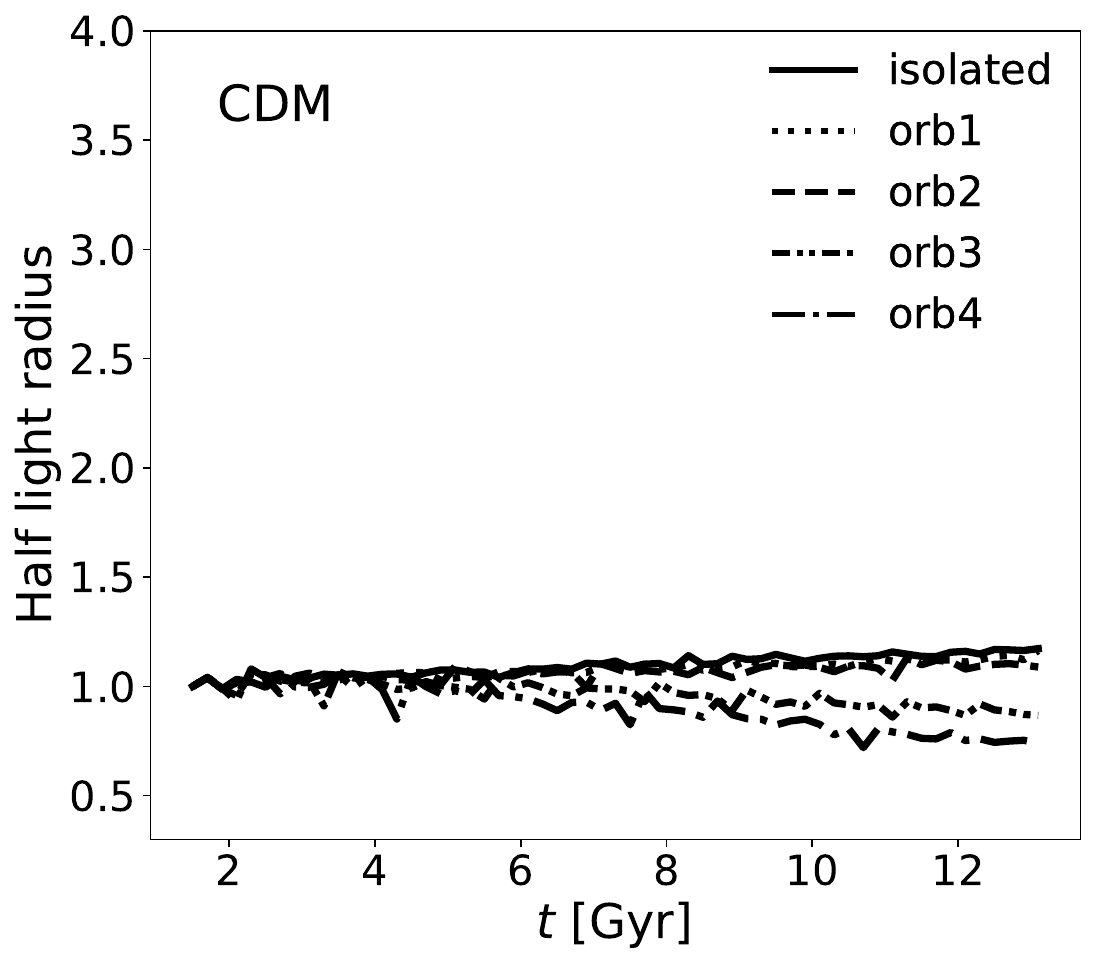}
        \caption{}
        \label{fig:rhalf-a}
    \end{subfigure}
    ~
    \begin{subfigure}[t]{0.48\textwidth}
        \centering
        \includegraphics[width=\textwidth, clip,trim=0.2cm 0cm 0.2cm 0cm]{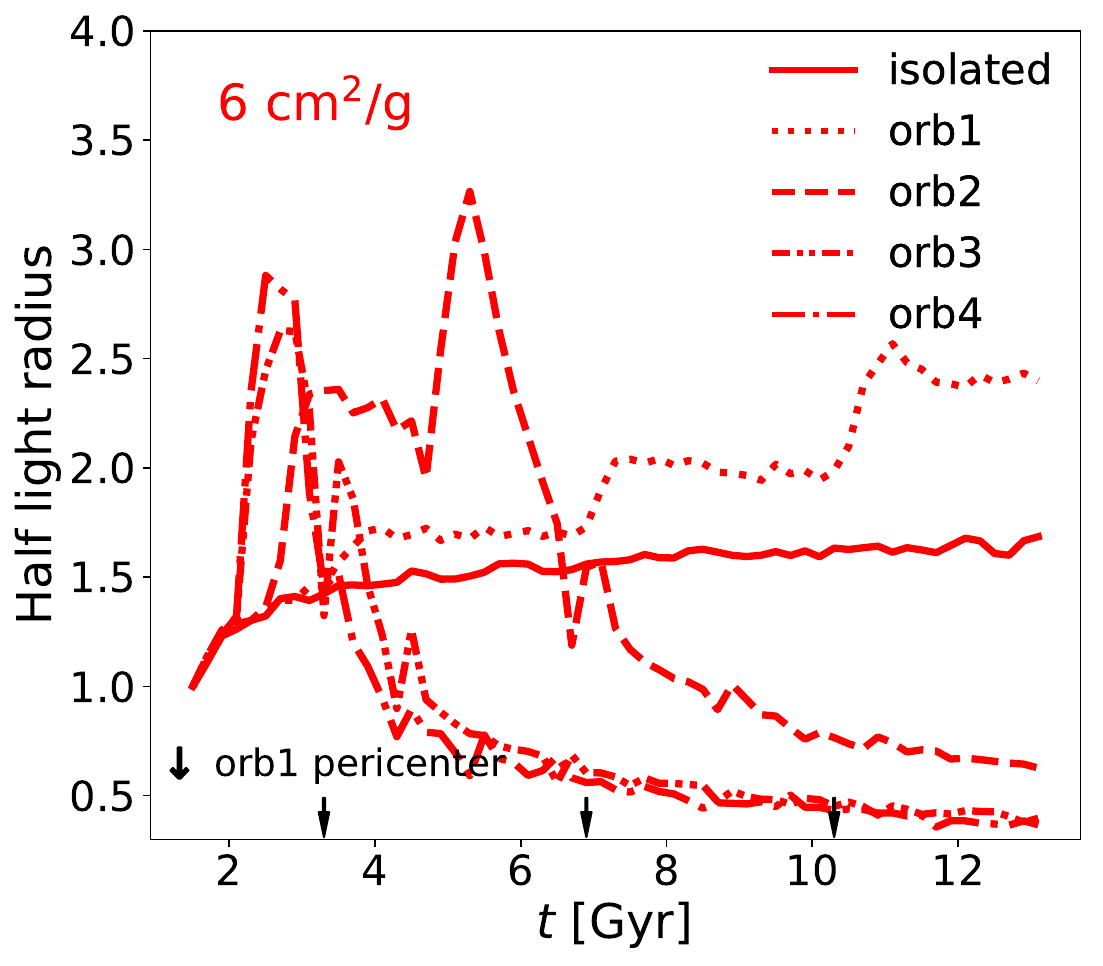}
        \caption{}
        \label{fig:rhalf-b}
    \end{subfigure}
    ~
    \begin{subfigure}[t]{0.48\textwidth}
        \centering
        \includegraphics[width=\textwidth, clip,trim=0.2cm 0cm 0.2cm 0cm]{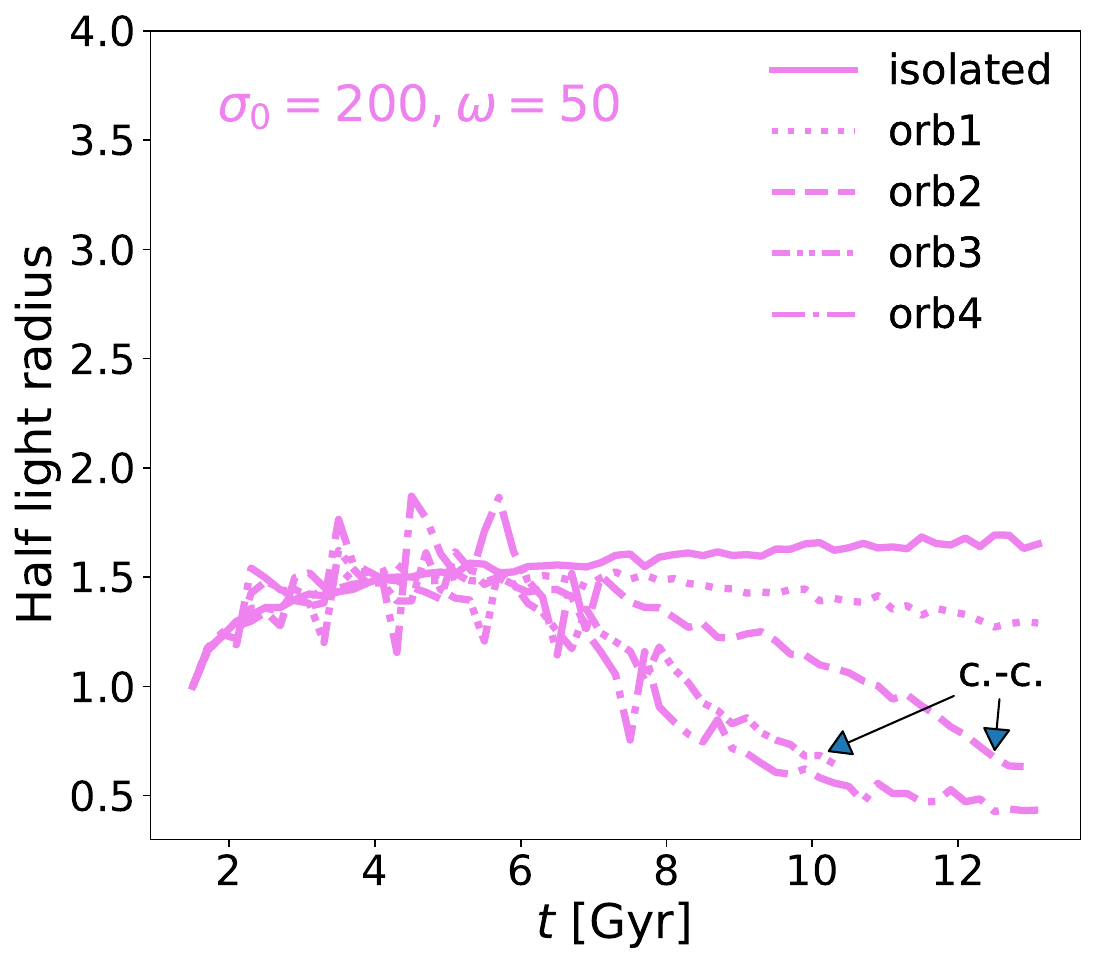}
        \caption{}
        \label{fig:rhalf-c}
    \end{subfigure}
    ~
    \begin{subfigure}[t]{0.48\textwidth}
        \centering
        \includegraphics[width=\textwidth, clip,trim=0.2cm 0cm 0.2cm 0cm]{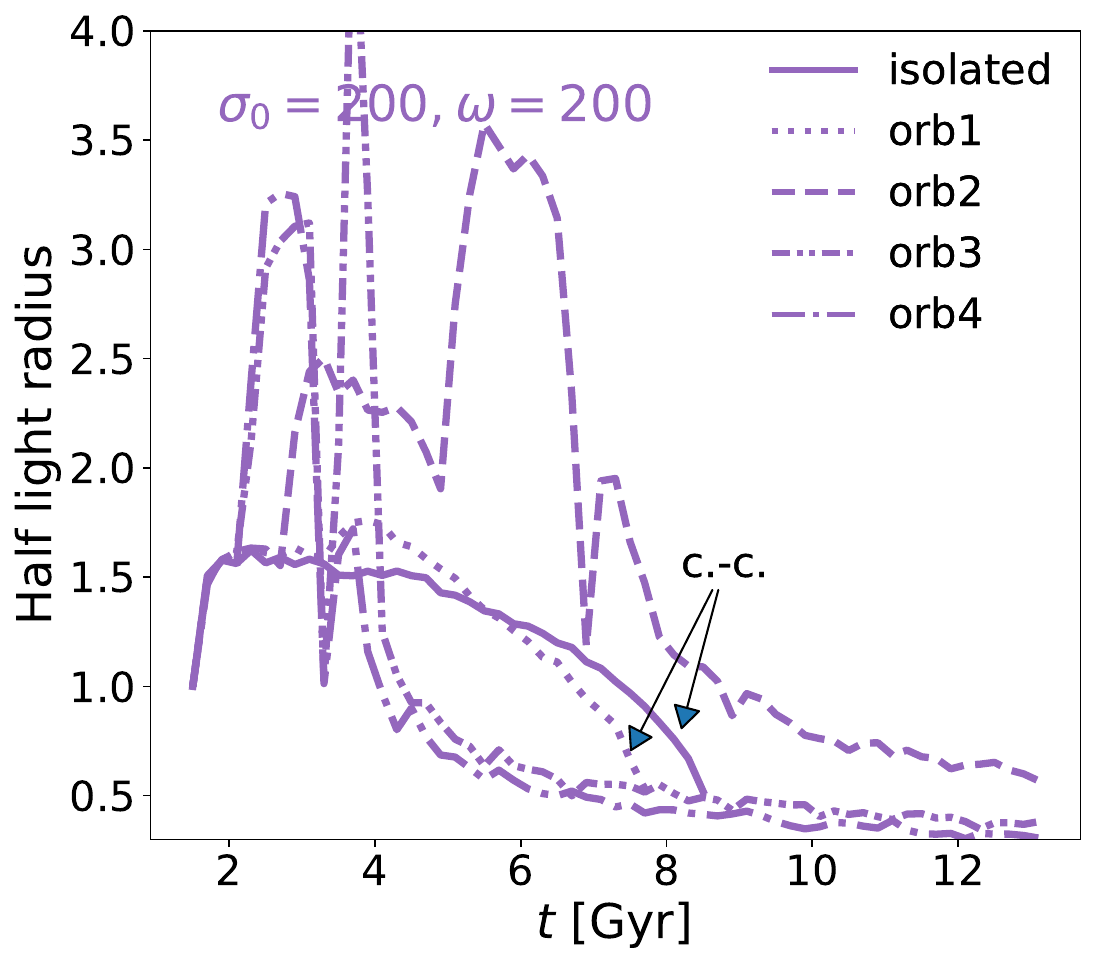}
        \caption{}
        \label{fig:rhalf-d}    
    \end{subfigure}
    \caption{The time evolution of the stellar half-light radius $r_{\rm half}$, normalized to its initial value, for the galaxies initialized with a Hernquist distribution. From a) to d) we show four DM models separately: CDM (black), constant cross section 6 $\rm cm^2/g$ (red),  $\{\sigma_0=200, \omega=50\}$ (pink) and  $\{\sigma_0=200, \omega=200\}$ (purple). Different line types represent different orbits, as labelled in each panel's legend. We also label the corresponding (sub)halos that reach our core-collapse threshold by $t=13.1$ Gyr with the `c.-c' notation in figures. In panel b) we also denoted the pericenter passage time of the orb1 satellites.}
    \label{fig:rhalf}
\end{figure*}

In this section, we investigate how the size of the satellite galaxy $r_{\rm half}$ evolves in time, as shown in Fig. \ref{fig:rhalf}. Through the combination of multiple SIDM models (see Fig. \ref{fig:sig-v}) and orbits (see Table \ref{table:orb}), we aim to explore the galaxy evolution in response to SIDM core-creation to core-collapse processes, and also with relatively weaker or stronger tidal/evaporation effects. The (sub)halos that reaches our core-collapse criterion are labelled as `c.-c.' in Fig. \ref{fig:rhalf}.

For the CDM case shown in Fig. \ref{fig:rhalf-a}, we show differences between isolated galaxies and satellites.  The isolated galaxy does not evolve (except for the $<10\%$ size increase that is caused by the lingering spurious diffusion we discussed in the previous section), as expected from the equilibrium initial conditions. 
The half-light radii of the galaxies with the largest pericenters (orb1 and orb2) hardly change, because of the outside-in nature of tidal stripping.  The tidal radii of these subhalos lie outside of the orbits of most of the star particles.  
For the more plunging orbits orb3 and orb4, the stellar mass loss becomes more obvious, $\sim\mathcal{O}(10\%)$, as the mass loss of the whole subhalo becomes extreme (see Fig.~\ref{fig:mp-cdm} for an example, where the subhalo has lost $>99\%$ of its initial mass on orb4), and the size of the stellar system also shrinks, as expected. Overall for these CDM cases, the galaxy half-light radius decreases with time, with a monotonically faster pace on orbits with smaller pericenters. Note that this behavior is slightly different when the star particles are initialized with a cored, Plummer profile, for which the tidal heating effect becomes more visible and causes the galaxy to puff up first before shrinking. We will discuss this further in Sec. \ref{sec:hevspl} (see also \cite{errani15, sanders18}).

However, the size evolution of the galaxy in SIDM scenarios is more complicated. Below we discuss galaxies in isolated SIDM halos first, then move on to satellites. When in isolation, SIDM halos go through the core-creation to core-collapse process without orbital effects. 
Of the three SIDM models we show in Fig. \ref{fig:rhalf}, two of the models (6 $\rm cm^2/g$ in Fig. \ref{fig:rhalf-b} and $\{\sigma_0=200, \omega=50\}$ in Fig. \ref{fig:rhalf-c}) have their isolated halos (solid lines) still in the core formation state at the end of the simulation.  The isolated halo with the $\{\sigma_0=200, \omega=200\}$ (solid line in Fig. \ref{fig:rhalf-d}) SIDM model reaches our core-collapse criterion at 8.7 Gyr (refer to Table \ref{table:ccp-orb} for all core-collapsed cases and the corresponding time $t_{\rm cc}$). 
In response, we can see that the sizes of the embedded galaxies in these three isolated halos (solid lines in Fig. \ref{fig:rhalf-b}, \ref{fig:rhalf-c} and \ref{fig:rhalf-d} respectively) all expand first, because the SIDM core-creation makes a less dense halo center and shallower potential, and the star particles migrate out in adiabatic expansion. The $\{\sigma_0=200, \omega=200\}$ isolated halo has a faster core-creation stage than the other two SIDM models, because the effective cross section for the typical velocities of our test halo (3$\times10^{10}M_\odot$) is larger in this model (see Fig. \ref{fig:sig-v}), hence the galaxy half-light radius increases faster and reaches its maximum at $t\sim2$ Gyr, while the galaxies in the other two models expand slowly to the maximal core toward the end of the simulation. After reaching the maximal core, for the isolated halo with the $\{\sigma_0=200, \omega=200\}$ model, the core-collapse process commences and the halo center core contracts, during which the galaxy size shrinks in accordance, until the halo hits our core-collapse criterion at $t=$ 8.7 Gyr and the simulation is terminated. We find that the galaxy expands to $\sim$1.6 times its initial size $r_{\rm half,0}$ at the maximally cored stage, and contracts to $\sim$0.5 time $r_{\rm half,0}$ when it is deep in the core-collapse phase. The galaxies in the cored isolated halos with 6 $\rm cm^2/g$ and $\{\sigma_0=200, \omega=50\}$ (solid lines in Fig. \ref{fig:rhalf-b} and \ref{fig:rhalf-c}) also expand $\sim 1.6$ times the original size, but with a much longer evolution time. This suggests that there may exist universal behavior in the galaxy size evolution during the core-expansion stage, as we discuss later in Sec. \ref{sec:track}.  

The core-collapsed isolated halo with the $\{\sigma_0=200, \omega=200\}$ model contracts to $\sim50\%$ its initial half-light radius, but we note that this should be viewed as the \textit{lower limit} of how compact a galaxy can be in this core-collapse case. This is because we have to cut off the simulation when a (sub)halo's central density grows to our core-collapse limit (see Sec. \ref{sec:method-overview}), but in reality the core-collapse should continue to proceed as a runaway process. The galaxy should keep contracting beyond that $\sim50\%$ point, unless/until the radius scale where core-collapse happens is much smaller than the galaxy size.  Such an example is discussed later in this section. Additional techniques/treatments beyond the particle-based N-body simulation, such as analytical frameworks \cite{sophia23} or fluid-like simulations, are needed for later stages of the core-collapse process when the scatterings are in the short-mean-free-path regime.

When SIDM physics within the subhalo is coupled with orbital effects from the host halo, the size evolution of the galaxy becomes much more complicated. Here we separate the discussion of SIDM satellite galaxies into two cases: with strong or weak evaporation effects.  We define the strong evaporation case as when the scattering cross section is at least $\sim \mathcal{O}(1)\ \rm cm^2/g$ at the host halo's orbital velocity scale $\mathcal{O}(10^2)\sim \mathcal{O}(10^3)\rm ~km/s$ (see Fig.~\ref{fig:sig-v} for the evaporation-relevant region in the parameter space; see also the Section IV C of \cite{zzc23}, where the authors varied the cross section only at the evaporation-velocity-scale to highlight its strength).  

The weak evaporation case is the case most commonly treated in the literature for subhalo evolution (e.g., \cite{sameie20}).  For the weak evaporation case with $\{\sigma_0=200, \omega=50\}$ in Fig. \ref{fig:rhalf-c}, the satellite galaxies in the subhalos' center first follow a universal expansion stage caused by SIDM core-creation in the  halo center, similar to the isolated galaxy-halo case.  As dark matter is gradually stripped from outer radii, the stellar system also starts to lose mass and shrink in size. There are two major differences with respect to the satellites in CDM subhalos: the maximum galaxy size before stellar mass loss is larger in SIDM cases, because of the SIDM core formation; and the speed of size shrinking is faster, because the SIDM core has a shallower potential and is less resistant to tidal stripping. As a result, we can see the time evolution curves of galaxy $r_{\rm half}$ in Fig. \ref{fig:rhalf-c} still line up in monotonic order of how close the orbits are from the host center, but with a larger spread than the CDM case in Fig. \ref{fig:rhalf-a}. Additionally, the $\{\sigma_0=200, \omega=50\}$ subhalos on orb2 and orb3 reach the core-collapse stage by the end of the simulation, causing contraction in the center of the satellite galaxy and thus accelerating the size shrinking in addition to the contribution of mass loss. The orb2 and orb3 subhalos core-collapse faster than the isolated or orb1 subhalo, because of the tidal stripping acceleration of the core-collapse process \cite{nishikawa20, zzc22, zzc23, neev23}. The orb2 and orb3 subhalos also core-collapse faster than the orb4 one, which inhabits an even closer orbit and experiences stronger tidal stripping. This seemingly counter-intuitive, non-monotonic result is because tidal heating also becomes more significant with closer orbits, injecting heat into the subhalo center and thereby countering the core-collapse process.  On orb4, the tidal-heating-deceleration outweighs the tidal-stripping-acceleration (see Fig. 6 of \cite{zzc22} for more examples and discussion). This delay in core-collapse for the orb4 case causes its mass loss to become large enough that the core never collapses, and it eventually results in a DM-deficient galaxy (see Sec. \ref{sec:time-evo-mlratio} for the mass-to-light ratio).  

When the subhalos are orbiting in a strong evaporation field, such as the 6 $\rm cm^2/g$ and $\{\sigma_0=200, \omega=200\}$ models (Fig. \ref{fig:rhalf-b} and \ref{fig:rhalf-d}), we observe a larger diversity in the size evolution history of satellite galaxies. Following the initial increase in size caused by SIDM core-creation, some of the galaxies show sharp periodic fluctuations in $r_{\rm half}$ (orb1 and orb2 in Fig. \ref{fig:rhalf-b} and orb2 in Fig. \ref{fig:rhalf-d}). This is because evaporation is strongest at pericenters (see Fig. 9 of \cite{zzc22} for an example of the correlation between evaporation strength and orbital distance), which leads to a nearly globally uniform DM mass loss from the whole subhalo including in the subhalo center.  This is unlike tidal stripping, which `peels' DM layer by layer.  During pericenter passages, the subhalo abruptly loses DM mass in its center, and the stellar particles migrate out, leading to larger $r_{\rm half}$ than in the isolated cases. This explains the step-function-like pattern for orb1 in Fig. \ref{fig:rhalf-b}, with each abrupt increase in half-light radius arising from a pericenter passage.  

For more plunging orbits in a strong evaporation field, the stellar size does not show such steady step-like features, only a temporary evaporation-induced size-boost followed by size-shrinking. For example, for orb2 in Fig. \ref{fig:rhalf-b} and \ref{fig:rhalf-d}, the evaporation mass loss is even more drastic than orb1 in Fig. \ref{fig:rhalf-b}, and the stellar system starts to be stripped by the tidal field soon, hence the drop in $r_{\rm half}$ after two pericenter passages. For even more plunging orbits, orb3 and orb4, we can see in Fig. \ref{fig:rhalf-b} and \ref{fig:rhalf-d} that the satellite galaxy only experiences one peak (first pericenter) in $r_{\rm half}$ from the evaporation size boost, before it decreases in size due to stellar mass loss. The speed of this size decrease slows down as $r_{\rm half}$ drops to $\sim 1/2$ of the initial value $r_{\rm half, 0}$. At this point, the DM component in the subhalo is nearly all lost and the DM-evaporation has no further influence, effectively producing a DM-free galaxy (see more discussion in Sec. \ref{sec:time-evo-mlratio}) which becomes more resistant to tidal stripping as the outer stars are stripped and only the relatively dense and cuspy distribution of stars in the galaxy center remains. A supplementary figure that correlates the co-evolution of $r_{\rm half}$, bound stellar mass and bound total mass is attached in Appendix \ref{appdx:suppfigs} as Fig. \ref{fig:rhalf-corr}, using the 6 $\rm cm^2/g$ satellites as an exmple. Note that Plummer satellites have slightly different evolution histories, as we will discuss in Sec. \ref{sec:hevspl}. 

\begin{figure*}
    \centering
    \begin{subfigure}[t]{0.48\textwidth}
        \centering        \includegraphics[width=\textwidth, clip,trim=0.2cm 0cm 0.2cm 0cm]{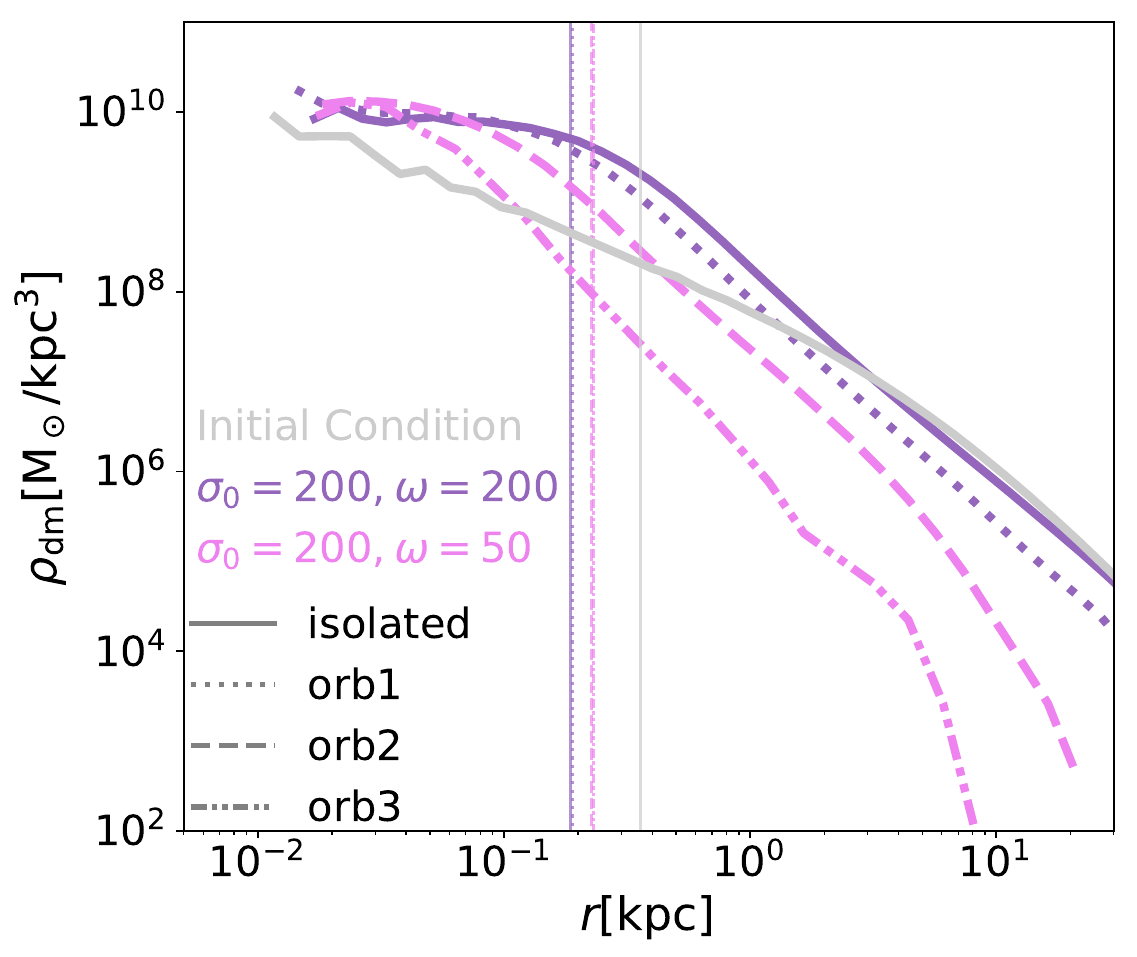}
    \caption{}
    \label{fig:ccp-dm-a}
    \end{subfigure}
    ~
    \begin{subfigure}[t]{0.48\textwidth}
        \centering        \includegraphics[width=\textwidth, clip,trim=0.2cm 0cm 0.2cm 0cm]{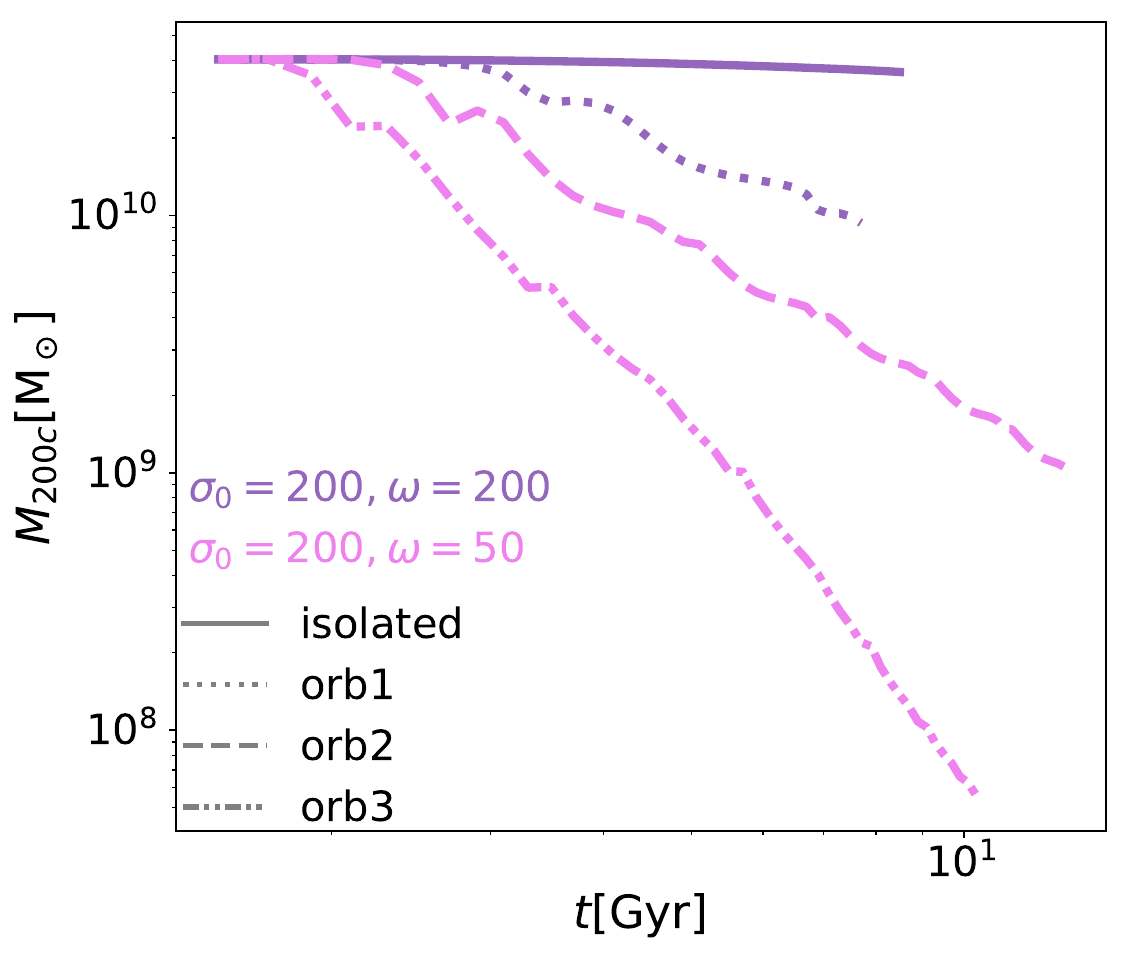}
        \caption{}
        \label{fig:ccp-dm-b}
        \end{subfigure}  
    ~
    \begin{subfigure}[t]{0.48\textwidth}
        \centering        \includegraphics[width=\textwidth, clip,trim=0.2cm 0cm 0.2cm 0cm]{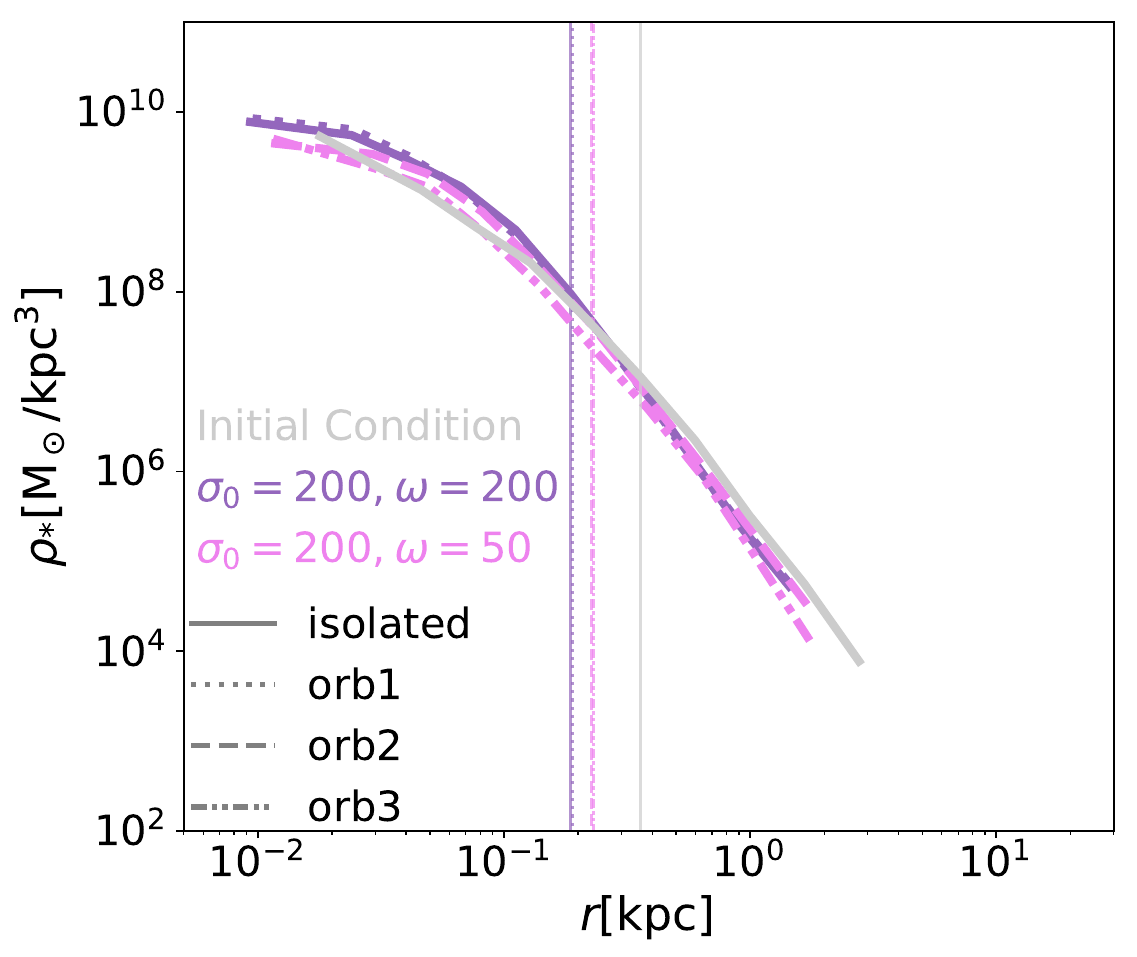}
        \caption{}
        \label{fig:ccp-dm-c}
        \end{subfigure}  
    ~
    \begin{subfigure}[t]{0.48\textwidth}
        \centering        \includegraphics[width=\textwidth, clip,trim=0.2cm 0cm 0.2cm 0cm]{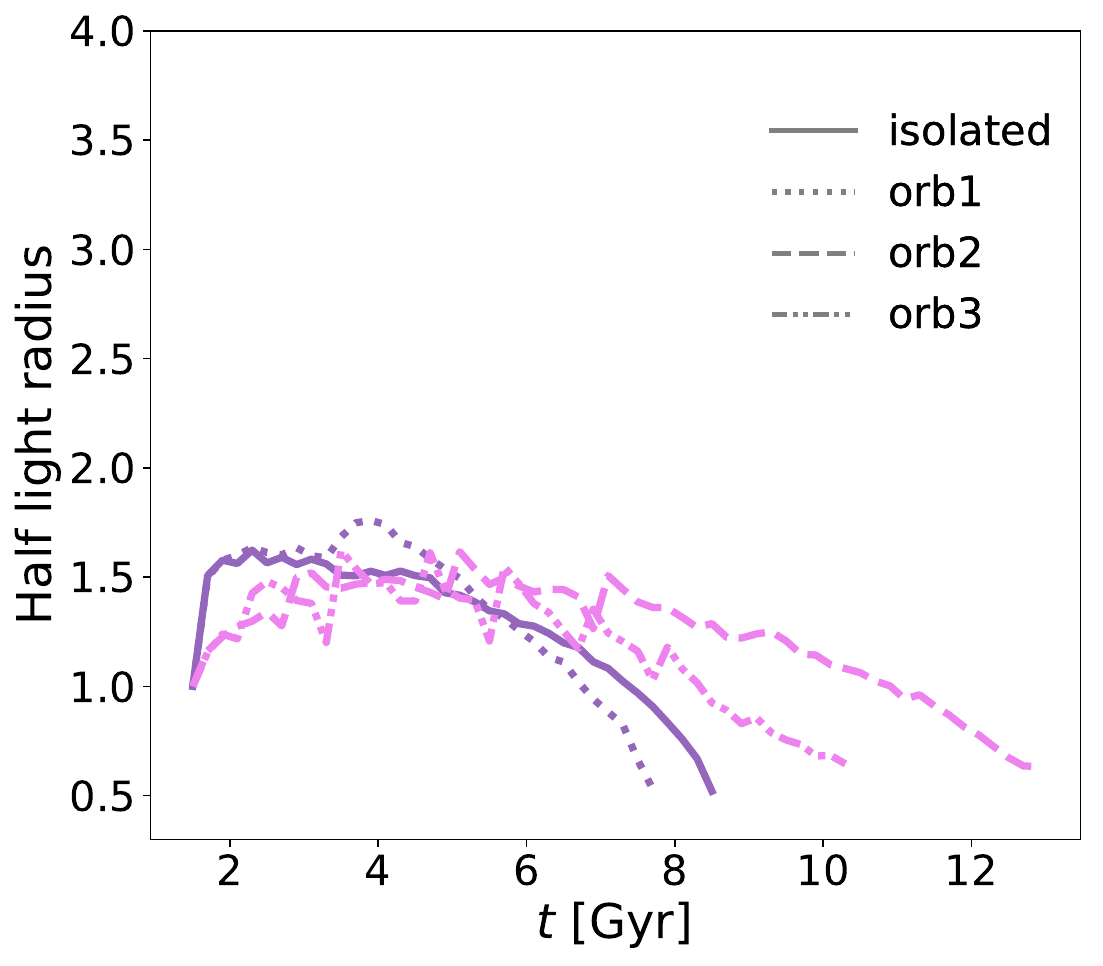}
        \caption{}
        \label{fig:ccp-dm-d}
        \end{subfigure}  
    \caption{For each of the four (sub)halos that we find core-collapsed (same as the `c.-c.' labelled ones in Fig. \protect\ref{fig:rhalf}): a) the DM density profile at the core-collapse time; b) time evolution of the bound mass; c) the stellar density profile at the core-collapse time; d) the time evolution of (normalized) half-light radius, which is the same as in Fig. \protect\ref{fig:rhalf} but replotted here for convenience. In a) and c) we also show the corresponding initial conditions in gray color, as well as the half-light radii in vertical lines. The colors representing SIDM models and line types for orbits are the same as in previous figures.}    
  \label{fig:ccp-dm}  
\end{figure*}

The path to core-collapse in subhalos can be diverse and complicated, because of the interaction with various host effects, as discussed in \cite{zzc22} (see their Fig. 8-11 for examples). Next we analyze the corresponding evolution of stellar size for these (sub)halos that reach our core-collapse criteria by the end of the simulation. There are in total four such cases, which we label as `c.-c.' in Fig. \ref{fig:rhalf} and list in Table \ref{table:ccp-orb}: orb2 and orb3 for $\{\sigma_0=200, \omega=50\}$ (weak evaporation), and isolated and orb1 for $\{\sigma_0=200, \omega=200\}$ (strong evaporation). In most of these core-collapsed cases, we observe significant speed-up in the reduction of $r_{\rm half}$, in response to the runaway, self-accelerating core-collapse process. The exception is the orb3 case of $\{\sigma_0=200, \omega=50\}$, where the runaway feature is less obvious. To investigate this exception, we plot the DM density profiles of these four core-collapsed subhalos at their last snapshots (each corresponds to the $t_{\rm cc}$ in Table \ref{table:ccp-orb}) in Fig. \ref{fig:ccp-dm-a}. We can see that the innermost density bins for all these (sub)halos have indeed grown to be denser than the initial CDM profile, which confirms the core-collapse.  However, the radial scale that is relevant for the core-collapse physics can be different among different cases.  For the three cases in which we observe the sharp reduction of $r_{\rm half}$, their DM core size is comparable to $r_{\rm half}$.  Thus, as the core collapses, the stellar component also become more compact because stars are gravitationally entrained with the DM. By contrast, the orb3 case of $\{\sigma_0=200, \omega=50\}$ has a core size of order $r\sim10^{-2}$ kpc, about an order of magnitude smaller than the stellar size. The distribution of the outer stars is thus not affected by the DM core contraction happening at much smaller radii.  This difference in scale arises from mass loss, as shown in Fig. \ref{fig:ccp-dm-b}. This halo loses much more mass during the orbital evolution than in the other three cases, before it eventually core-collapses. Therefore, from this set of comparisons we can conclude that when core-collapse happens in a (sub)halo, there can be corresponding features reflected in the evolution of the embedded galaxy, such as the sharply decreasing $r_{\rm half}$ (or the sharply increasing mass-to-light ratio, as we will show in the next section), but only when the radius scale of core-collapse is comparable to or larger than the size of the galaxy.

Here we briefly summarize the main takeaways that we find in the time evolution of satellites' $r_{\rm half}$: a) as the SIDM core forms and expands, $r_{\rm half}$ gradually increases to a similar maximal amount for all the SIDM models and orbits we have tested (when host-evaporation is not present); b) in a non-negligible evaporation field, the satellite galaxy further receives a boost in size, which is correlated with pericenter passages; this boost in $r_{\rm half}$ may be only temporary, and turn into a sharp decrease of $r_{\rm half}$ later, if the evaporation mass-loss is significant; c) the reduction of $r_{\rm half}$ can be caused by either core-collapse or stellar mass loss, with different shapes of the evolution curves: for the mass loss cases, the rate of decrease in $r_{\rm half}$ gradually slows down and eventually forms stable DM-free galaxies; for the core-collapse cases, the rate of decrease in $r_{\rm half}$ is accelerated over time, unless there is significant mass loss in the inner halo that results in the core-collapse region being much smaller than the galaxy size.   

\subsection{Time evolution: mass-to-light ratio}\label{sec:time-evo-mlratio}

\begin{figure*}
    \centering
    \begin{subfigure}[t]{0.48\textwidth}
        \centering
        \includegraphics[width=\textwidth, clip,trim=0.2cm 0cm 0.2cm 0cm]{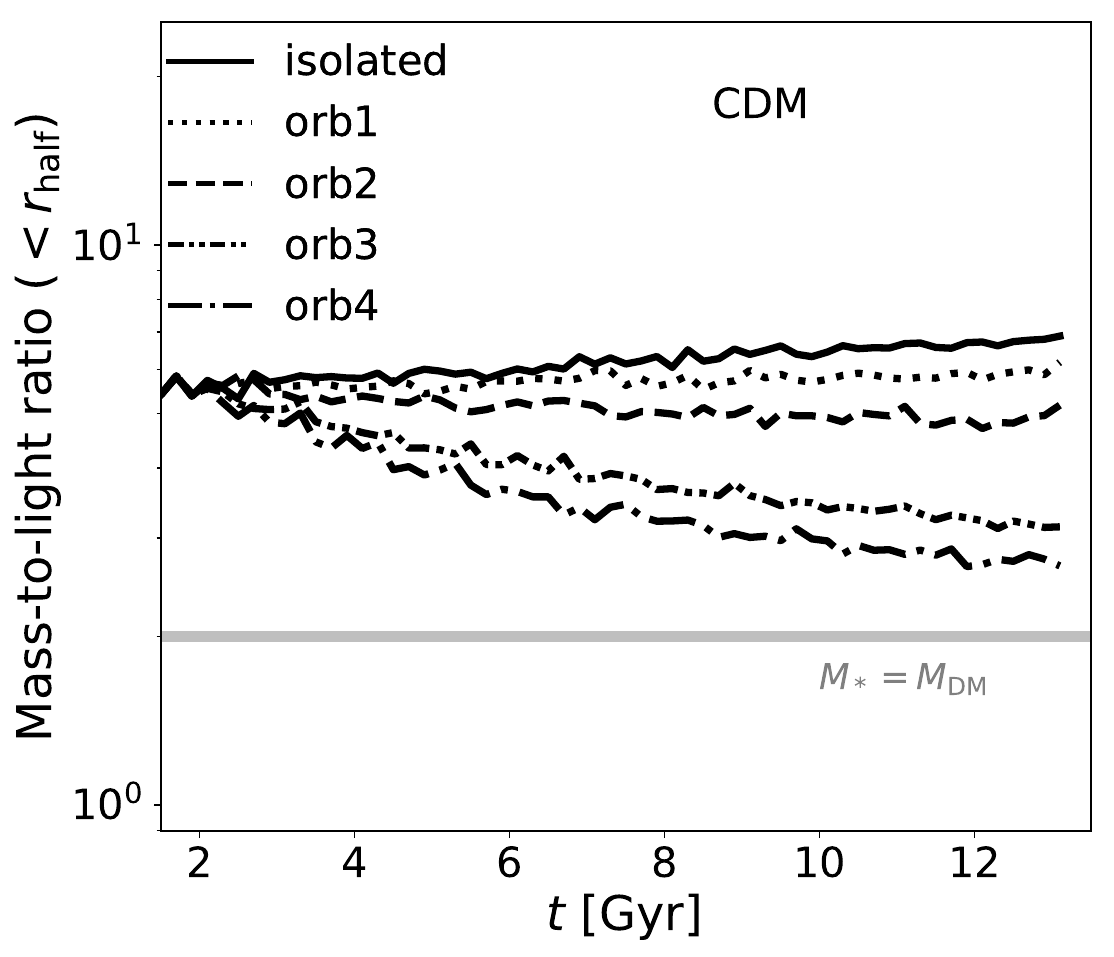}
        \caption{}
        \label{fig:ml-a}
    \end{subfigure}
    ~
    \begin{subfigure}[t]{0.48\textwidth}
        \centering
        \includegraphics[width=\textwidth, clip,trim=0.2cm 0cm 0.2cm 0cm]{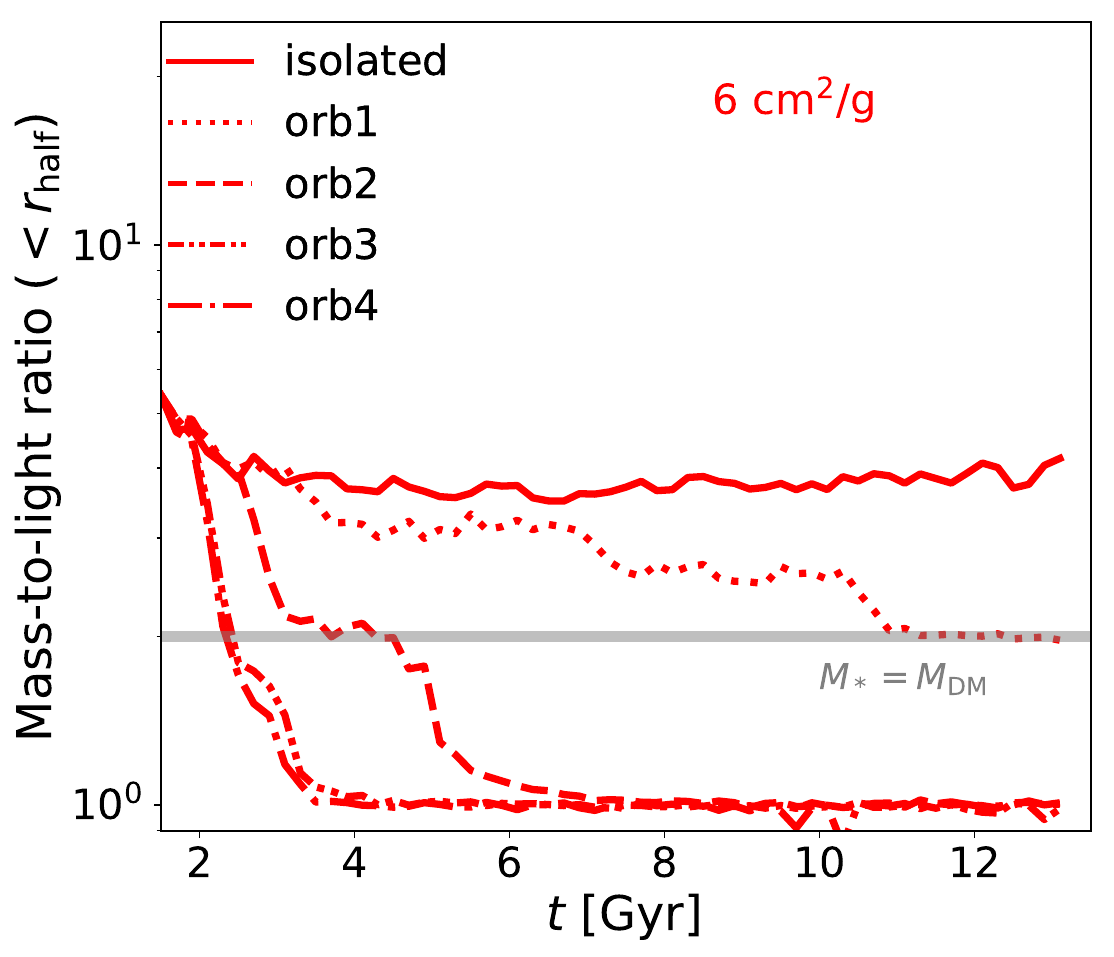}
        \caption{}
        \label{fig:ml-b}
    \end{subfigure}
    ~
    \begin{subfigure}[t]{0.48\textwidth}
        \centering
        \includegraphics[width=\textwidth, clip,trim=0.2cm 0cm 0.2cm 0cm]{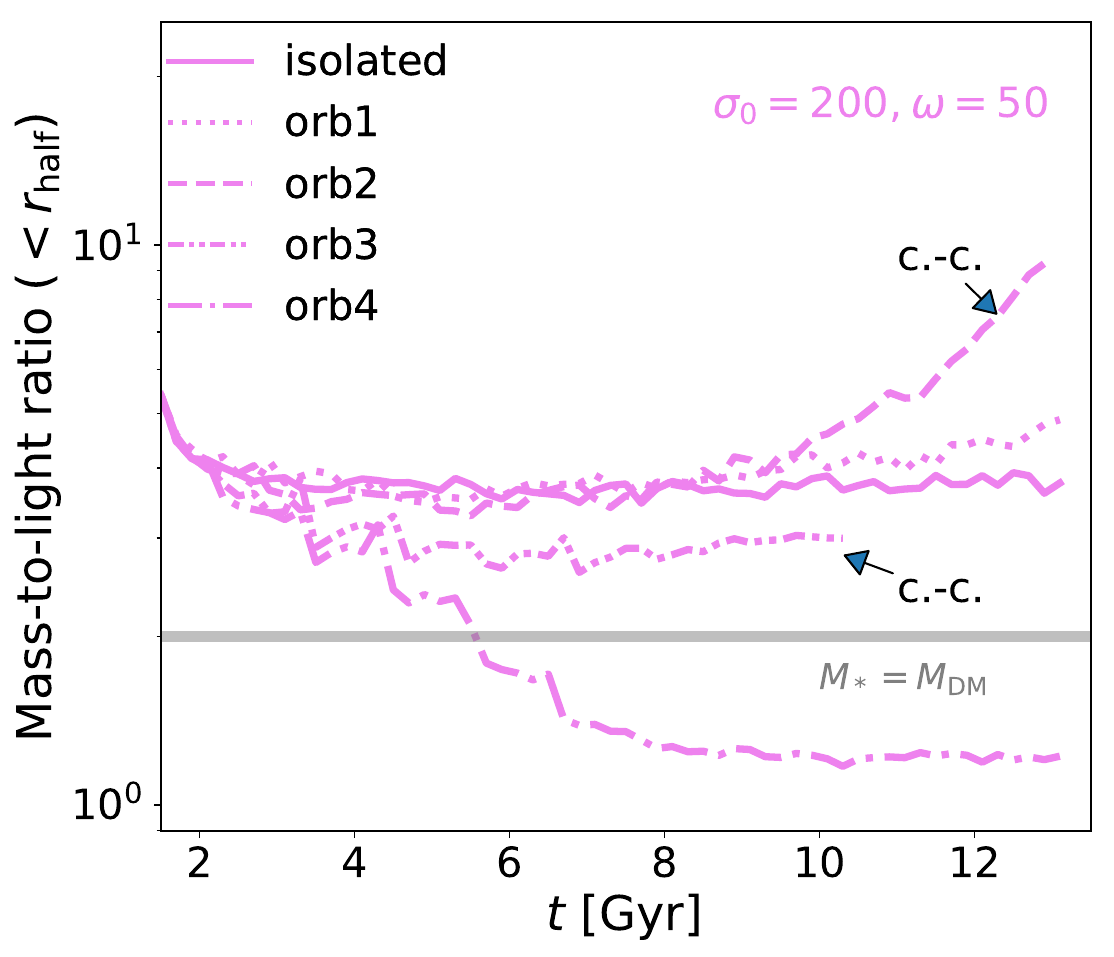}
        \caption{}
        \label{fig:ml-c}
    \end{subfigure}
    ~
    \begin{subfigure}[t]{0.48\textwidth}
        \centering
        \includegraphics[width=\textwidth, clip,trim=0.2cm 0cm 0.2cm 0cm]{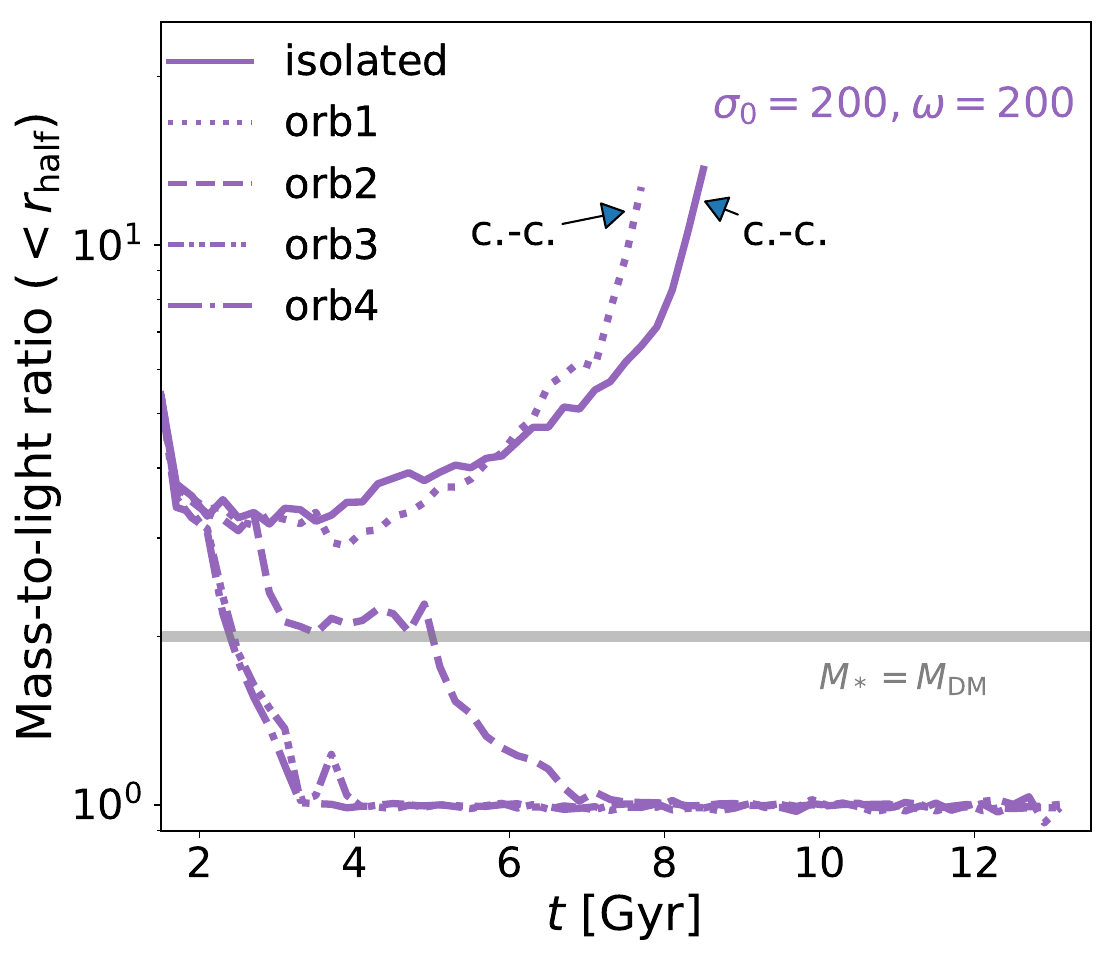}
        \caption{}
        \label{fig:ml-d}    
    \end{subfigure}
    \caption{Similar to Fig. \ref{fig:rhalf}, but the time evolution of the mass-to-light ratio within $r_{\rm rhalf}$ for the galaxies initialized with a Hernquist distribution. The mass-to-light ratio is defined as the ratio between the total mass (star+DM) and the stellar mass (within the assigned aperture). We label the corresponding (sub)halos that reach our core-collapse threshold by $t=13.1$ Gyr with the `c.-c' notation in figures. We also mark the horizontal line of mass-to-light ratio = 2, or $M_*(<r_{\rm half})=M_{\rm DM}(<r_{\rm half})$, below which the stellar component outweighs DM, and the satellite can be viewed as a DM-deficient galaxy. When mass-to-light ratio drops to $\sim$ 1, the satellite is nearly DM-free and becomes a self-bound galaxy/star-cluster.}
    \label{fig:ml}
\end{figure*}

In this section, we focus on the time evolution of the satellite's mass-to-light-ratio as another possible observable which has direct relevance to recently discovered DM-deficient galaxies \cite{df2, df4, df4shen} and DM-dominated ultra-faint dwarfs \cite{simon2019}. We define the mass-to-light  ratio as the ratio between the total mass (DM+stars) and the stellar mass within the $r_{\rm half}$ of the satellite galaxy.

For the CDM (sub)halos in Fig. \ref{fig:ml-a}, we observe a trend similar to the evolution of $r_{\rm half}$: the galaxy system in isolation stays at the initial state, which is expected given our numerical convergence tests in Sec. \ref{sec:res},  and the satellites have mass-to-light ratio declining over time, at a faster pace for more plunging orbits from orb1 to orb4. This is because as star particles are gradually peeled off, the satellite's $r_{\rm half}$, within which we calculate the mass-to-light ratio, shrinks to smaller radii (see Fig. \ref{fig:rhalf-a} for a recap), where the stellar density contributes more to the total mass density than at larger radii.   

The stellar component in SIDM systems always evolves with time, regardless of whether or not the system is isolated. During the core-creation phase, both the DM and stellar components in the center expand to larger radii, and also $r_{\rm half}$ increases. The compound result of these effects appears to be a $20\sim30\%$ drop in the mass-to-light ratio within $r_{\rm half}$, as we can see in the solid lines in Fig. \ref{fig:rhalf-b} to \ref{fig:rhalf-d}. The isolated galaxies whose SIDM halos do not core-collapse by the end of the simulation maintain the mass-to-light ratio at the core-creation stage (solid lines in Fig. \ref{fig:rhalf-b} and \ref{fig:rhalf-c}), while in the core-collapsing case (solid line in Fig. \ref{fig:rhalf-d}), we can see that the mass-to-light ratio is sharply increasing as core-collapse happens. This is the result of DM collapse toward the halo center during this runaway process.  Stars are gravitationally entrained toward the center, but in a less intensive manner than the DM, which we previously saw in the half-light radius evolution (see the previous section).

For the SIDM subhalos that remain cored, the satellite mass-to-light ratio decreases over time as $r_{\rm half}$ shrinks (see Fig. \ref{fig:rhalf-c}) when the evaporation effect is weak (Fig. \ref{fig:ml-c}). When the evaporation is non-negligible (orb1 to orb4 in Fig. \ref{fig:ml-b} and orb2 to orb4 in \ref{fig:ml-d}), the reduction of mass-to-light ratio becomes even faster as DM is evaporated from the center of the subhalos. Thus we see more efficient formation of DM-deficient galaxies in Fig. \ref{fig:ml-b} and \ref{fig:ml-d} than Fig. \ref{fig:ml-c}, with the mass-to-light ratio falling below 2 (gray lines). Furthermore, the  satellites under strong evaporation eventually become DM-free, as the mass-to-light ratio falls to 1, and are completely bound by the self-gravity of stellar components. Note that, in our simulations, this is a result only for galaxies with a Hernquist initial profile, which is cuspy enough to be self-bound against the tidal field. We will see in Sec. \ref{sec:hevspl} and Fig. \ref{fig:hevspl-c} that galaxies with the Plummer stellar profiles will rapidly dissolve in the host once their DM is depleted.

For the SIDM subhalos that undergo core-collapse, the mass-to-light ratio within $r_{\rm half}$ decreases during the core-creation then rises as core-collapse is triggered, similar to the core-collapsed halos in isolation, in most cases. The unusual case that we discussed in the previous section, orb3 for $\{\sigma_0=200, \omega=50\}$, still presents peculiar behavior in the mass-to-light ratio, which does not have the sharp increase near core-collapse that we see for the other core-collapsing subhalos. This is again because the radius scale on which core-collapse physics is happening for this subhalo is much smaller than $r_{\rm half}$, as we have shown in Fig. \ref{fig:ccp-dm}. Hence, the core-collapse process of this perculiar satellite is not revealed in the mass-to-light ratio.

We briefly summarize our findings in the study of the mass-to-light ratio as follows: a) the mass-to-light ratio within $r_{\rm half}$ decreases during the SIDM core-formation stage; b) both CDM and cored SIDM subhalos have the mass-to-light ratio deceasing as they lose mass in the tidal field. This decrease can be strengthened by a non-negligible evaporation field. c) the only case where we find an increasing mass-to-light ratio is when core-collapse is triggered.    Note that the results we present here are for the galaxies initialized with a Hernquist profile. We discuss the Plummer profile galaxies further in Sec. \ref{sec:hevspl} and show their mass-to-light ratio evolution in Fig. \ref{fig:mlp}.

\subsection{Tidal tracks}\label{sec:track}

\begin{figure*}
    \centering
    \begin{subfigure}[t]{0.48\textwidth}
        \centering
        \includegraphics[width=\textwidth, clip,trim=0.2cm 0cm 0.2cm 0cm]{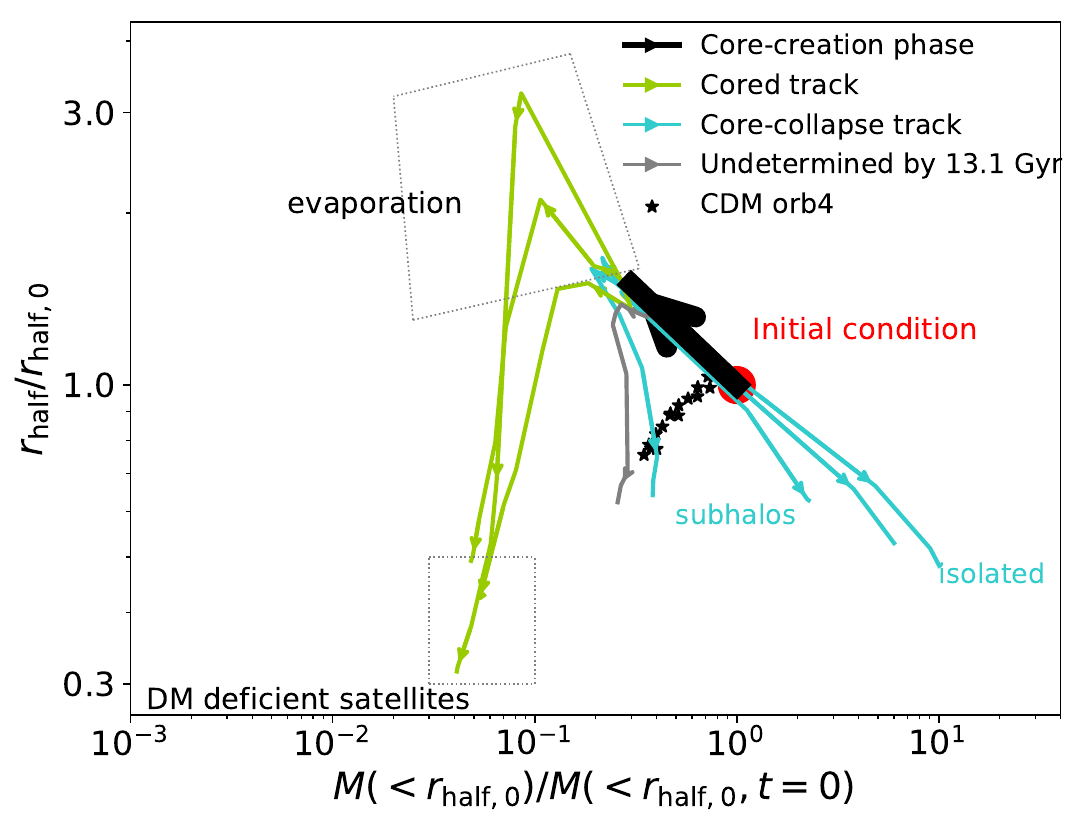}
        \caption{}
        \label{fig:track-a}
    \end{subfigure}
    ~
    \begin{subfigure}[t]{0.48\textwidth}
        \centering
        \includegraphics[width=\textwidth, clip,trim=0.2cm 0cm 0.2cm 0cm]{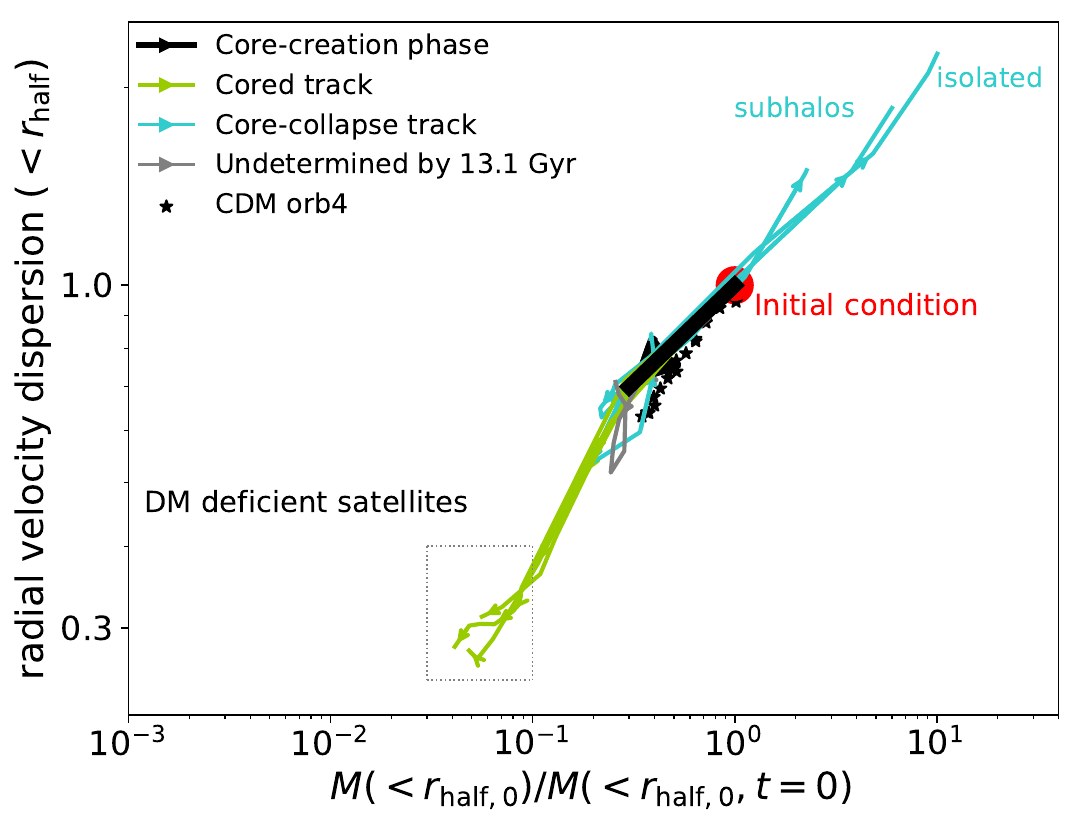}
        \caption{}
        \label{fig:track-b}
    \end{subfigure}
    ~
    \begin{subfigure}[t]{0.48\textwidth}
        \centering
        \includegraphics[width=\textwidth, clip,trim=0.2cm 0cm 0.2cm 0cm]{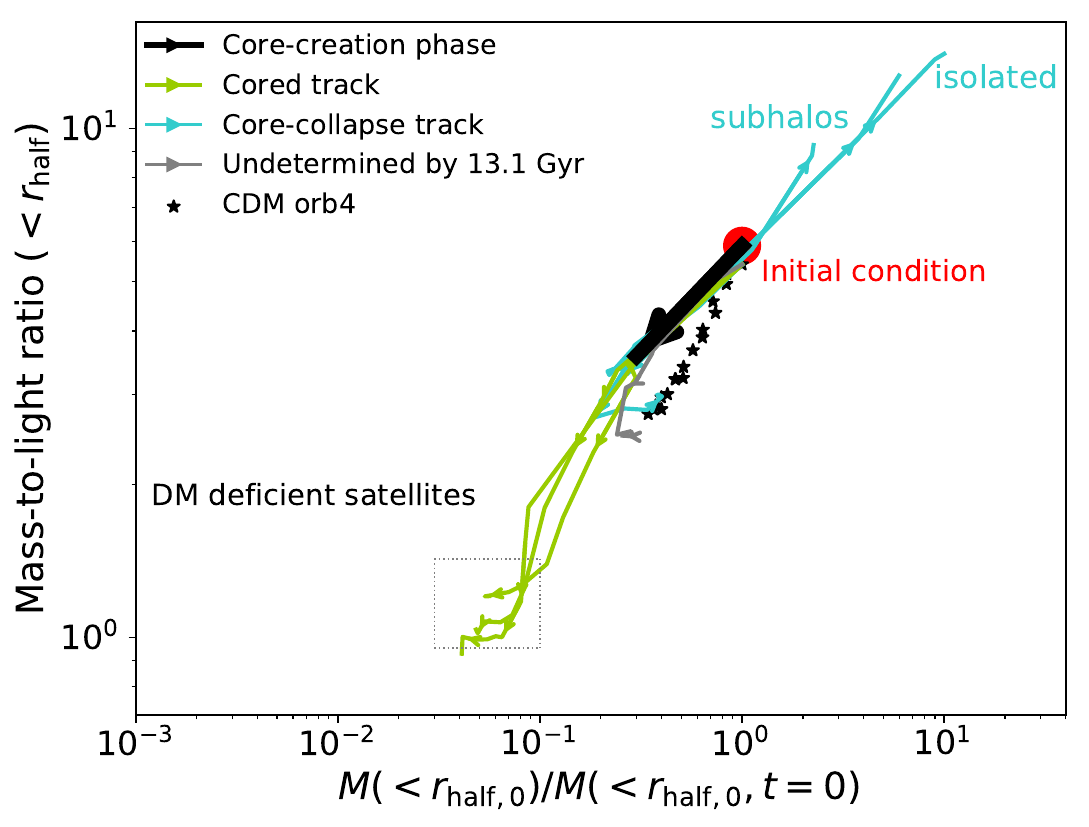}
        \caption{}
        \label{fig:track-c}
    \end{subfigure}
    ~
    \begin{subfigure}[t]{0.48\textwidth}
        \centering
        \includegraphics[width=\textwidth, clip,trim=0.2cm 0cm 0.2cm 0cm]{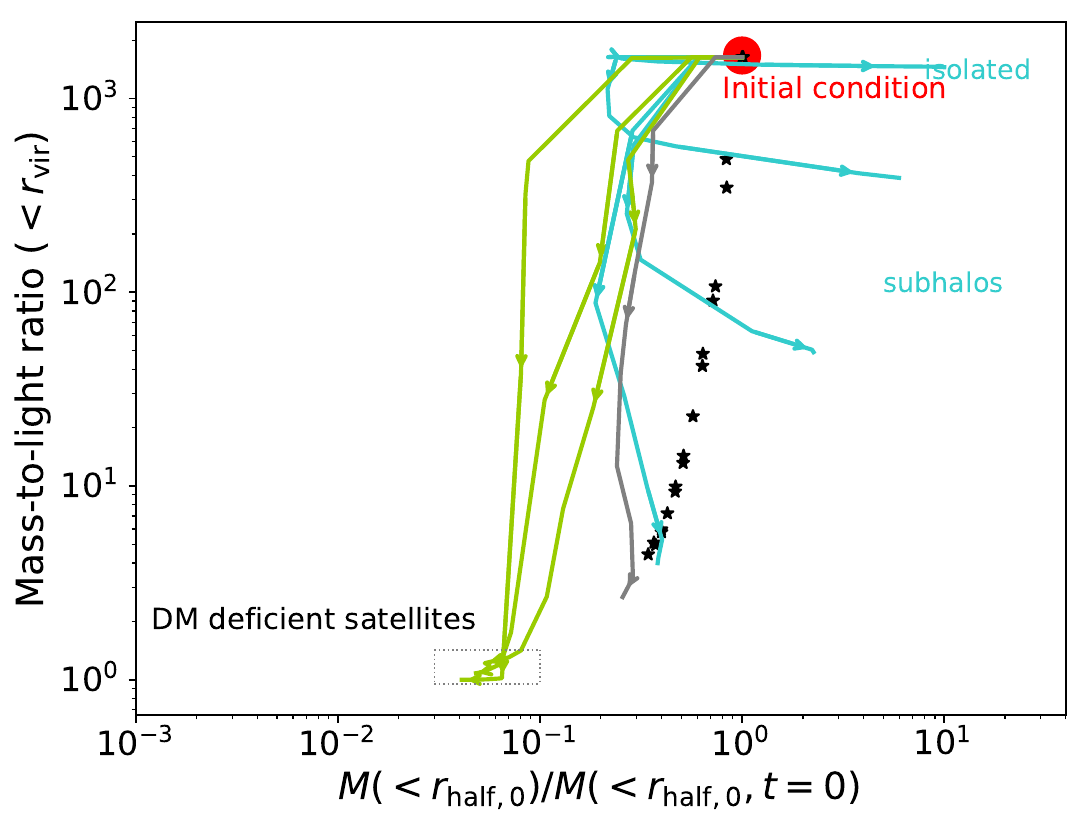}
        \caption{}
        \label{fig:track-d}
    \end{subfigure}    
    \caption{Tidal evolution tracks of: a) the (normalized) half-light radius $r_{\rm half}$; b) the (normalized) stellar radial velocity dispersion within $r_{\rm half}$; c) the mass-to-light ratio within $r_{\rm half}$; d) the the mass-to-light ratio within $r_{\rm 200c}$ of the (sub)halo. The x-axis tracer/anchor used for these tidal tracks is the total mass within the initial $r_{\rm half}$, thus $M(<r_{\rm half, 0})$, which is also normalized to its initial value. The initial condition in each diagram is labelled by a red dot. The black arrow represents a universal first stage of stellar component evolution during the SIDM core-creation phase. The green and cyan lines are evolution tracks of satellite galaxies that eventually fall into the cored or core-collapsed branches respectively. The black stars show the tidal track of the CDM satellite on orb4, our most plunging orbit. We also note the regions corresponding to the evaporation-induced size-boost and the formation of DM-deficient galaxies/star-clusters in the figure. Tidal tracks in this figure are from selected satellite galaxies in our simulations, for the purpose of demonstrating the main features. }
    \label{fig:trackmain}
\end{figure*}

In this section, we analyze the evolution of satellite galaxies' size, mass-to-light ratio, and radial velocity dispersion as a function of the total mass enclosed in the initial half-light radius $r_{\rm half,0}$ instead of time. We choose this metric for two reasons: First, we can have a better view of the correlation/co-evolution of galaxy properties and SIDM physics, as the change in total mass at small radii is mainly caused by DM evolution. Second, using our high-resolution, controlled simulations, we can examine the potential universality of these ``tidal evolution tracks'', and thus the possibility of building a relation between various galaxy properties and $M(<r_{\rm half,0})$, with the latter being less demanding in resolution. Such parametrized tidal tracks can then be used in larger-scale simulations with lower resolution (e.g., cosmological simulations) or analytical frameworks \cite{galacticus, dnyang23, sashimi-sidm} to anchor/predict the evolution of galaxies. These tidal-track types of studies have been conducted in the past for satellite galaxies with CDM \cite{Penarrubia07, Penarrubia10, errani15, carleton19, abenson22, xdu24}, with the parametrized relations applied in CDM cosmological simulations to infer statistics of dwarf galaxies \cite{sales19, carleton19}. Exploration of these tidal tracks is of even more importance for SIDM, as SIDM simulations incur a higher computational cost, especially in scenarios of core-collapse.   In this section, we showcase a few representative examples of tidal-tracks of SIDM satellites in Fig. \ref{fig:trackmain}, with a more complete version of all simulations shown in Appendix \ref{appdx:tidal-track-full}. We defer a more detailed analysis including the parameterization of these tidal tracks to future work, where a larger set of simulations will be needed.

In Fig. \ref{fig:track-a}, we show the evolution track of $r_{\rm half}$. All satellites start from the same initial condition labelled by the red dot, at $(1,1)$, since both $r_{\rm half}$ and the enclosed mass within the initial half-light radius $M(<r_{\rm half,0})$ are normalized to their initial values. All satellites then share a universal first-stage path in the $r_{\rm half}$ vs. $M(<r_{\rm half,0})$ plane, as highlighted by the thick black arrow, showing the self-similarity of SIDM satellites during the core-creation phase. 

Afterwards, these satellites bifurcate into either the cored branch (green), where eventually $r_{\rm half}$ decreases and $M(<r_{\rm half,0})$ decreases; or core-collapse branch (cyan), where eventually $r_{\rm half}$ decreases and $M(<r_{\rm half,0})$ increases. Depending on the evaporation strength, satellites on the cored branch can expand further during core-creation and experience an increase in  $r_{\rm half}$  before decreasing due to stellar mass loss, as we highlight with the `evaporation region' in Fig. \ref{fig:track-a}. This evaporation-boosted size corresponds to the peaks in $r_{\rm half}$ in Fig. \ref{fig:rhalf-b} and \ref{fig:rhalf-d}. 

Satellites on the core-collapse (cyan) branch reverse the evolution direction after reaching the maximal core,  marching towards the lower-right corner of the $r_{\rm half}$ vs. $M(<r_{\rm half,0})$ diagram. Of all the core-collapsing satellites (see Table \ref{table:ccp-orb}), we notice that the isolated case always has the largest $M(<r_{\rm half,0})$, since it experiences no mass loss and has the largest central region for core-collapse (see also Fig. \ref{fig:ccp-dm-a}). Moreover, in this diagram, this isolated satellite's core-collapsing phase is nearly completely overlapping with its core-creation phase. This suggests that the stars' response to the DM evolution can indeed be modelled as an adiabatic-expansion-then-adiabatic-contraction process in future analytical works (see also Zavala et al in prep.). The timescales are $\mathcal{O}(1)$ Gyr and $\mathcal{O}(10)$ Gyr for SIDM core-formation and core-collapse respectively, while the typical orbital timescales for stars are $2\pi r_{\rm half}/\sigma_v \sim \mathcal{O}(0.1)$ Gyr, further supporting the feasibility of analytically modeling this response process.

The remainder of the satellites on the core-collapse branch are aligned from right to left in the ascending order of mass loss within $r_{\rm half,0}$. The two core-collapsed `typical' subhalos next to the isolated one (the two middle cyan lines in Fig. \ref{fig:track-a}) behave similarly to it, with greater inner total mass and smaller $r_{\rm half}$ than the initial condition at the end of the evolution. For the most extreme case of mass loss where the subhalo still manages to core-collapse, the left-most cyan line in Fig. \ref{fig:track-a} which corresponds to orb3 of $\{\sigma_0=200, \omega=50\}$, we find  $M(<r_{\rm half,0})$ not increasing as drastically as other core-collapsing cases. Again, this is because the radius scale of core-collapse is much smaller than $r_{\rm half,0}$, as shown in Fig. \ref{fig:ccp-dm-a}. Just next to this left-most cyan line in Fig. \ref{fig:track-a}, we find a satellite (orb3 of $\{\sigma_0=100, \omega=50\}$, plotted with the gray line) that does not reach our core-collapse criterion by the end of the simulation, but also behaves unlike other `normal' cored satellites that land in the `DM-deficient galaxies' region. Instead, it appears rather similar to the left-most cyan line (orb3 of $\{\sigma_0=200, \omega=50\}$), still proceeding with an early core-collapse phase even with a large amount of mass loss, and may either eventually core-collapse given a longer evolution time, or further lose DM and become a DM-deficient galaxy.  We thus consider its evolutionary state  `undetermined'. This suggests that this marginal case of satellite may be near the boundary that separates the cored branch and the core-collapse branch in the diagram of Fig. \ref{fig:track-a}.

We show the evolution track of the star particles' radial velocity dispersion within $r_{\rm half}$ in Fig. \ref{fig:track-b}. Compared to the tidal evolution track of $r_{\rm half}$, we can see that the main features are similar: a universal core-creation phase for all satellites, and then the bifurcation into the cored branch or the core-collapse branch.  A major difference, however, is that the overall diversity in the tidal tracks of the velocity dispersion appears to be far smaller than that of the $r_{\rm half}$, with the cored branch and core-collapse branch nearly aligned along the same line. This is because the velocity dispersion is more directly determined by the enclosed mass. We also find that the (sub)halos on the core-collapse branch all have a late-time steepening in their slope in the velocity dispersion vs. enclosed mass plane in Fig. \ref{fig:track-b}. This is because the aperture for measuring velocity dispersion, $r_{\rm half}$, gets noticeably smaller than $r_{\rm half,0 }$, the aperture for enclosed mass in the x-axis. As core-collapse reaches late stages, the enclosed mass grows faster at the smaller radius $r_{\rm half}$ than at the larger, fixed radius $r_{\rm half, 0}$.  Hence, the velocity dispersion within $r_{\rm half}$, as a proxy of $M(<_{\rm half})$, grows faster than $M(<_{\rm half, 0})$ in the late stages of core collapse. 

We show the evolution track of the satellites' mass-to-light ratio within $r_{\rm half}$ in Fig. \ref{fig:track-c} and within the (sub)halo's $r_{\rm 200c}$ (only the bound materials are counted) in Fig. \ref{fig:track-d} respectively. The former is more relevant for observations, while the latter more relevant for halo-wide statistics that do not depend on accurate measurement of the stellar $r_{\rm half}$ and can thus be more robustly measured even in lower-resolution simulations. In Fig. \ref{fig:track-c}, we find a similar trend as for the tidal track of velocity dispersion in Fig. \ref{fig:track-b}. After the universal core-creation phase, the mass-to-light ratio of the cored subhalos  monotonically decreases until the floor of forming DM-free galaxies is reached. This is because the mass loss rate of cored DM near the center of subhalos is higher than that of stars, with or without evaporation, plus $r_{\rm half}$ continues to shrink to where the mass-to-light ratio is initially lower.   In the core-collapsing (sub)halos, on the contrary, the DM concentrates towards the center at a much faster rate than the stars, with the latter only dragged to the center as a response to the deepening gravitational potential caused by DM.  Thus, the mass-to-light ratio grows monotonically as core-collapse proceeds. 

As shown in Fig. \ref{fig:track-d}, the mass-to-light ratio within $r_{\rm 200c}$, however, behaves differently. This is because the mass-to-light ratio within the tidal radius is less directly correlated with the mass change in the inner radius $r_{\rm half, 0}$. There is no longer a universal first stage of evolution during core creation. The mass-to-light ratio within $r_{\rm 200c}$ for an isolated system remains at the initial value since there is no mass loss on the whole-halo scale. For subhalos, the mass-to-light ratio starts to drop as the tidal (and evaporation) effects kick in and DM is stripped away, when the subhalos can be at different stages in the core-creation phase.  Thus the subhalos' tracks in Fig. \ref{fig:track-d} have different ``peeling off'' points from the isolated case of the top horizontal line, which mark the corresponding inner mass $M(<r_{\rm half, 0})$ when tidal effects start to become significant.  The cored satellites then evolve to become DM-deficient galaxies, for which the mass-to-light ratio for the whole subhalo is $\sim1$, similar to Fig. \ref{fig:track-c}. For the core-collapsing subhalos, as core-collapse begins and inner mass $M(<r_{\rm half,0})$ increases, the mass-to-light ratio within $r_{\rm 200c}$ does not increase as in the case for small radii $<r_{\rm half}$, but keeps decreasing. This is because the satellites/subhalos always lose mass regardless of whether core-collapse happens. 
Comparing Fig. \ref{fig:track-c} and \ref{fig:track-d}, we can see that our proposed proxy or `anchor' for fitting/parameterization of tidal tracks, the enclosed mass within the satellites' initial $r_{\rm half}$, does show a tight correlation with properties of stellar components which are generally located at similar radii, but is not an ideal tracer for the halo-wide properties at large radii. Or, vice versa, halo-wide properties such as the total mass loss, which has been used as the tracer in modelling CDM subhalos \cite{hayashi03, Penarrubia07, Penarrubia10, xdu24}, may not be a good tracer for what happens in the inner subhalo (thus including the satellite galaxies) in SIDM cases. 

Apart from the evolution tracks of SIDM (satellite) dwarfs, we have also shown the corresponding tidal tracks of the CDM satellite on our most plunging orbit orb4 as black stars in Fig. \ref{fig:track-a} to \ref{fig:track-c}. From the much larger span of SIDM stellar properties than CDM in all three panels, we can again emphasize on the larger diversity of (satellite) dwarfs in SIDM models. The universality in CDM tidal tracks, as previously reported in \cite{Penarrubia07, errani15, abenson22}, thus also display different behaviors from the universality we find in SIDM dwarfs.

In summary, this section showcases the tidal evolution tracks of SIDM dwarf satellites. Despite the large span in the values of satellites' stellar properties during the evolution (diversity), we find self-similar behaviors in their tidal tracks that can potentially be parametrized into fitted relations (universality). Each tidal track bifurcates into two branches, depending on whether the DM (sub)halo eventually core-collapses or not. For the tidal tracks of the mass-to-light ratio and stellar velocity dispersion within $r_{\rm half}$, we find the two branches are tightly aligned along a single line. For the tidal tracks of the stellar size $r_{\rm half}$, the shapes of the two branches are less self-similar and additional parameters must be introduced in future works of parameterization: evaporation strength for the cored branch and inner mass loss for the core-collapse branch.

\subsection{Hernquist vs. Plummer initial conditions for stars}\label{sec:hevspl}

In previous subsections, we focused on analyzing systems initialized with a cuspy, Hernquist stellar density profile. In this section, we explore differences in results produced by Plummer initial distributions of stars. Observationally, dwarf galaxies are found to have either cored or cuspy stellar density profiles \cite{moskowitz20, cerny23}, hence our choice of simulating both Hernquist and Plummer initial conditions for stars aims to cover both of these populations. 
Note that there is a known, intrinsic issue with setting up the initial condition of a Plummer galaxy embedded in an NFW halo in equilibrium, which we discuss in more detail in Appendix \ref{appdx:plummer}, including our 
current solution to the problem.

\begin{figure*}
    \centering
    \begin{subfigure}{0.48\textwidth}
        \centering
        \includegraphics[width=\textwidth]{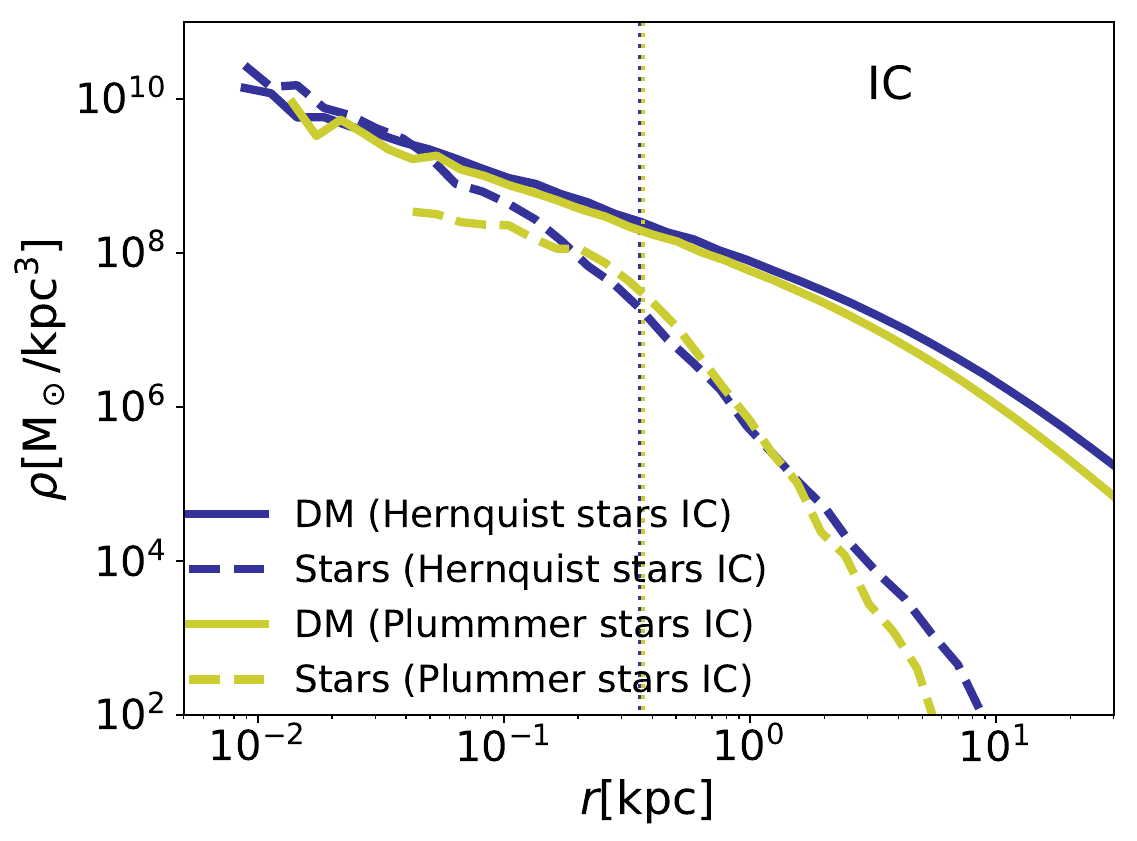} 
        \caption{}
        \label{fig:dm+star-pro-a}
    \end{subfigure}
    ~
    \begin{subfigure}{0.48\textwidth}
        \centering
        \includegraphics[width=\textwidth]{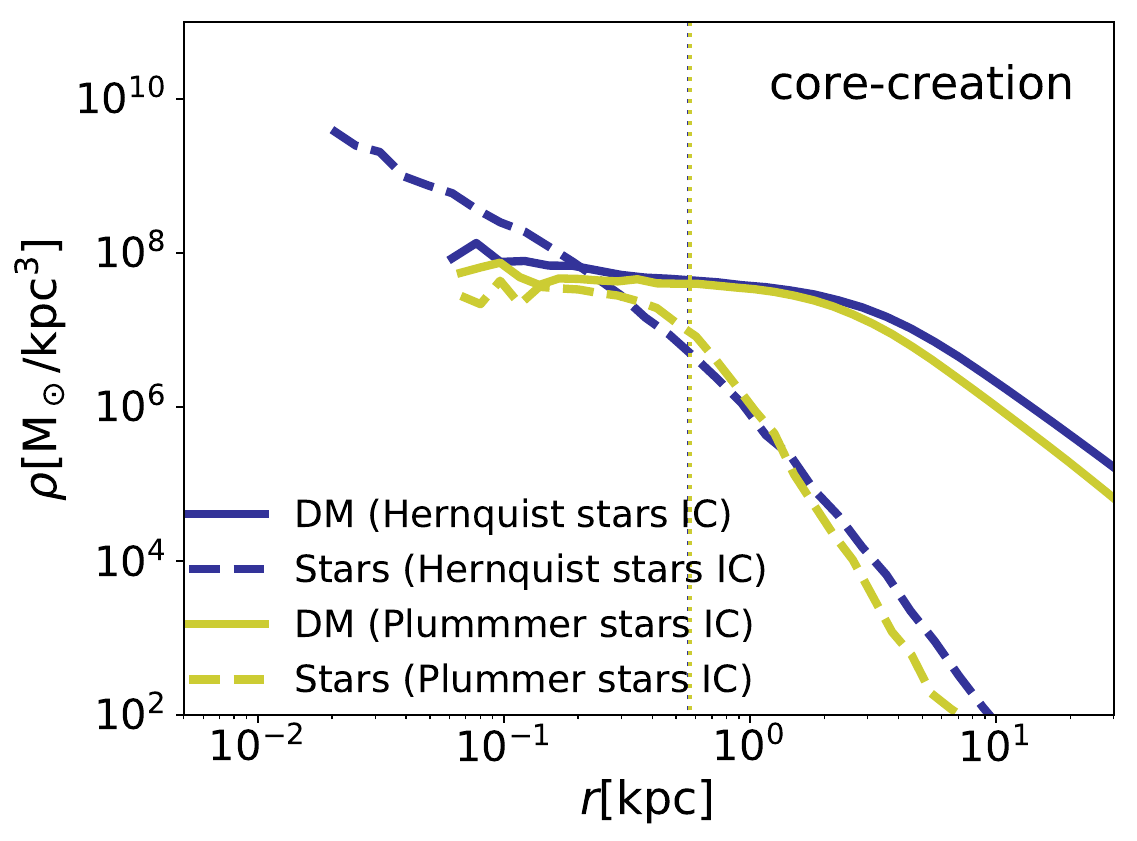} 
        \caption{}
        \label{fig:dm+star-pro-b}        
    \end{subfigure}
    ~
    \begin{subfigure}{0.48\textwidth}
        \centering
        \includegraphics[width=\textwidth]{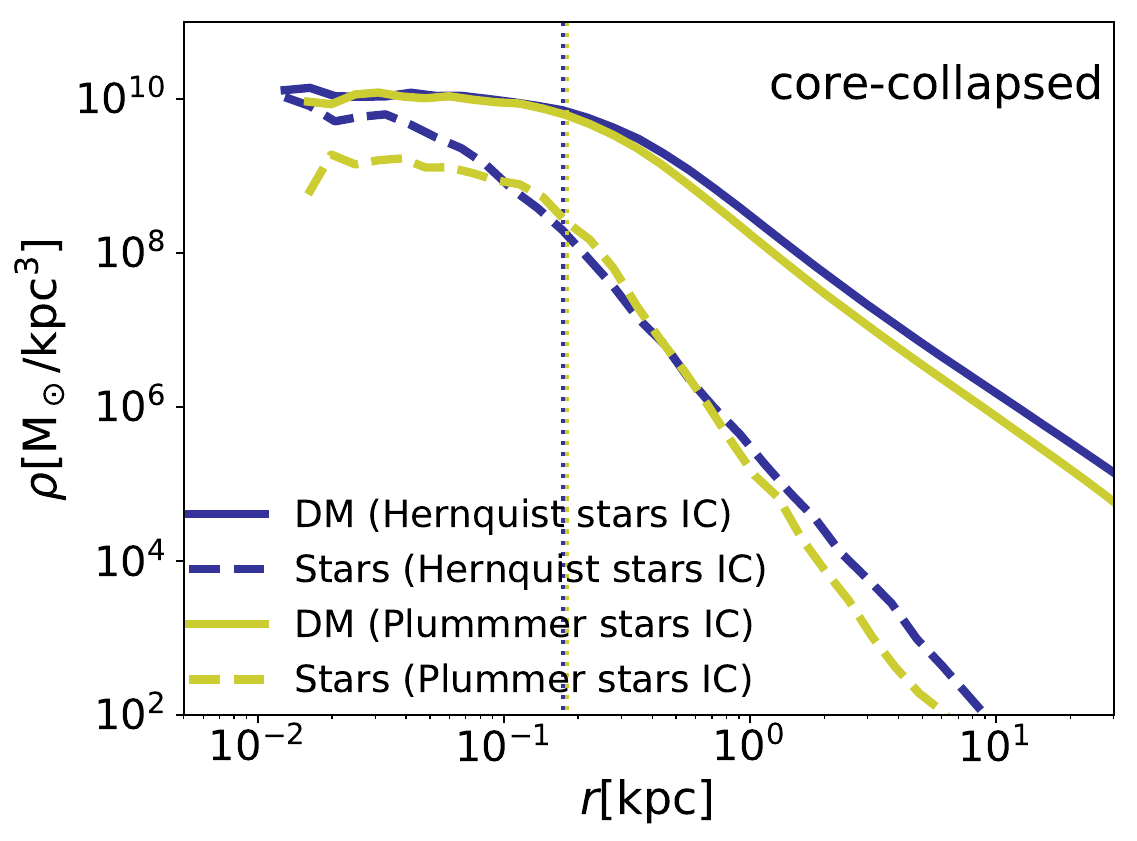} 
        \caption{}
        \label{fig:dm+star-pro-c}
    \end{subfigure} 
    ~
    \begin{subfigure}{0.48\textwidth}
        \centering
        \includegraphics[width=\textwidth]{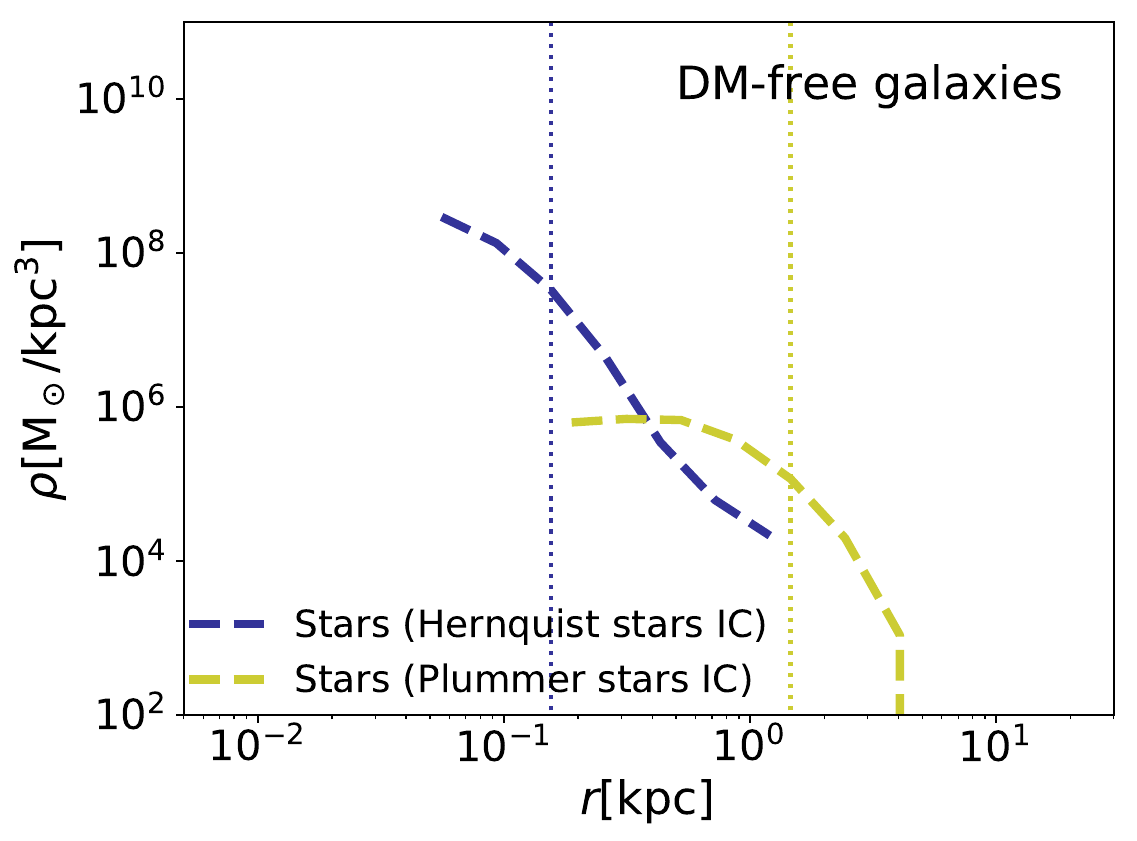} 
        \caption{}
        \label{fig:dm+star-pro-d}
    \end{subfigure} 
    \caption{Comparison of Hernquist vs. Plummer galaxies' DM and stellar density profiles. Panels a) to c) show the snapshots for: initial conditions, core-creation stage and core-collapsed stage of the isolated galaxy-halo system with SIDM model $\{\sigma_0=200, \omega=200\}$. Panel d) shows the final remnant of cored subhalos when the DM components are fully depleted by the tidal field---the Hernquist galaxy can remain self-bound while the Plummer galaxy is unstable in the tidal field. The solid lines show DM density profiles and the dashed lines show stellar density. The lines in blue show satellites generated with a Hernquist profile, while the yellow lines show Plummer cases.}
    \label{fig:dm+star-pro}
\end{figure*}

In Fig. \ref{fig:dm+star-pro}, we show the evolution of density profiles of both DM and stellar components of a galaxy-halo system with SIDM model $\{\sigma_0=200, \omega=200\}$, for both Hernquist and Plummer galaxies. From Fig. \ref{fig:dm+star-pro-a} to \ref{fig:dm+star-pro-c}, we show the evolution from the initial conditions to core-creation, then core-collapse for isolated galaxies. The Hernquist galaxy is always cuspier and more concentrated than the Plummer galaxy throughout the dark halo evolution, although the half-light radii are nearly equal despite the difference in profile shape.

Differences in the evolution of the stellar component are significantly more pronounced for satellites.  As the DM is gradually stripped away by tidal forces and evaporation, the distribution of the stellar components determines whether the galaxy can survive in the tidal field (see also \cite{errani24b}). The Hernquist galaxies, as we discussed in previous sections, are able to become self-bound due to their cuspy potentials even after the DM is completely removed, thus forming DM-free galaxies. Galaxies with cored Plummer stellar distributions, which tend to be more common among low-mass dwarfs that are less baryon-dominated  \cite{moskowitz20}, do not have such a cuspy inner density to counter the tidal field, and will be dissolved in the host halo once their DM component is depleted. We show a comparison of the remnants of Hernquist (at the end of simulation) and Plummer galaxies (at the last snapshot before complete dissolution) that are after DM-depletion in Fig. \ref{fig:dm+star-pro-d}. 

\begin{figure*}
    \centering
    \begin{subfigure}{0.45\textwidth}
        \centering
        \includegraphics[width=\textwidth]{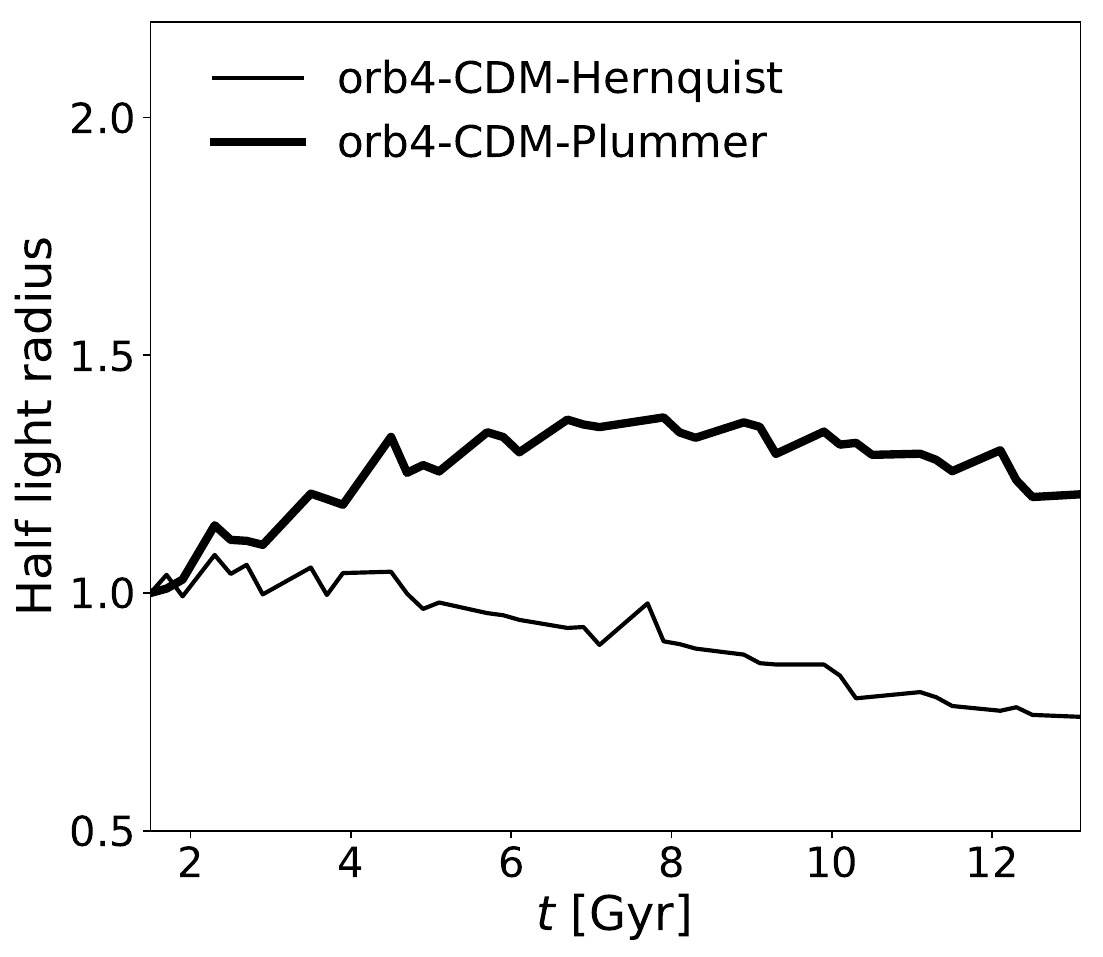} 
        \caption{}
        \label{fig:hevspl-a}
    \end{subfigure}
    ~
    \begin{subfigure}{0.44\textwidth}
        \centering
        \includegraphics[width=\textwidth]{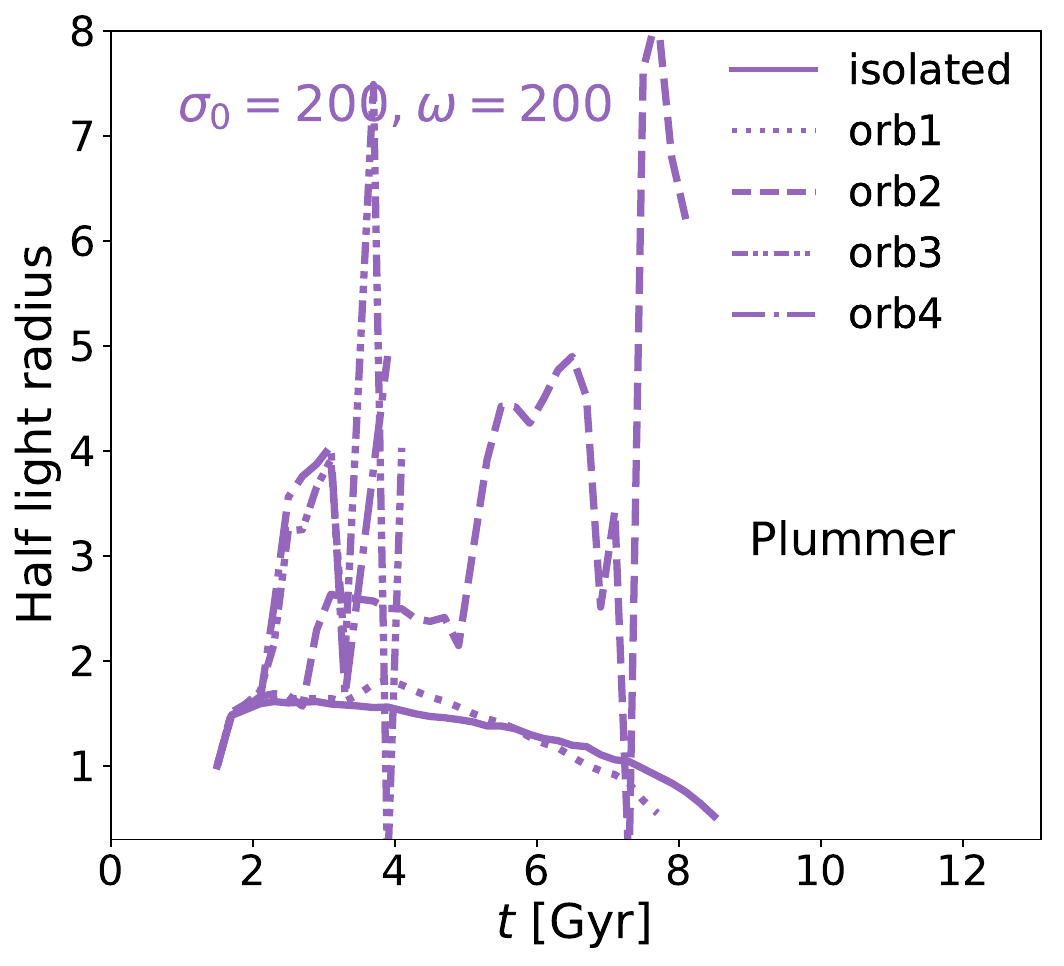} 
        \caption{}
        \label{fig:hevspl-b}
    \end{subfigure}
    ~
    \begin{subfigure}{0.43\textwidth}
        \centering
        \includegraphics[width=\textwidth]{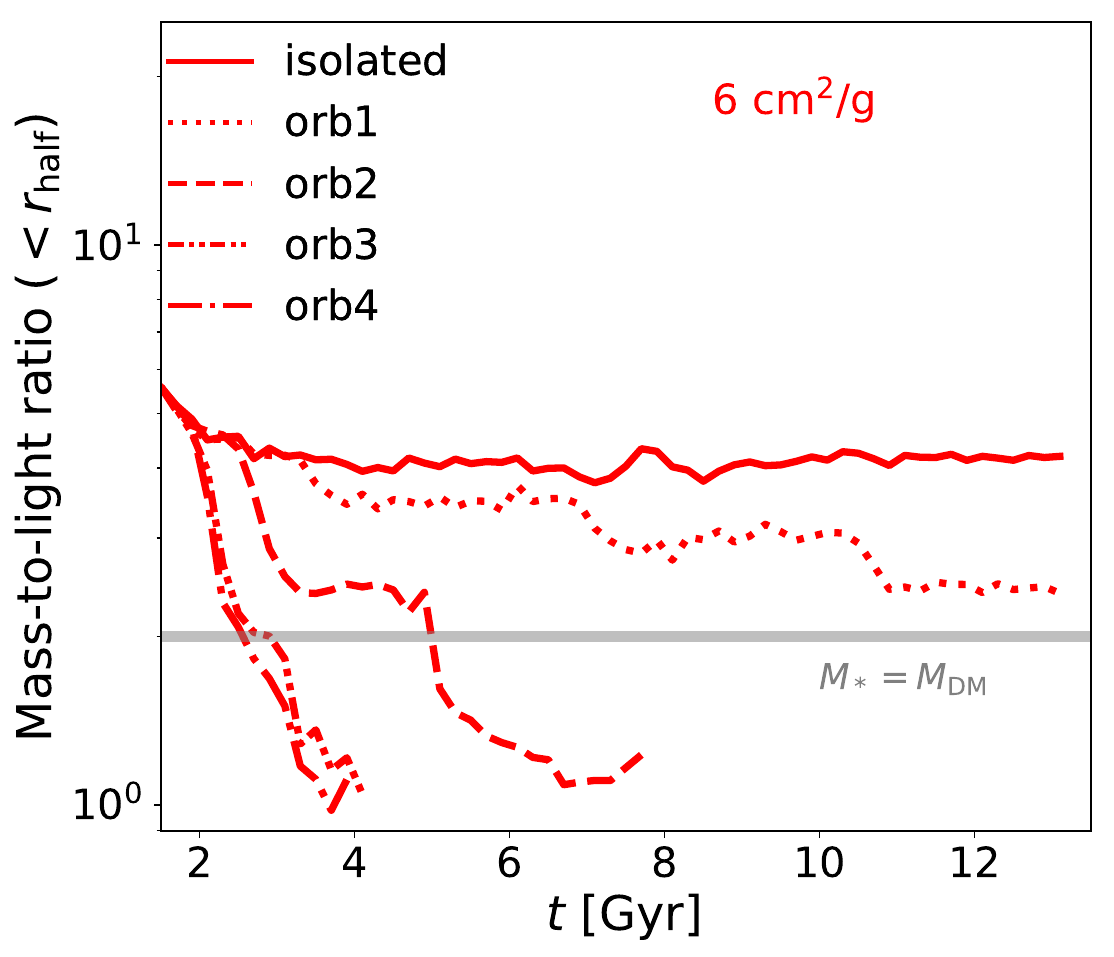} 
        \caption{}
        \label{fig:hevspl-c}
    \end{subfigure}
    ~
    \begin{subfigure}{0.48\textwidth}
        \centering
        \includegraphics[width=\textwidth]{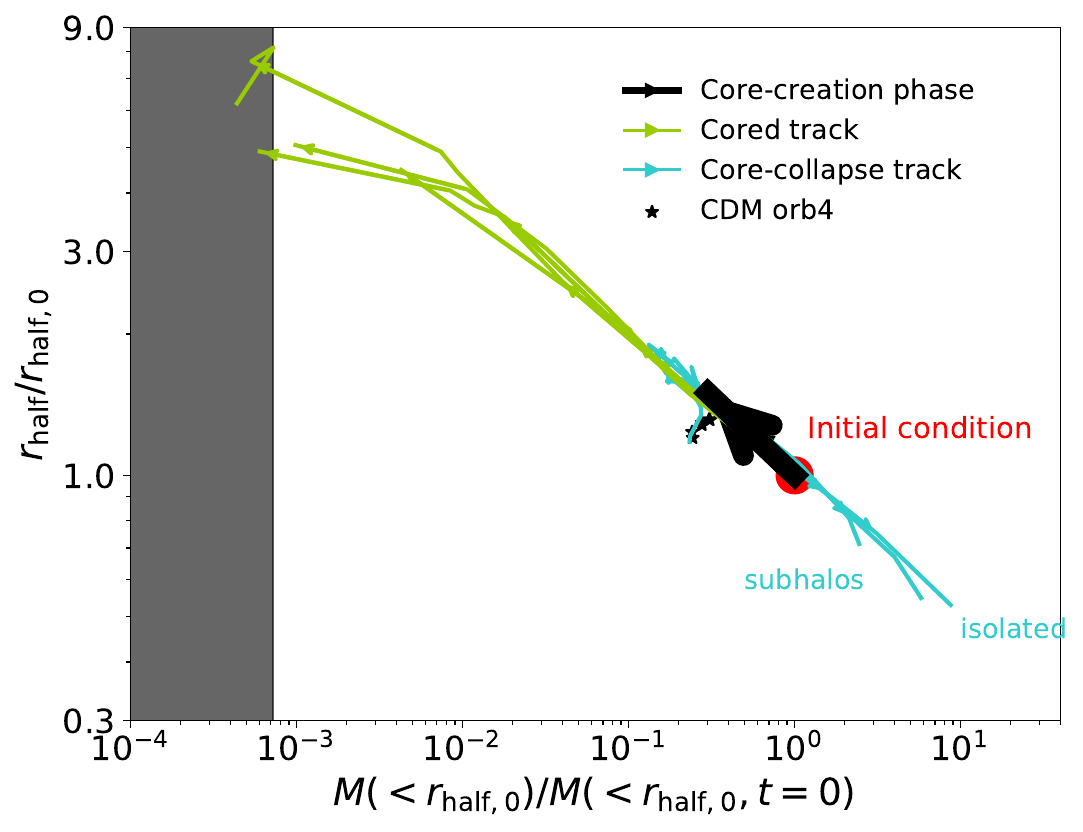} 
        \caption{}
        \label{fig:hevspl-d}
    \end{subfigure}
    ~
    \begin{subfigure}{0.48\textwidth}
        \centering
        \includegraphics[width=\textwidth]{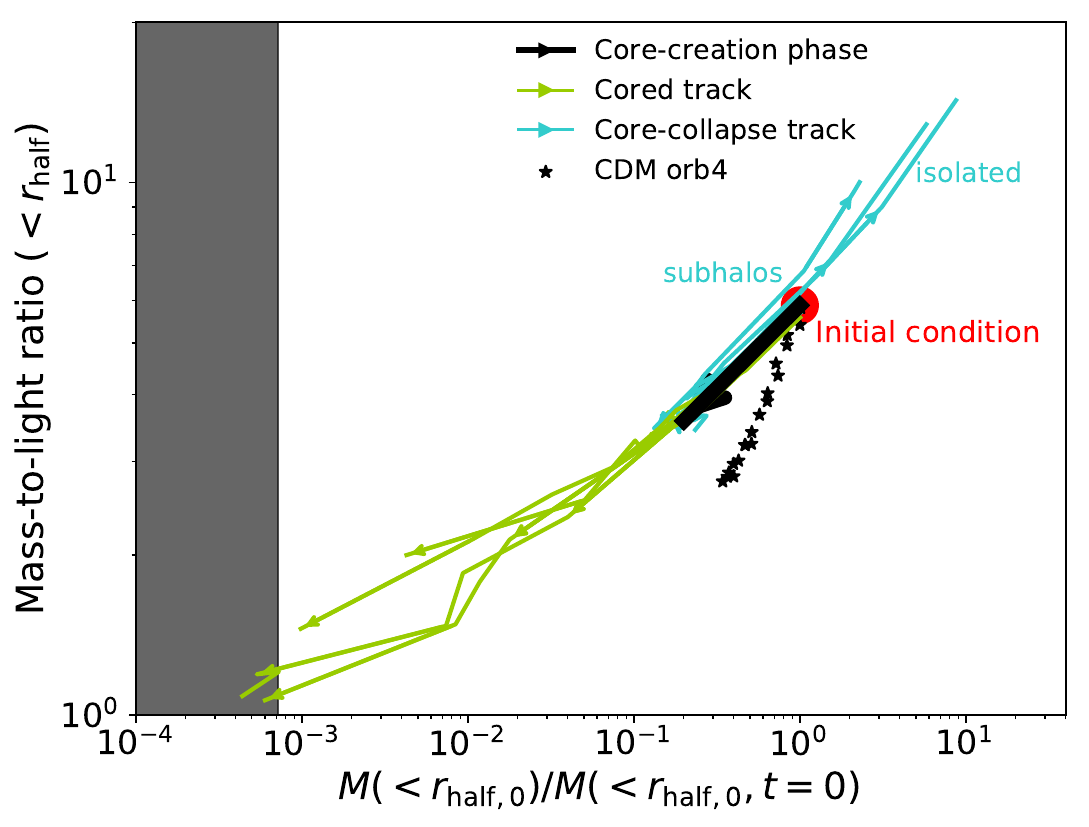} 
        \caption{}
        \label{fig:hevspl-e}
    \end{subfigure}
    ~
    \begin{subfigure}{0.47\textwidth}
        \centering
        \includegraphics[width=\textwidth]{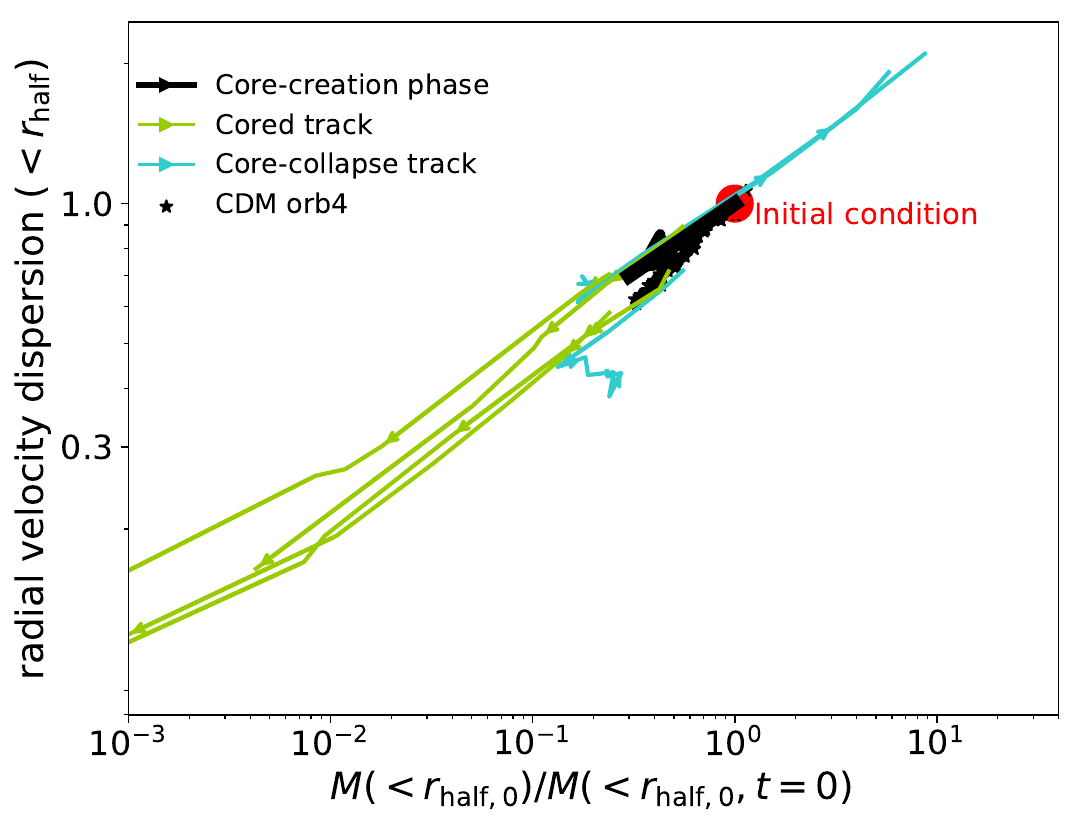} 
        \caption{}
        \label{fig:hevspl-f}
    \end{subfigure}
    \vspace{-0.3cm}
    \caption{Time evolution and tidal tracks of Plummer satellites. a) Time evolution of $r_{\rm half}$ for the CDM case on orb4 (the most plunging orbit); b) Time evolution of $r_{\rm half}$ for $\{\sigma_0=200, \omega=200\}$ subhalos; c) Time evolution of the mass-to-light ratio for 6 $\rm cm^2/g$ subhalos; d) Tidal tracks of $r_{\rm half}$ for the same subhalos in Fig. \ref{fig:track-a}; e) Tidal tracks for the mass-to-light ratio; f) Tidal tracks of stellar radial velocity dispersion within $r_{\rm half}$.  The shaded region in d) and e) are where there are less than 10 simulation particles (DM+stars) within $r_{\rm half,0}$.} 
    \label{fig:hevspl}
\end{figure*}

We next examine differences in the evolution of satellite properties resulting from the to a Plummer profile instead of the default Hernquist profile in generating stellar components, as summarized below and in Fig. \ref{fig:hevspl}:
\begin{itemize}
    \item \textit{More visible tidal heating (CDM)} In Fig. \ref{fig:hevspl-a}, we show the time evolution of $r_{\rm half}$ for  satellite galaxies in CDM subhalos, choosing the most plunging orbit orb4 to highlight the effects. For the Hernquist case, $r_{\rm half}$ slowly decreases with time, as the star particles are stripped away. For the Plummer case, however, we can see that $r_{\rm half}$ first increases before decreasing again. This is because tidal heating causes the Plummer satellite galaxy to puff up. The same heating also applies to the Hernquist galaxy, but the puffing-up is far less prominent because of its cuspy central potential. This tidal heating effect is also reported in \cite{errani15}, whose results are in good agreement with our Fig. \ref{fig:hevspl-a}. Note that here we use the CDM case to demonstrate the tidal heating effect, because for SIDM there is additional physics causing $r_{\rm half}$ to increase, the adiabatic expansion by core-creation and heat injection by evaporation, which can be more significant than tidal heating in puffing up the galaxy. 
    \item \textit{Increased maximal boost on $r_{\rm half}$} Because the Plummer satellite has a shallower central potential, its stellar component is less bound and receives a larger boost in $r_{\rm half}$ by the evaporation field compared to the Hernquist case. In Fig, \ref{fig:hevspl-b} and \ref{fig:hevspl-d}, we find that the maximum of $r_{\rm half}$ is about six to eight times the initial value, a factor of two boost from the Hernquist case. 
    \item \textit{Not forming stable DM-free galaxies} 
    Since the Plummer galaxy has a much shallower stellar potential, it is not able to hold itself together in a strong tidal field after the DM is almost entirely stripped, and thus cannot form stable DM-free galaxies, in contrast to galaxies with  Hernquist stellar profiles. We can see the complete dissolution of the Plummer galaxies in the density plot Fig. \ref{fig:dm+star-pro-d}, and also the time-evolution plots Fig. \ref{fig:hevspl-b} and \ref{fig:hevspl-c}, where the dissolved Plummer satellites are missing from halo finders (thus are no longer present in the simulation).
    \item \textit{Longer time to core-collapse} As has been reported in \cite{wxfeng21, sq23, ymzhong23, dnyang24}, the existence of a baryonic potential can accelerate the core-collapse process of SIDM (sub)halos.  This acceleration is more significant for Hernquist galaxies that are cuspier and have deeper potentials than Plummer galaxies. Hence, we observe a comparatively longer core-collapse time $t_{\rm cc}$ in the Plummer cases, especially for those that intrinsically collapse later and thus rely more on the acceleration effect from stars, such as the orb2 and orb3 cases of $\{\sigma_0=200,\omega=50\}$ in Table \ref{table:ccp-orb}.
    
\end{itemize} 

\subsection{Tidal tracks with varying initial stellar mass or size}\label{sec:trackvar}

\begin{figure*}
    \centering
    \begin{subfigure}{0.52\textwidth}
        \centering
        \includegraphics[width=\textwidth]{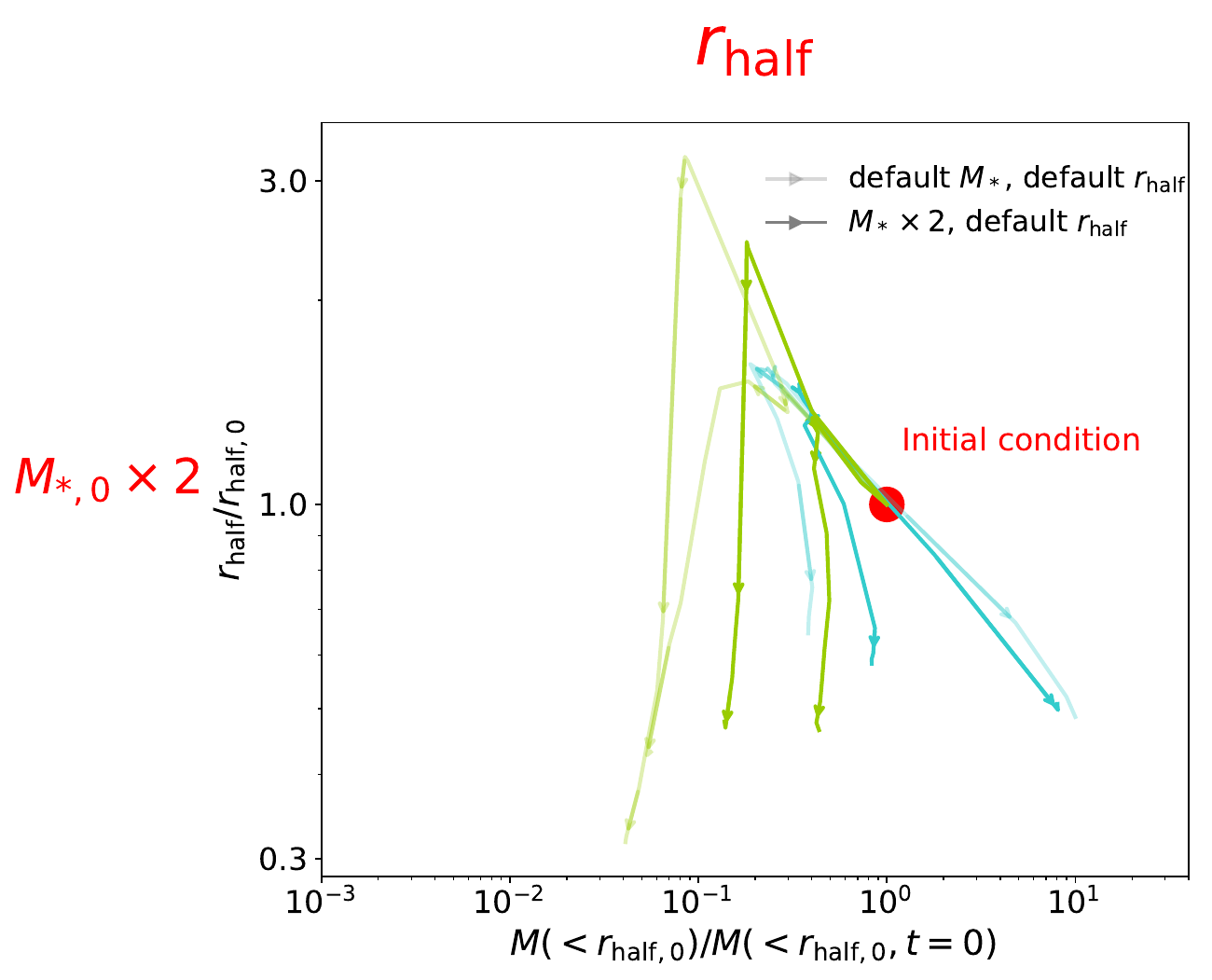} 
        \caption{}
        \label{fig:track-var-a}
    \end{subfigure}
    ~
    \begin{subfigure}{0.45\textwidth}
        \centering
        \includegraphics[width=\textwidth]{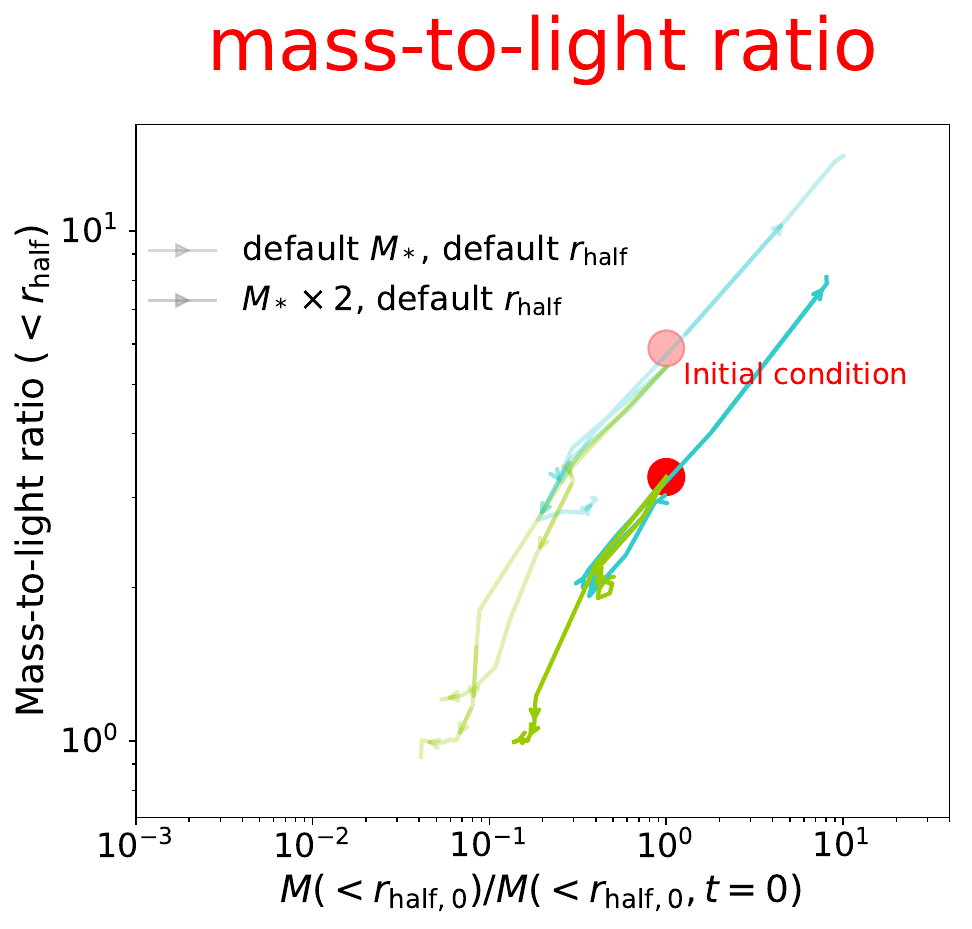} 
        \caption{}
        \label{fig:track-var-b}
    \end{subfigure}
    ~
    \begin{subfigure}{0.51\textwidth}
        \centering
        \includegraphics[width=\textwidth]{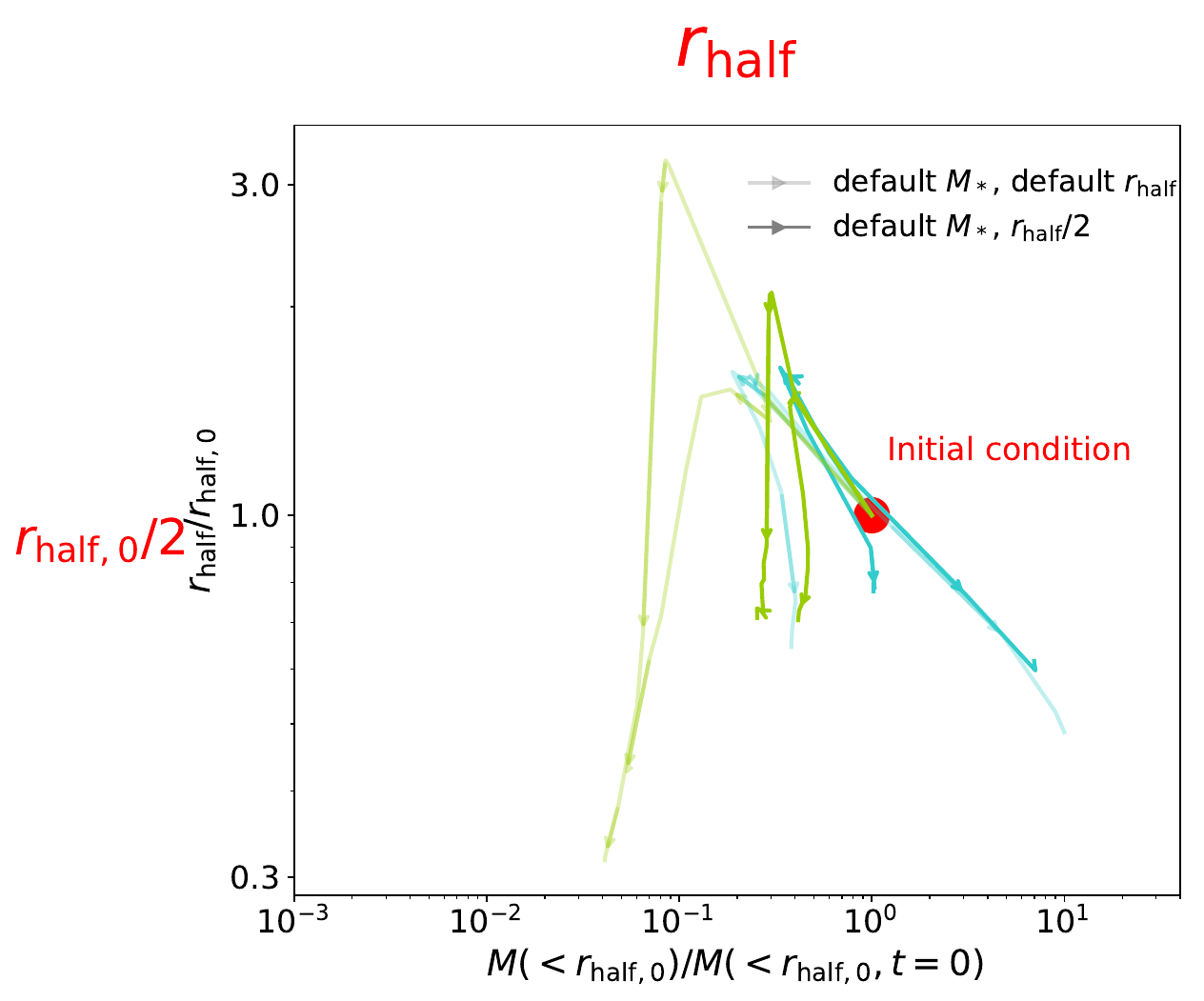} 
        \caption{}
        \label{fig:track-var-c}
    \end{subfigure}
    ~
    \begin{subfigure}{0.45\textwidth}
        \centering
        \includegraphics[width=\textwidth]{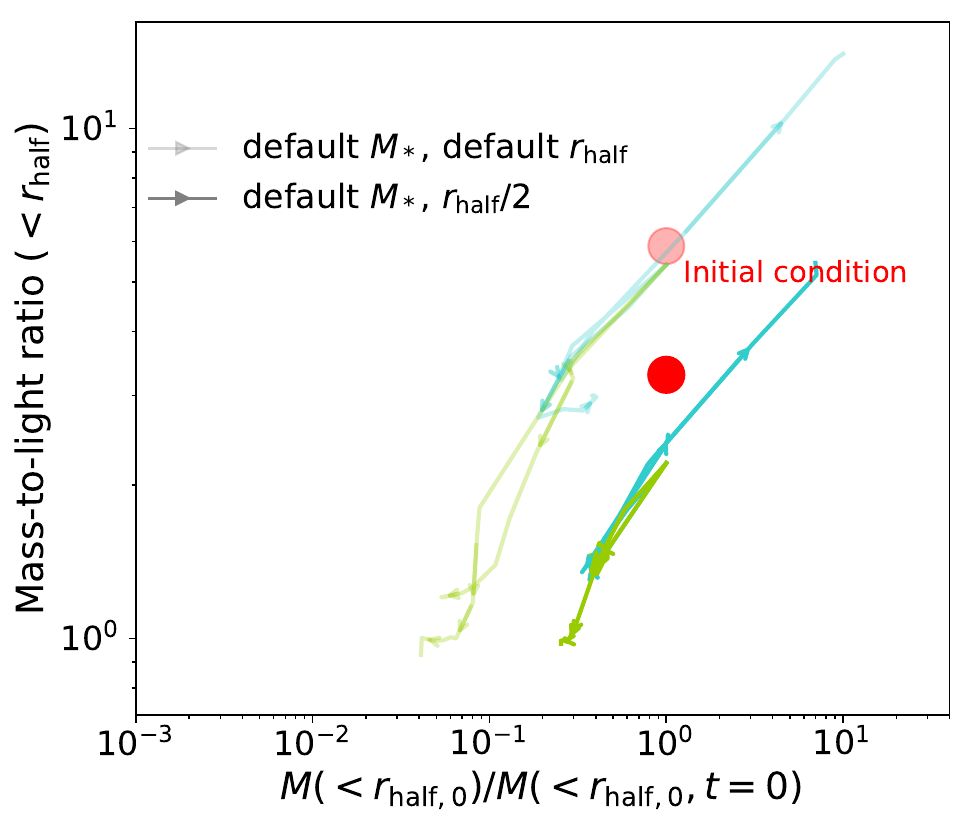} 
        \caption{}
        \label{fig:track-var-d}
    \end{subfigure}
    ~
    \begin{subfigure}{0.52\textwidth}
        \centering
        \includegraphics[width=\textwidth]{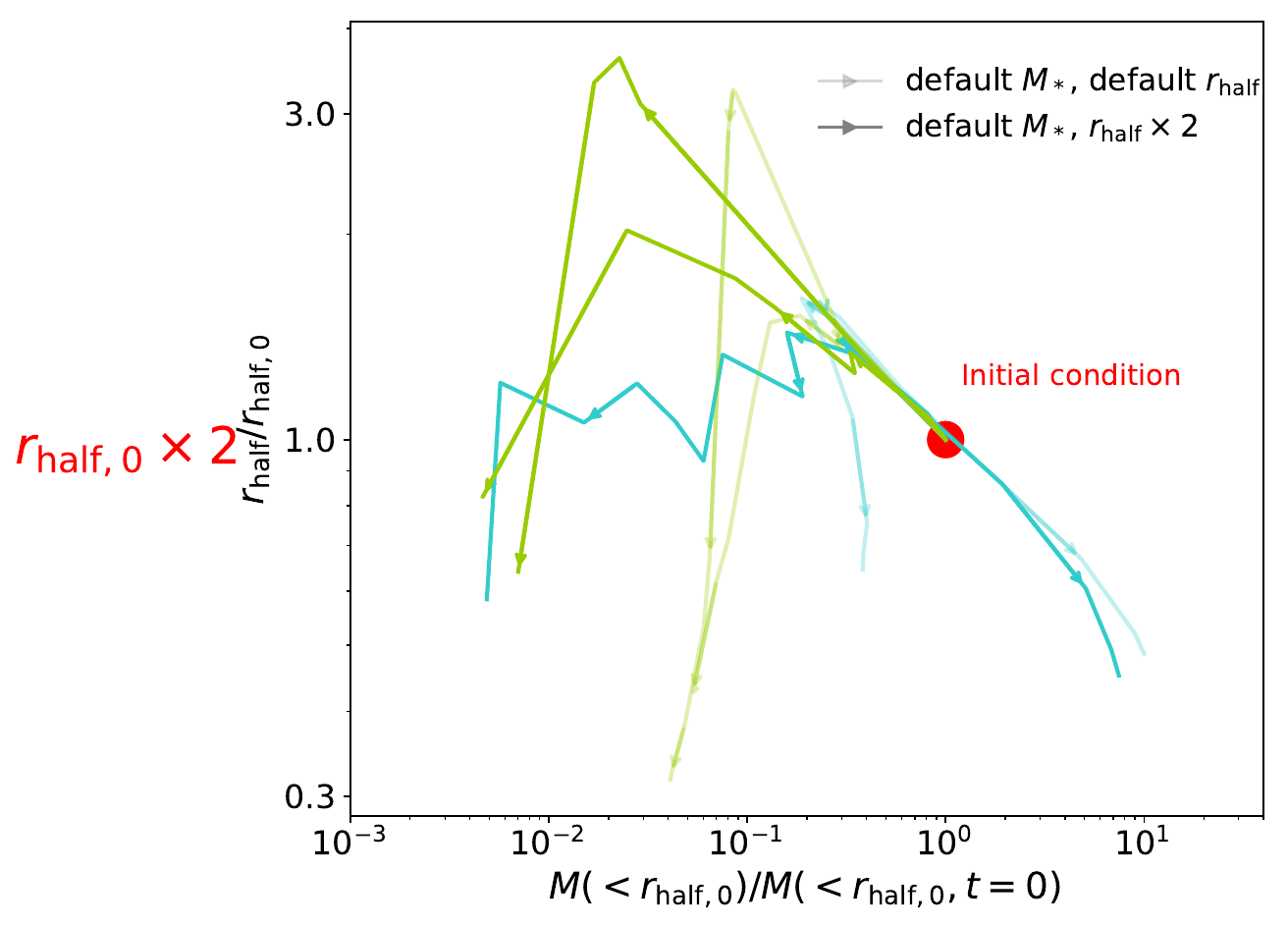} 
        \caption{}
        \label{fig:track-var-e}
    \end{subfigure}
    ~
    \begin{subfigure}{0.45\textwidth}
        \centering
        \includegraphics[width=\textwidth]{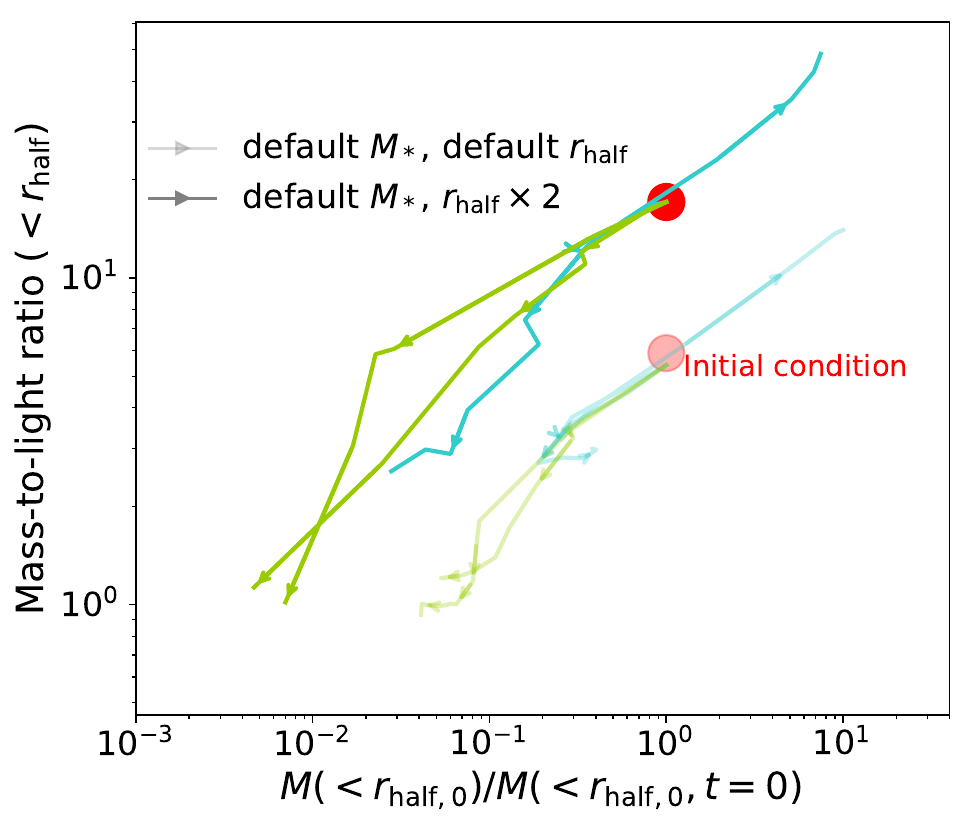} 
        \caption{}
        \label{fig:track-var-f}
    \end{subfigure}
    \caption{ Tidal tracks similar to those in Fig. \protect\ref{fig:trackmain}, but with varying initial stellar properties, including doubled stellar mass, halved and doubled stellar size. The default parameters (which are the same as for our main tidal track results in \protect\ref{fig:trackmain}) are shown in transparent lines and the variants in opaque lines. The left column shows the tidal tracks for the stellar size, and the right column shows the mass-to-light ratio.}
    \label{fig:track-var}
\end{figure*}

In Section \ref{sec:track}, we showed the tidal evolution tracks of a single satellite galaxy evolving with different combinations of SIDM models and orbits, but with the same initial stellar properties. In this section, we vary the initial conditions of the galaxy with three test cases: doubling the stellar mass $M_*$ while using the default initial stellar size $r_{\rm half}$; or fixing $M_*$ while changing $r_{\rm half}$ by a factor of 2 or 1/2. The aim of this section is to test the robustness of the tidal tracks that we show in Fig. \ref{fig:trackmain}. Here we show to what extent the previously demonstrated universality in tidal tracks is affected when studying different galaxies, which is necessary to understand before applying the tidal track relations in a more realistic/diverse context of galaxy evolution.

In Fig. \ref{fig:track-var-a}, \ref{fig:track-var-c} and \ref{fig:track-var-e}, we show the tidal tracks of $r_{\rm half}$ for the default galaxy (transparent lines) and its three variants (opaque lines): doubled initial $M_*$, halved initial $r_{\rm half}$, and doubled $r_{\rm half}$, which we denote as $M_{*,0} \times 2$, $r_{\rm half,0}/2$ and $r_{\rm half,0}\times2$, respectively. Note that for the readability of the figures, in this set of comparisons we show only two cases of cored subhalos in green, $\{\sigma_0=200, \omega=50\}$ on orb4 and $\{\sigma_0=200, \omega=200\}$ on orb4, and two cases of core-collapsed subhalos (`core-collapsed' when in the case of default initial galaxy) in cyan, $\{\sigma_0=200, \omega=200\}$ in isolation and $\{\sigma_0=200, \omega=50\}$ on orb3. 

We see that for the isolated galaxy-halo system (the right-most cyan lines), varying the initial $M_*$ or $r_{\rm half}$ barely shifts the tidal track during the whole process of core-formation to core-collapse, as the opaque and transparent evolution tracks nearly overlap with each other. 

For satellite galaxies with orbital effects, shown as the other three paired cases in comparison in Fig. \ref{fig:track-var-a}, \ref{fig:track-var-c} and \ref{fig:track-var-e}, we observe a noticeable shift in the tidal tracks of $r_{\rm half}$. When the initial stellar potential is more prominent, either with a larger stellar mass ($M_{*,0} \times 2$) or a more compact distribution ($r_{\rm half,0}/2$), the stellar particles are more bound and thus less responsive to the (adiabatic) expansion during core-creation, also less susceptible to the late-time tidal stripping. Therefore, the range of the fractional $r_{\rm half}$ during its whole evolution cycle is reduced, appearing as a compression in the $r_{\rm half}$ tidal tracks along the y-axis. For the same reason, the mass loss in the inner region $<r_{\rm half,0}$ is slower, and the tidal tracks of subhalos are thus shifted toward larger inner mass (rightward along the x-axis). 

In contrast, when the stellar potential is less deep ($r_{\rm half,0}\times2$), we see the opposite effect, that the maximal $r_{\rm half}$ is increased and the inner mass loss is more extreme. Another noticeable difference caused by this $r_{\rm half,0}\times2$ variant is that there are no stable DM-free galaxies as remnants of the tidal evolution. Even if the stars have a cuspy initial profile, their distribution is too dispersed to bind together the stellar remnants, at least at our current resolution limit of the satellite's innermost region. To summarize, we find that variations in satellite galaxies' initial stellar properties, such as the mass and size, do affect the tidal evolution tracks of the stellar size. However, this effect is highly explainable, with the potential of being included in future parameterizations of the tidal tracks, perhaps given an additional parameter to describe the initial stellar potential. We postpone the detailed analysis of this to future works when more simulations are available.

The tidal tracks of the mass-to-light ratio within $r_{\rm half}$, unlike the stellar size, are more robust when varying the initial stellar properties, as we show in Fig. \ref{fig:track-var-d} to \ref{fig:track-var-f}. Note that in these plots, same as in Fig. \ref{fig:track-c}, we do not normalize the mass-to-light ratio to its initial value. Hence, the tidal tracks of the mass-to-light ratio for the varied stellar ICs appear in parallel with the original tracks instead of overlapping with them. We find that the tidal tracks of the mass-to-light ratio are much less affected by using different stellar ICs than those of the stellar size, suggesting robustness in future works for tidal track parameterization. The tidal tracks for the stellar velocity dispersion are also tested with these variants of the stellar ICs, which we show in Appendix \ref{appdx:suppfigs}.

Apart from the difference in the evolution tracks of the stellar component itself, varying the initial stellar properties sets a different stellar potential at the halo center, thus affecting the core-collapse process of the DM component \cite{ymzhong23, sq23, dnyang24}. In Table \ref{table:ccp-var}, we list the core-collapse time of these variant cases in comparison to the default stellar IC. Our results qualitatively agree with physical intuition and previous research \cite{ymzhong23, sq23, dnyang24}, that the core-collapse process is accelerated when the stellar potential is deeper, and vice versa. We also find that subhalos are more sensitive in their core-collapse time to the varied stellar potential than are isolated halos. For example, the $[\sigma_0=200, \omega=50]$ case on orb3 reaches the core-collapse threshold at 10.3 Gyr with the default stellar IC, but with the ``$r_{\rm half,0}\times 2$'' variant it is unable to core-collapse before the end of the simulation. This is also shown in Fig. \ref{fig:track-var-e} and \ref{fig:track-var-f}, where a cyan-colored tidal track, originally determined to be on the core-collapse branch, is shifted to the cored branch with the ``$r_{\rm half,0}\times 2$'' variant. The same variants comparison for the isolated $[\sigma_0=200, \omega=200]$ does not see a noticeable difference in the core-collapse time. This is because the strength of the stellar potential helps bind the DM in the inner region, countering the disruption to the subhalo core-collapse from tidal heating and evaporation (see detailed discussion in \cite{zzc22}). Although this stellar-potential-back-reaction brings more complexity/diversity to the statistics of subhalo core-collapse in the presence of stellar particles, we remark that it should not require additional fitting parameters for tidal tracks. This is because whether or not core-collapse occurs is itself always a necessary input to the fitted tidal tracks, dictating which branch the tracks fall onto, regardless of how much the core-collapse itself is affected by the baryons.

\begin{table*} 
	\centering
	\begin{tabular}{lcccc} 
        \hline
        \hline
	SIDM model \& orbit & default stellar IC & $M_{*,0} \times 2$ & $r_{\rm half,0} / 2$ & $r_{\rm half,0} \times 2$  \\
		\hline
        $[\sigma_0=200, \omega=200]$ isolated & $t_{\rm cc}=8.7$ Gyr & 8.3 Gyr & 8.5 Gyr & 8.7 Gyr  \\   \hline
        $[\sigma_0=200, \omega=50]$ orb3 & 10.3 Gyr & 8.1 Gyr   
 & 8.5 Gyr & - \\   \hline
        $[\sigma_0=200, \omega=50]$ orb4 & - & 7.9 Gyr   
 & - & - \\   \hline     
	\end{tabular}
	\caption{ The core-collapse time, $t_{\rm cc}$, of the default (sub)halo-galaxy system and its variants with different initial stellar properties, for a variety of SIDM models and orbits. The cases that do not reach the core-collapse threshold are labelled as `-' in this table. }
    \label{table:ccp-var}
\end{table*}


\section{Comparison with observed dwarfs}\label{sec:obs}

In this section, we present comparisons between the satellite and isolated SIDM dwarfs in our simulations and observed dwarfs, as an attempt to explore the potential of SIDM models in explaining observational outliers. We show the comparison in the stellar mass vs. stellar size plane in Sec. \ref{sec:obs1}, and in the mass-to-light ratio vs. stellar size plane in Sec. \ref{sec:obs2}. 

\subsection{Stellar mass--size plane}\label{sec:obs1}

\begin{figure}
    \centering
    \begin{subfigure}[t]{0.48\textwidth}
        \centering
        \includegraphics[width=\textwidth, clip,trim=0.2cm 0cm 0.2cm 0cm]{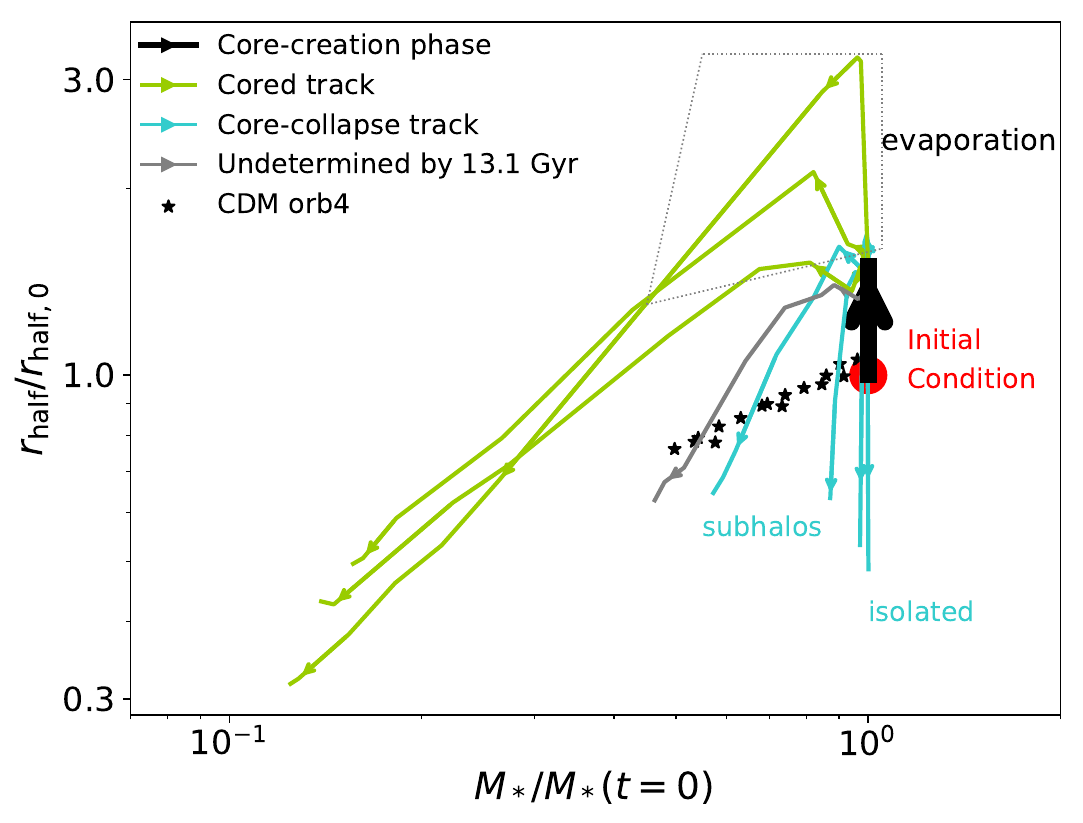}
        \caption{}
        \label{fig:mass-size-a}
    \end{subfigure}
    ~
    \begin{subfigure}[t]{0.48\textwidth}
        \centering
        \includegraphics[width=\textwidth, clip,trim=0.2cm 0cm 0.2cm 0cm]{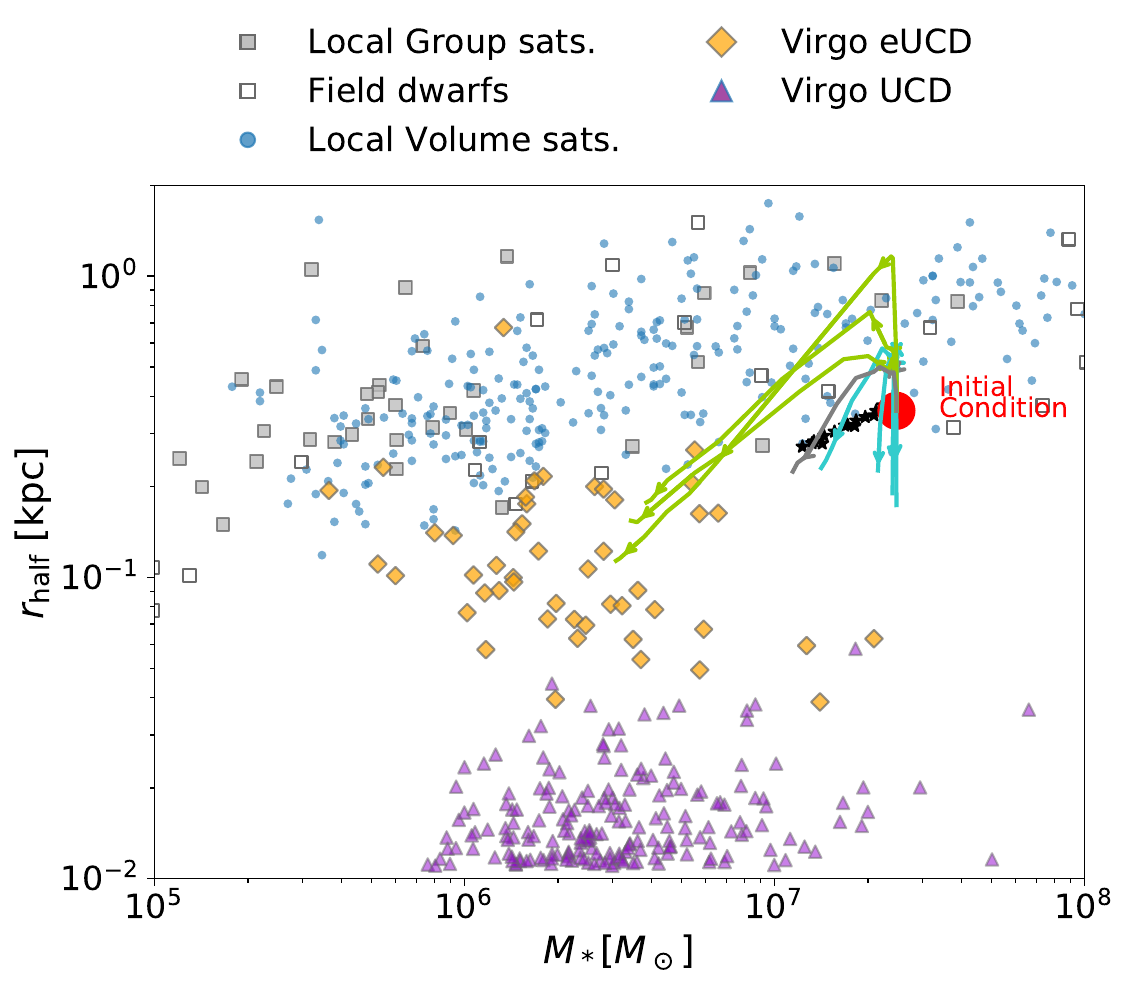}
        \caption{}
        \label{fig:mass-size-b}
    \end{subfigure}    
    \caption{The evolution of selected field and satellite SIDM dwarfs in the stellar mass vs. stellar size plane. Color schemes and notations are the same as in Fig. \protect\ref{fig:trackmain}. To place the evolutionary tracks in context, we also show Local Group satellites (gray squares) and local field dwarfs (white squares) from the Local Volume Database, satellites of Milky Way analogs in the Local Volume (blue dots; \cite{carlsten22}), ultracompact dwarfs (UCDs) in Virgo (purple triangles; \cite{liu2020}), and extended ultracompact dwarfs (eUCD) in Virgo (orange diamonds; \cite{kxwang23}). }
    \label{fig:mass-size}
\end{figure}

In this section, we focus on the stellar mass--stellar size relation for these SIDM (satellite) dwarfs during their evolution. We show the evolution tracks of the ``representative'' SIDM dwarfs from the previous section in the stellar mass vs. stellar size plane in Fig. \ref{fig:mass-size-a}, then overplot observed dwarf galaxies in Fig. \ref{fig:mass-size-b}. This aims to take a first demonstrative look at how SIDM evolution might explain the lingering diversity problem in dwarf galaxies' size, covering both the ultradiffuse and ultracompact populations \cite{drinkwater03, McConnachie12, czliu15, vDk15, koda15, carlsten22, sales22, jxli23, kxwang23}.

For this demonstrative purpose, the observational samples we show in Fig. \ref{fig:mass-size-b} include dwarf galaxies in a variety of environments, from field dwarfs, to satellites of the Milky Way and Milky Way-like systems, to the Virgo Cluster. For the dwarf galaxies in local environments, we include Local Group satellites (gray squares) and local field dwarfs (white squares) from the Local Volume Database\footnote{\url{https://github.com/apace7/local_volume_database }} \citep{Pace2024} and the satellites of Milky Way analogs in the Local Volume (blue dots) \citep{carlsten22}. For the Virgo Cluster dwarf satellites, we include both ultracompact dwarfs (UCDs, purple triangles) and extended ultracompact dwarfs (eUCDs, orange diamonds) \cite{liu2020,kxwang23}. We find that the evolution tracks of these SIDM field and satellite dwarfs can easily span the range of the local diffuse dwarfs, even though they all start from the same initial condition, without considering the variations of the initial stellar distribution prior to any SIDM and tidal evolution (see also Sec. \ref{sec:trackvar} for more discussion). 

The comparison with UCDs and eUCDs is less direct in Fig. \ref{fig:mass-size-b}, since our choice of the dwarf galaxy is initially $\sim 1$ dex more massive than the UCDs and eUCDs. However, we can still make the attempt. Assuming that the response of stellar size to SIDM core-collapse has approximately self-similar behavior across different masses of dwarf galaxies, then UCD-mass SIDM dwarfs on the core-collapsed branch would experience a similar decrease in $r_{\rm half}$ of about a factor of 2, by the point the simulation is terminated. These UCD-mass SIDM dwarfs would then partially overlap with the eUCDs in the stellar mass--size plane, while it still requires a drop in $r_{\rm half}$ of as much as nearly an order of magnitude to further accommodate the UCDs. However, we propose a few arguments about the feasibility of producing UCDs via the mechanism of SIDM subhalo core-collapse, although without further simulating into the late core-collapse stages (which is intrinsically difficult to do):

\begin{itemize}
    \item The contraction of stellar components should continue after the termination point of the simulation, because the DM core-collapse will continue. This is also expected from the late-acceleration feature in the decrease of $r_{\rm half}$ of the core-collapsing dwarf satellites, as we have shown and discussed in Fig. \ref{fig:rhalf} (the c.-c. labeled lines).
    \item However, this continued $r_{\rm half}$ contraction, driven by late-stage DM core-collapse, should not proceed forever as to make  $r_{\rm half}$ arbitrarily small. As we have discussed with the ``exception'' case in Fig. \ref{fig:ccp-dm}, once the collapsing core shrinks to be much smaller than  $r_{\rm half}$, then $r_{\rm half}$ is no longer sensitive to the change in the core-collapse density evolution, since the potential at $r_{\rm half}$ no longer deepens.  Therefore, we expect that $r_{\rm half}$ should further contract by only a finite amount. At this point, it is inconclusive whether $r_{\rm half}$ can reach the UCDs in Fig. \ref{fig:mass-size-b} even after the further contraction by late-stage core collapse is taken into account.
    \item Another ingredient that we have only briefly discussed in Sec. \ref{sec:res} is the two-body gravitational relaxation among stellar particles. When the subhalo experiences core-collapse, or when the DM is depleted and only a self-bound stellar component remains, the stellar particles have small $r_{\rm half}$ and high density. This causes the relaxation timescale to reduce to $\sim1$ Gyr, as we show in Fig. \ref{fig:stellar_trelax}, which is much shorter than the simulation timescale $\sim10$ Gyr. Thus the relaxation is enhanced as core-collapse proceeds, and serves as a numerical effect to counter the shrinking of $r_{\rm half}$. Hence we expect the $r_{\rm half}$ to be smaller and closer to the UCD range in simulations with higher resolution. 
    \item Another question that needs to be addressed if SIDM core-collapse is to explain the UCDs is why the UCDs are only found in dense clusters. Since our work uses a galaxy-group host halo, the tidal stripping on stars may be weaker than in cluster environments, and the reduction of $r_{\rm half}$ may have been underestimated than in the case of cluster UCDs. 
\end{itemize}

Overall, through this first demonstrative comparison in the stellar mass--size plane, we find that the diversity driven by SIDM physics (and with orbital effects) can easily cover the range of normal and diffuse dwarfs, but may have difficulty in accommodating the UCDs. We comment that dedicated studies into possible formation scenarios of UCDs are needed in the future, as an essential prerequisite of a comprehensive evaluation of SIDM's potential to resolve the diversity problem. This will also likely require better understanding and numerical treatments of the late-stage core-collapsing SIDM (sub)halos in simulations.


\subsection{DM-deficient DF2/DF4-like ultradiffuse dwarfs}\label{sec:obs2}

\begin{figure}
    \centering
    \begin{subfigure}{0.48\textwidth}
        \centering
        \includegraphics[width=\textwidth]{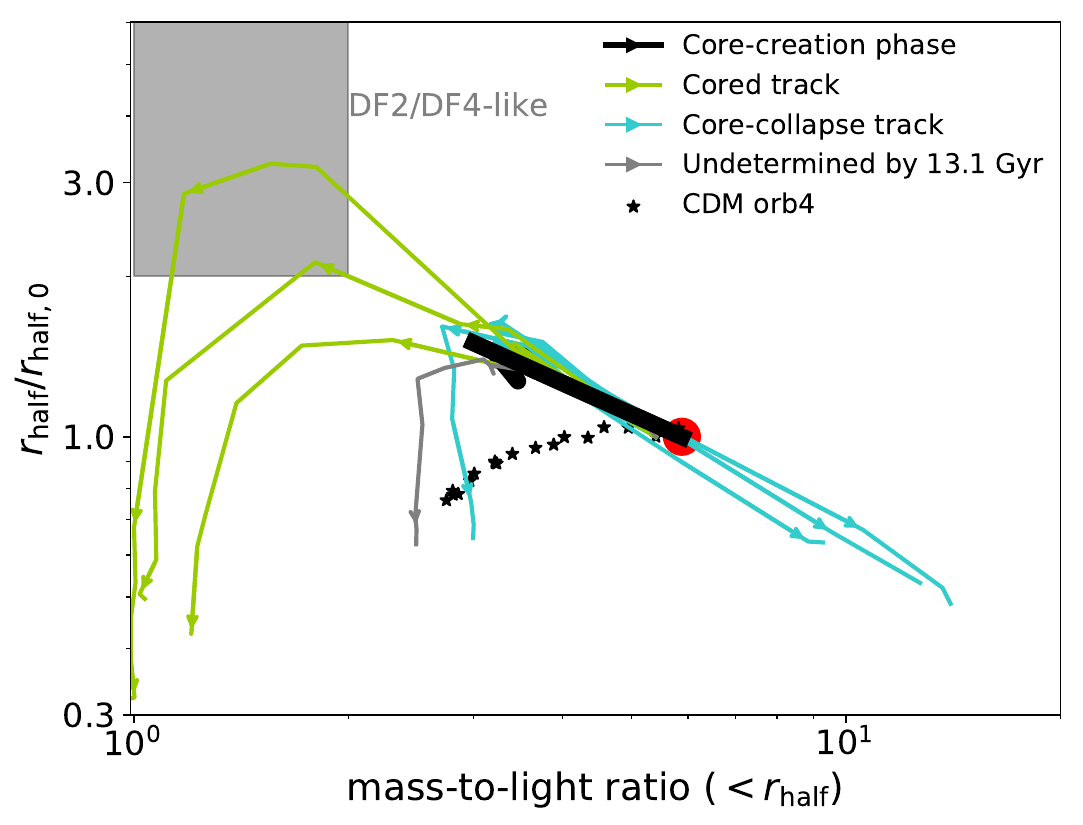} 
        \caption{}
        \end{subfigure}
    ~
    \begin{subfigure}{0.48\textwidth}
        \centering
        \includegraphics[width=\textwidth]{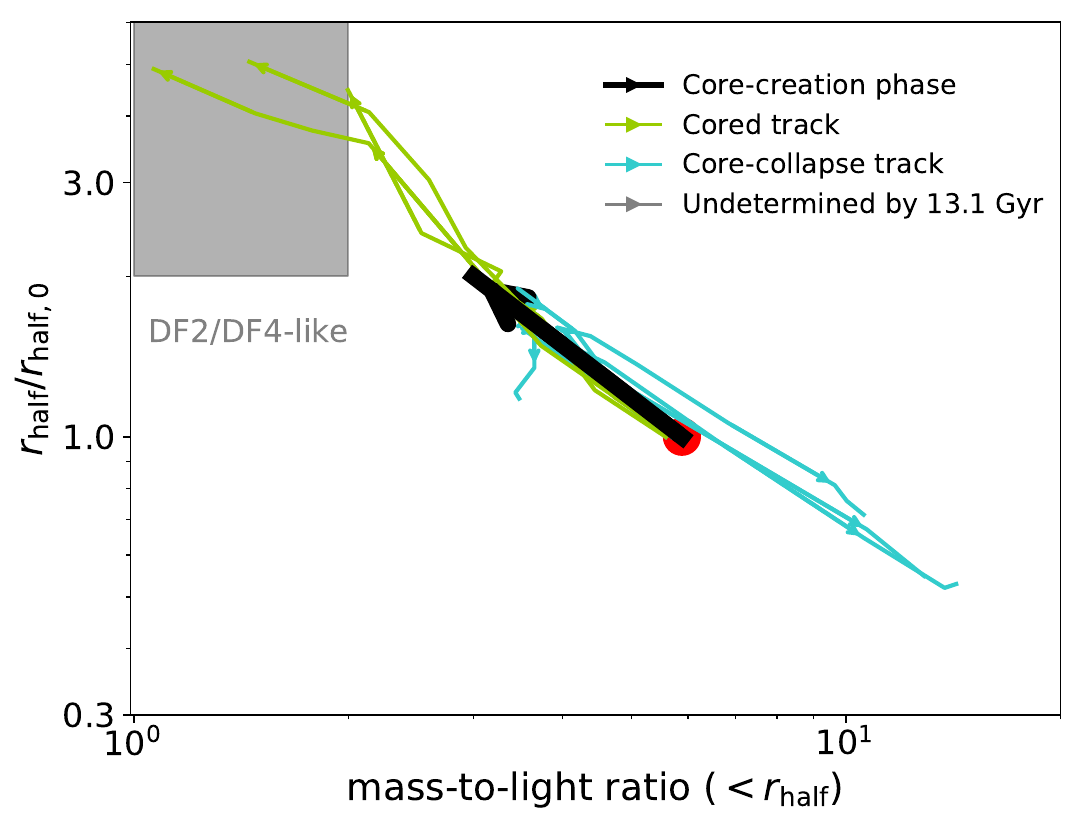} 
        \caption{}
    \end{subfigure}
    \caption{ Evolution tracks of SIDM dwarf galaxies in the mass-to-light ratio vs. stellar size plane. The gray shaded region approximately marks the parameter space that resembles the observed DM-deficient galaxies NGC-1052 DF2 and DF4. The top panel shows the dwarfs with a Hernquist initial density profile while the bottom shows the case with a Plummer profile.}
    \label{fig:track-ml-size}
\end{figure}

We began a discussion in Sec. \ref{sec:time-evo-rhalf} to \ref{sec:track} of the formation of DM-free galaxies through enhanced tidal stripping on SIDM cores and evaporation, during which the DM component is depleted. However, the remaining self-bound DM-free galaxies will be either stable and compact (Hernquist), or short-lasting and diffuse (Plummer). This may appear to be incompatible with the confirmed DM-deficient dwarfs DF2 ($2.0\times10^8 M_\odot,2.7$ kpc \cite{danieli19}) and DF4 ($1.5\times10^8 M_\odot,2.1$ kpc \cite{df4shen}), which are reported to have half-light radii $\gtrsim2$ times than the mean stellar mass--size relation. Therefore, here we map out the evolution of SIDM dwarfs in the mass-to-light ratio vs. $r_{\rm half}$ plane in Fig. \ref{fig:track-ml-size}, to see if there are any SIDM dwarfs that can accommodate both features of such galaxies: DM-deficient and diffuse.

In Fig. \ref{fig:track-ml-size}, we see a few cored Hernquist satellites crossing the DF2/DF4-like space (gray shaded region) during their evolution, with the help of evaporation, while the cored Plummer satellites all end up in the DF2/DF4-region before they are completely disrupted (which happens rapidly after this point, see Fig. \ref{fig:hevspl}). Thus we can say that SIDM \emph{does} show the potential of explaining the formation of DF2/DF4-like dwarfs, as previously proposed by \cite{dnyang20-df2, zczhang24}. The frequency of such SIDM tidally-formed DF2/DF4-like satellites is yet to be systematically evaluated, which as we have demonstrated are highly dependent on the orbit/environment. This, or in general, any similar tidal-formation scenarios, cannot be eventually confirmed or eliminated until we have more knowledge of the currently poorly-constrained distance and relationship (bound or not) between DF2/DF4 and the nearby galaxy group NGC1052 \cite{zlshen21}. However, the recently reported radial distance of 2.1 Mpc between DF2 and DF4 (but not their respective distances to the NGC1052 group \cite{zlshen21}) suggests that it would be unlikely that both DM-deficient dwarfs are satellites of NGC1052. This may therefore slightly favor another popular formation scenario, the `bullet-dwarf', which is the head-on collision of gas-rich galaxies \cite{silk19, vdk22, shin20, jhlee21, jhlee24, ymtang24}, over the tidal-disruption scenario we discussed in this work. One possible remedy to coordinate the tidal-formation of DF2/DF4-like dwarfs with their large radial distance, is to change the bound satellites we explore here to high-speed flybys with a close pericenter, which is beyond the scope of this work. In this way, the short-surviving problem of Plummer satellites can also be automatically reconciled.    

Another interesting point, although slightly off the track of the main topic of this manuscript, is that the bullet-dwarf scenario with CDM is also reported to produce compact instead of diffuse DM-deficient dwarfs \cite{jhlee21, jhlee24}. It may thus be of potential interest to investigate the bullet-dwarf case in SIDM models, where the collision progenitors have intrinsically shallower central potential and may give birth to DM-deficient dwarfs that are more diffuse.

\section{Summary and discussion}\label{sec:summary}

In this work, we trace the evolution of dwarf galaxies with different SIDM models and orbits, using controlled simulations. Our goal is to explore the response of the stellar component of a galaxy to the evolution of the DM (sub)halo, from core-creation to core-collapse.  Our test (sub)halos are set up with two species of particles, dark matter (DM) and stars, with the DM component initialized with an NFW profile and the stellar component with either a Hernquist (cuspy) or Plummer (cored) profile.  We first (Sec. \ref{sec:res}) quantify the resolution and particle mass ratio between DM and stars required to carry out this kind of ``star+DM'' controlled simulations. We find that an equal mass ratio between DM and star particles is recommended in order to avoid the energy equipartition effect caused by mass segregation, which would lead to a spuriously increased galaxy size, as also reported in \cite{ludlow19, ludlow23}. As for the particle resolution, we find that our highest resolution (with $\sim 10^7$ particles in total, see Table \ref{table:reso}) is needed in scenarios where a subhalo experiences extreme mass loss and has $\lesssim10^{-3}$ of its initial mass remaining bound, which is relevant for the formation of DM-free galaxies in some SIDM cases. 

We then generate production runs of these controlled simulations of satellite galaxies, combining seven DM models and five orbits. We examine the time evolution of these satellites' half-light radii and mass-to-light ratios (defined by the ratio between total mass and stellar mass) within $r_{\rm half}$ in Sec. \ref{sec:time-evo-rhalf} and \ref{sec:time-evo-mlratio}. The size evolution of dwarf galaxies displays a much larger diversity in SIDM cases than in CDM: the core-creation process of SIDM halos puffs up the galaxy by $\sim 50\%$ in $r_{\rm half}$, which is further boosted to as much as three times (Hernquist; $\sim$ eight times for Plummer galaxies) its initial value if there is non-negligible evaporation of the subhalo by the host. The cored SIDM subhalos are more susceptible to tidal fields, resulting in the stellar components being more easily tidally stripped, leading to a drop in $r_{\rm half}$ following the previous expansion, and eventually to the formation of DM-free galaxies/star-clusters  for Hernquist initial conditions.  Galaxies with cored (Plummer) stellar profiles are not able to be self-bound and will dissolve in the host once the DM component is depleted. By contrast, core-collapsing subhalos experience an accelerating decrease in their galaxies' $r_{\rm half}$, because the stellar components undergo adiabatic contraction in correspondence to the collapsing/contracting DM core. The mass-to-light ratio within $r_{\rm half}$ decreases (becomes more stellar-dominated) with DM core-creation, tidal stripping and evaporation, which all lead to more DM than stellar mass loss in the (sub)halo center. When core-collapse happens, however, the DM concentrates towards the center through the gravothermal process more drastically than do the stellar components, which are only pulled by the deepening gravitational potential. Thus, the mass-to-light ratio increases.  The accelerated size-shrinking and mass-to-light ratio increase can be regarded as signs of ongoing core-collapse, but they are not as obvious when core-collapse happens at radii smaller than $r_{\rm half}$ (see Fig. \ref{fig:ccp-dm}). This happens when a subhalo has experienced a large amount of mass loss before core-collapsing. 

In addition to exploring their time evolution, we also link the stellar properties to the evolution of the total mass within the initial half-light radius $M(<r_{\rm half,0})$. This aims to showcase the ``tidal tracks'' \cite{Penarrubia10, errani15} for SIDM cases, exploring the possibility of establishing parametrized relations for reconstructing dwarf galaxies' properties in larger-scale simulations where the resolution may not be sufficient or in analytical galaxy evolution codes \cite{galacticus, sashimi-sidm}. As shown in Fig. \ref{fig:trackmain} and \ref{fig:hevspl}, apart from the large diversity produced by SIDM core-formation to core-collapse, we do find some universality among the tidal evolution tracks of satellite galaxies' size, mass-to-light ratio and radial velocity dispersion, which are physically interpretable and seemingly parameterizable. These SIDM (satellite) dwarfs thus follow similar evolution tracks, although very different (and with larger span) than the CDM tidal tracks.

It is important to understand the limitations of our controlled star+DM simulations, and other intriguing extensions we may be able to explore in future works, which we list below:

\begin{itemize}
    \item \textit{More test cases for parameterizing tidal tracks} In this work, we demonstrate the potential universality of the tidal evolution tracks of dwarf (satellite) galaxies in SIDM models, with one test case of subhalo and its galaxy orbiting a host. More variants, such as different subhalo masses, different host masses, more SIDM models and orbits, need to be tested for robustness and generalization, which we postpone to future works. 
    
    \item \textit{Core-collapse treatment as a lower limit} Since we define a (sub)halo as core-collapsed and terminate the simulation once its central density $\rho_{\rm cen50}$ exceeds 5 times its initial value, our results in this work should be viewed as a lower limit on an SIDM (sub)halo's inner density. The DM mass contraction should continue into the late core-collapse stage until triggering the relativistic instability and then black hole formation (see \cite{sophia23} for more discussion). However, the stars may follow a different path. As we have shown with the ``exception'' case $\{\sigma_0=200, \omega=50\}$ on orb3 (see Fig. \ref{fig:rhalf-c} and \ref{fig:ml-c}), the stellar properties within $r_{\rm half}$ become ``decoupled'' from the core-collapse process, once the core-collapse radius falls much below the scale of  $r_{\rm half}$. Following the same logic, the stellar evolution within all core-collapsing (sub)halos will eventually become less affected beyond a certain point, since the collapsing DM core keeps shrinking to smaller radii.
    

    \item \textit{Stellar stream} In this work, we focus on analyzing star particles that remain bound to the subhalo, which is approximately a spherical overdensity region determined by the halo finder \texttt{AHF}. Tidally stripped DM and stars form leading and trailing arm-like structures, stretching from and co-moving with the subhalo. The morphology of these stellar streams can be used to infer the structure of their progenitors \cite{varghese11} and constrain DM models. Perturbations to stellar streams caused by nearby small subhalos are observed as gaps and off-stream spurs associated with stellar streams \cite{bonaca18}, which reveal the presence of invisible subhalos and may also be useful for inferring their structures. With the growing number of stellar streams observed in current and upcoming surveys \cite{tsli19, bonaca24}, it would be important and intriguing to theoretically predict properties of stellar streams under different SIDM models \cite{xyzhang24-gd1}, for which the ``star+DM'' controlled simulations we present in this work may be a good start.  A demonstrative, visualized animation of such stellar streams from our simulations can be found \href{https://youtu.be/4kOeuhtVzhc}{here}, where we include four different dark matter models and put the subhalo on orb2. 
    
    \item \textit{Star formation} The expansion and contraction of star particles in response to SIDM core-creation and core-collapse, as we show in this work, in principle should also apply to gas, which is missing in our simulations. This may have interesting implications for the star formation rate. For example, when core-collapse happens, the gas should undergo contraction toward the center together with DM and stars, increasing the star formation rate. But, on the other hand, the gas would also be heated up during this process, countering the formation of stars. The overall effect of SIDM physics on star formation can thus be complicated but intriguing, and may display a diversity as well, coupling with the subhalo's orbital history. 
    
    \item \textit{Dark matter deficient galaxies} We have shown that SIDM dwarf satellites can easily become DM-deficient, due to the relatively fast depletion of the DM component in the tidal (and evaporation) field (see also \cite{dnyang20-df2, zczhang24}). However, the stellar remnants need to have a cuspy, compact distribution to survive the tidal field, after the DM is removed. These stable, DM-free stellar remnants are thus different from the recently observed DM-deficient galaxies NGC-1052 DF2 and DF4 \cite{df2, df4}, which are reportedly ultradiffuse. Nevertheless, during the evolution path, we do still find a window in the mass-to-light ratio vs. stellar size plane that can satisfy both the diffuse and DM-deficient features of DF2/DF4 (see Fig. \ref{fig:track-ml-size}).  Such DF2/DF4-like dwarfs formed through this SIDM-tidal scenario, however, are only expected to have these features for a very brief period during their life-evolution. They should thus be relatively rare in observations, with strong orbit/eniroment dependence. Further dedicated studies into the SIDM-driven formation of such DF2/DF4-like systems are thus needed, testing multiple variants such as the progenitor mass, initial stellar density profile, and the dwarf galaxy as a bound satellite or a flyby. 
\end{itemize}

This work serves as part of a series to systematically study the stellar response to SIDM halo evolution and subhalo-host interactions, revealing both the diversity and the universality in SIDM dwarfs' stellar evolution. Here, we quantitatively show that SIDM has the potential to explain the observed diversity in dwarf galaxies.  In future work, we will  parameterize the tidal tracks of SIDM dwarfs for general use in the community.

\begin{acknowledgments}
We thank Stacy Kim, Daneng Yang, Ethan Nadler, Raphael Errani, Akaxia Cruz, Hai-bo Yu, Louis Strigari, Joohyun Lee and Alexander Knebe for useful discussions.

Z.C. Zeng is partially supported by the Presidential Fellowship of the Ohio State University Graduate School. We thank the host of the Kavli Institute for Theoretical Physics during the darkmatter24 workshop, where most of the authors of this work were invited, and the grant NSF PHY-2309135 to KITP for this program.

The simulations in this work were conducted on Ohio Supercomputer Center \cite{osc}, mostly on the CCAPP condo. 
\end{acknowledgments}

\appendix


\section{Initial stellar distribution as a Plummer profile}\label{appdx:plummer}

\begin{figure}
    \includegraphics[width=\columnwidth]{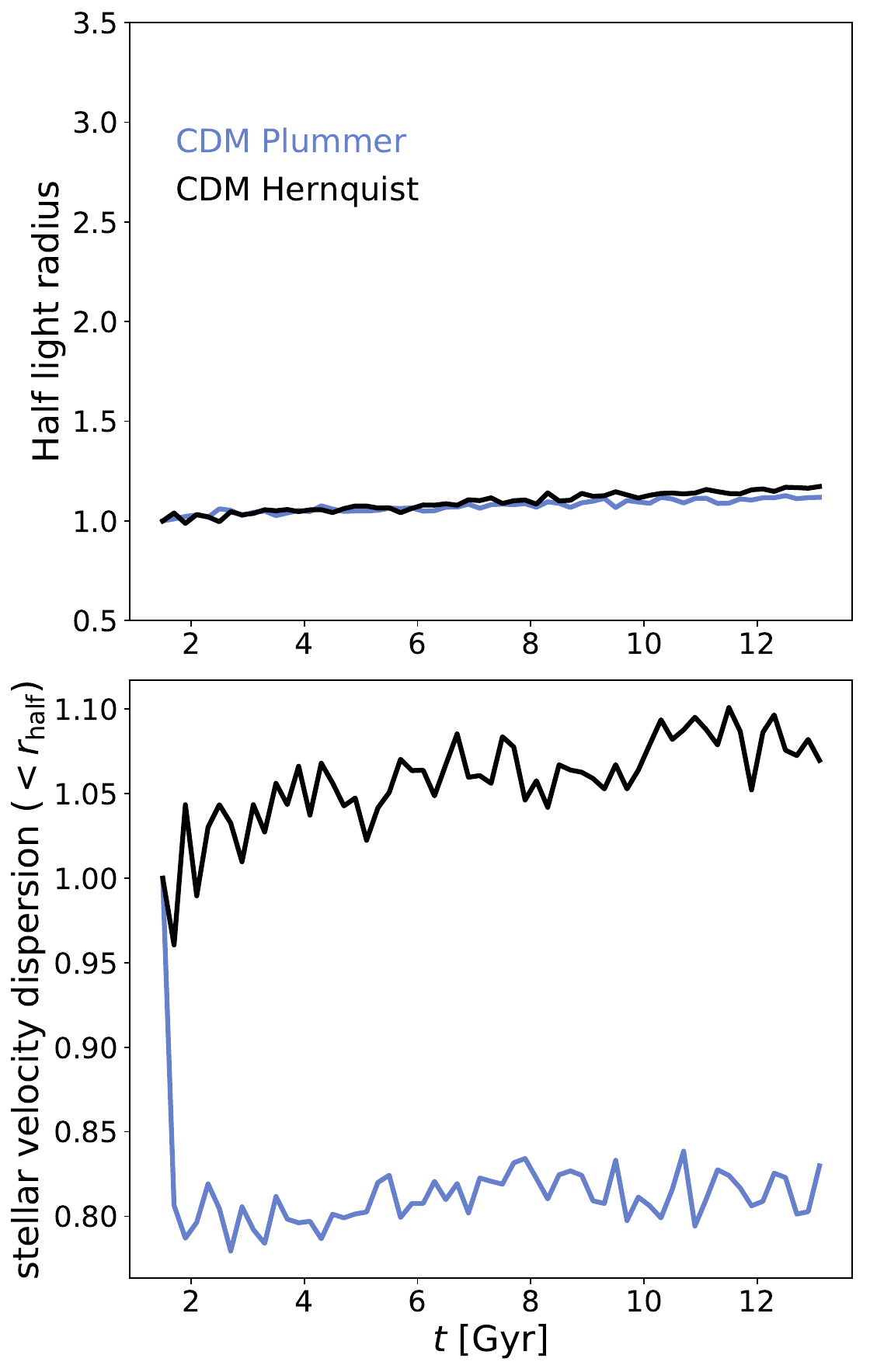} 
    \caption{ The initial condition issue for the Plummer galaxy.  The top panel shows the time evolution of the $r_{\rm half}$ for the Plummer (blue) and Hernquist (black) galaxies, where both cases are stable in time. The bottom panel shows the stellar velocity dispersion within $r_{\rm half}$, where we can clearly see a sharp drop at the first snapshot for the Plummer case. }
    \label{fig:plummer-vdisp}
\end{figure}

In setting up the Plummer galaxies in Sec. \ref{sec:hevspl}, we encounter an issue that the initial conditions of the stellar components are not generated in equilibrium as is intended. Here we describe the problem, explain the physical reason behind it, and describe our temporary fix to it.

In Fig. \ref{fig:plummer-vdisp}, we show the time evolution of the Plummer and Hernquist galaxies in isolation with a CDM halo component, using our highest, converged particle resolution and 1:1 mass ratio between DM and star particles (see Sec. \ref{sec:res}). In this scenario, both cases are expected to stay stable at the initial value. We can see in Fig. \ref{fig:plummer-vdisp} that the stellar size indeed only slightly grows over time ($<10\%$ by the end of the simulation) for both the Plummer and Hernquist cases, which is not a physical effect but lingering numerical issues we discussed in Sec. \ref{sec:res}. However, the stellar velocity dispersion within $r_{\rm half}$ sees a sharp drop of $\sim 20\%$ at the very first snapshot for the Plummer galaxy, which then stays at a plateau afterwards, while the Hernquist galaxy has an approximately stable velocity dispersion. This strongly suggests that the initial velocity distribution of the Plummer case is not compatible with its density distribution, leading to this obvious non-equilibrium behavior. 

This issue is known and acknowledged by the initial condition generator \texttt{SpherIC}, with a physically intuitive explanation that the NFW profile for DM is too steep to hold a cored Plummer stellar component. A quantitative understanding of this problem starts with the Eddington Inversion method~\cite{eddington16, lacroix18, errani20}, which \texttt{SpherIC} uses to calculate the velocity distribution, including 1-component (DM-only) or 2-component (DM+stars) systems in equilibrium:

\begin{equation}
    f(E) = \frac{1}{\sqrt{8}\pi^2} \int_E ^0 \frac{d^2 \nu}{d \Phi^2} (\Phi -E)^{-1/2}d\Phi.
\end{equation}

Here $\nu$ is the number density of a species of particles (either DM or star), $\Phi$ is the gravitational potential and $E$ is the energy of a particle. The distribution function $f(E)\equiv dN/d\Omega$ then determines the number density of simulation particles in the phase space $\Omega=(\vec{v}, \vec{r})$. For the case with Plummer stars and NFW dark matter, we find that the second order derivative $\frac{d^2 \nu}{d \Phi^2}$ can become negative at very small radii ($\lesssim$ 0.1 kpc for our test galaxy-halo system in this work), which then leads to the distribution function having negative values at small radii.  Since the probability of assigning a specific velocity (or energy $E$) cannot be negative, \texttt{SpherIC} complains and uses $f(E)=0$ at these small radii instead of the calculated negative values.

We remark that this issue of a negative distribution function is  intrinsic to setting up a Plummer galaxy in an NFW halo. 
In fact, this incompatibility has been used to rule out cuspy DM density profiles in cored low-mass dwarfs \cite{almeida23, almeida24} (see also \cite{battaglia08, strigari10, strigari17}). Hence, we choose a temporary fix for this work. First, note that in Fig. \ref{fig:plummer-vdisp}, the stellar size of the Plummer case remains stable in time, despite the problem of non-equilibrium. This suggests that the effect of the issue is limited, mostly on the calculation relevant to velocity dispersion, such as our tidal tracks in Fig. \ref{fig:hevspl-f}. Second, as we can see in the bottom panel of Fig. \ref{fig:plummer-vdisp}, the system can equilibrate soon after the simulation begins, with the stellar velocity dispersion barely evolving after the first output snapshot. Therefore, we take the first snapshot of the Plummer galaxy embedded in an isolated CDM halo as the new initial condition for the rest of the tidal track studies of the Plummer cases in Sec. \ref{sec:hevspl}.

\section{Tidal tracks for all simulations}\label{appdx:tidal-track-full}

\begin{figure*}
    \centering
    \begin{subfigure}{0.42\textwidth}
        \centering
        \includegraphics[width=\textwidth]{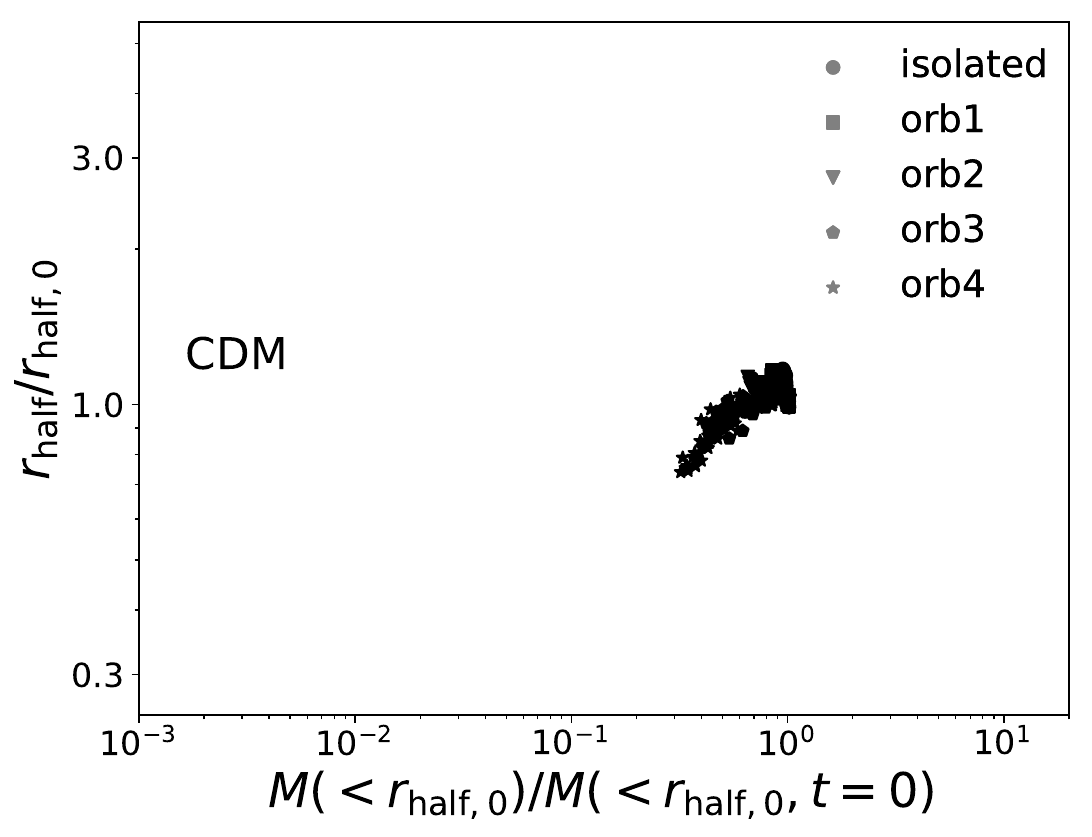} 
        \label{fig:trackall-size-a}
    \end{subfigure}
    ~
    \vspace{-0.5cm}
    \begin{subfigure}{0.42\textwidth}
        \centering
        \includegraphics[width=\textwidth]{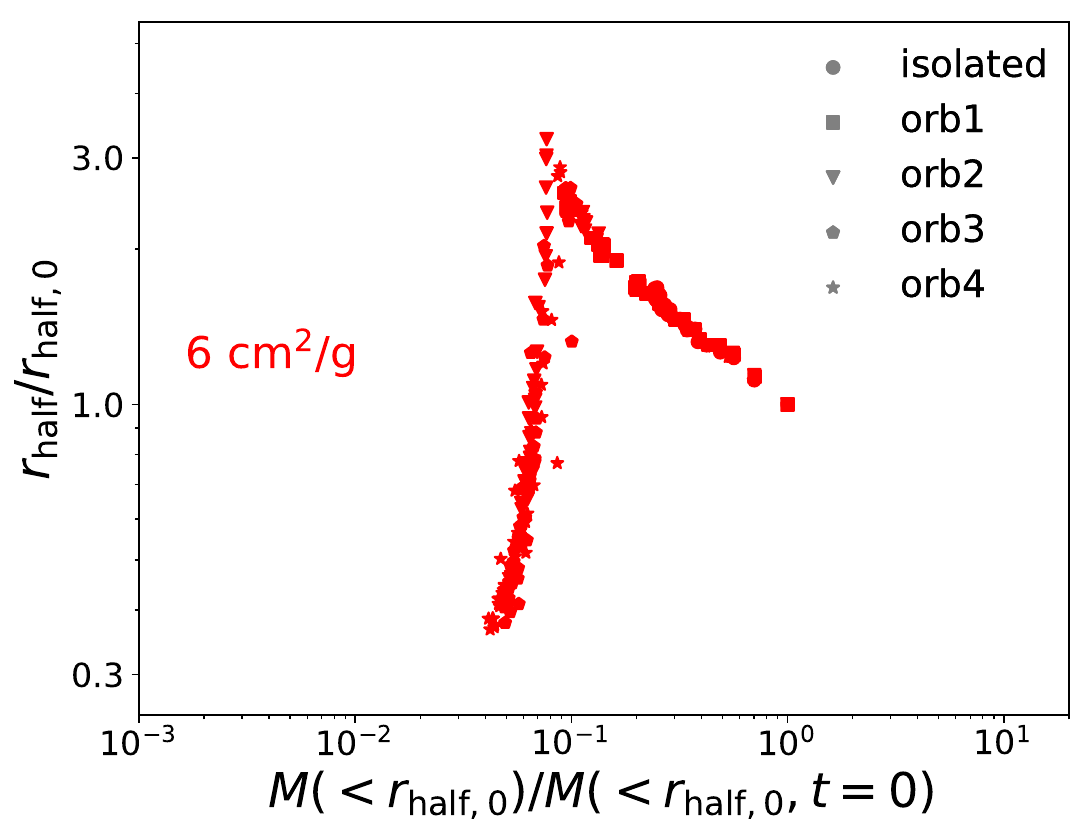} 
        \label{fig:trackall-size-b}
    \end{subfigure}
    ~
    \vspace{-0.5cm}
    \begin{subfigure}{0.42\textwidth}
        \centering
        \includegraphics[width=\textwidth]{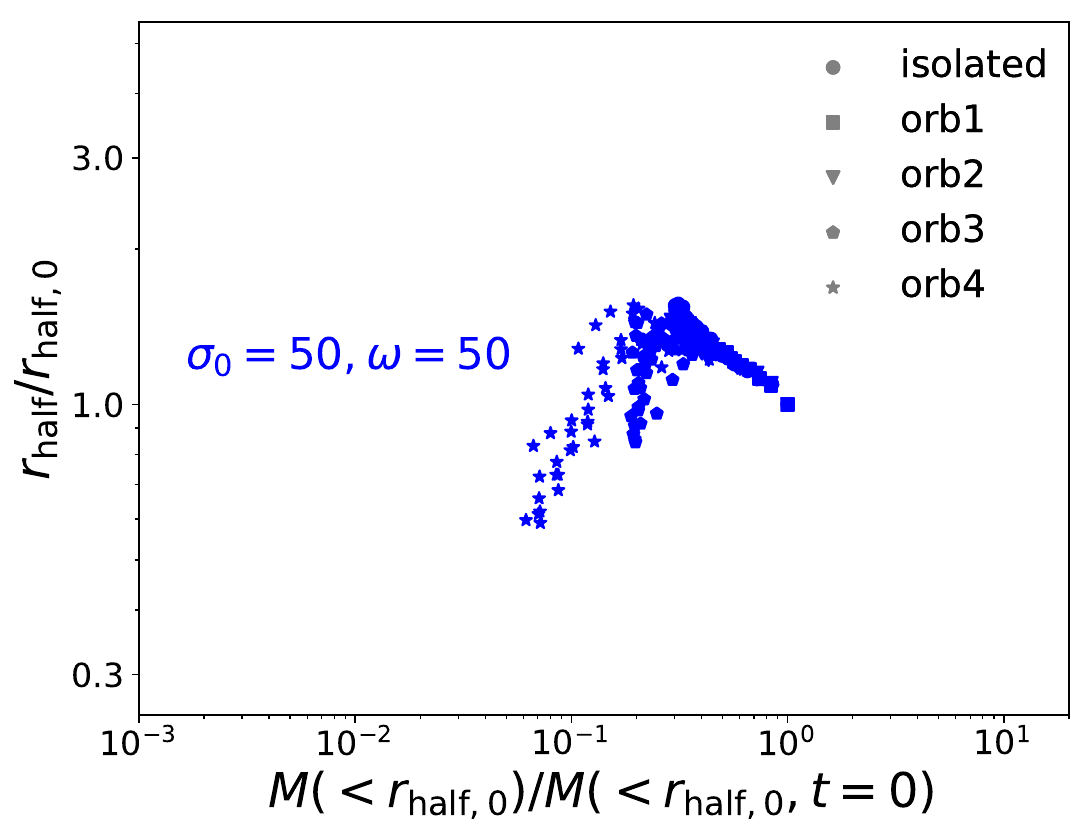} 
        \label{fig:trackall-size-c}
    \end{subfigure}
    ~
    \begin{subfigure}{0.42\textwidth}
        \centering
        \includegraphics[width=\textwidth]{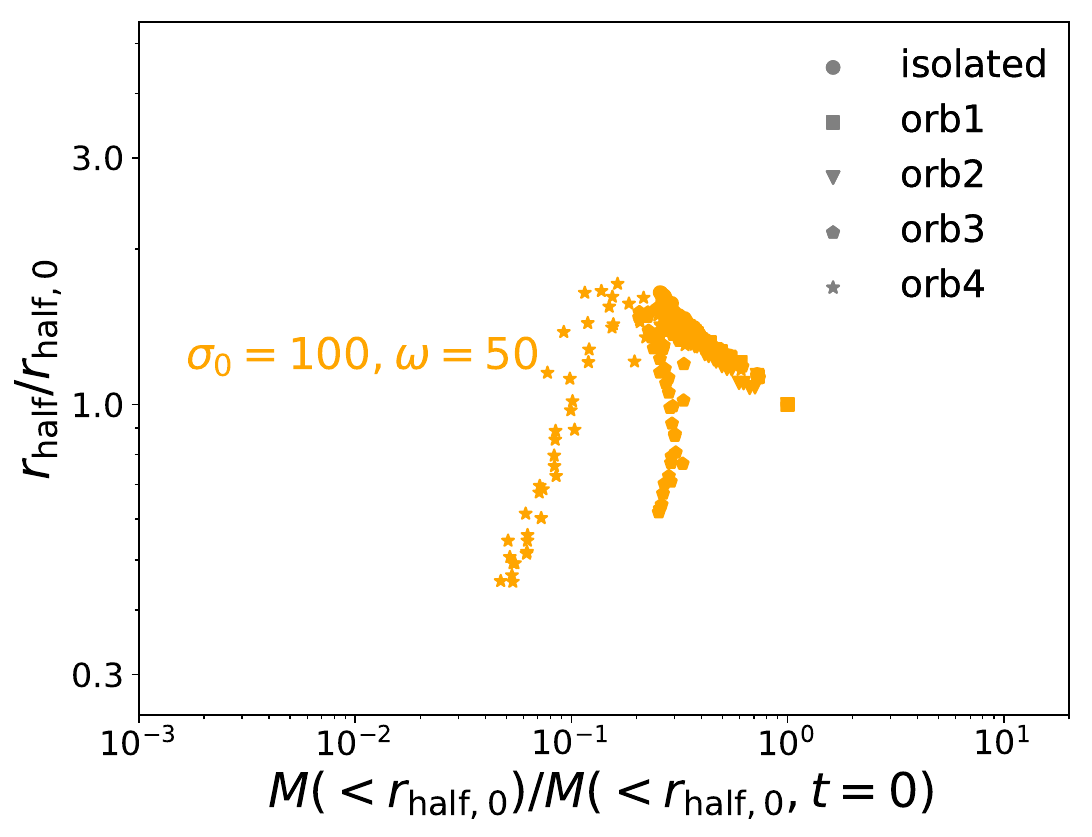} 
        \label{fig:trackall-size-d}
    \end{subfigure}
    ~
    \vspace{-0.5cm}
    \begin{subfigure}{0.42\textwidth}
        \centering
        \includegraphics[width=\textwidth]{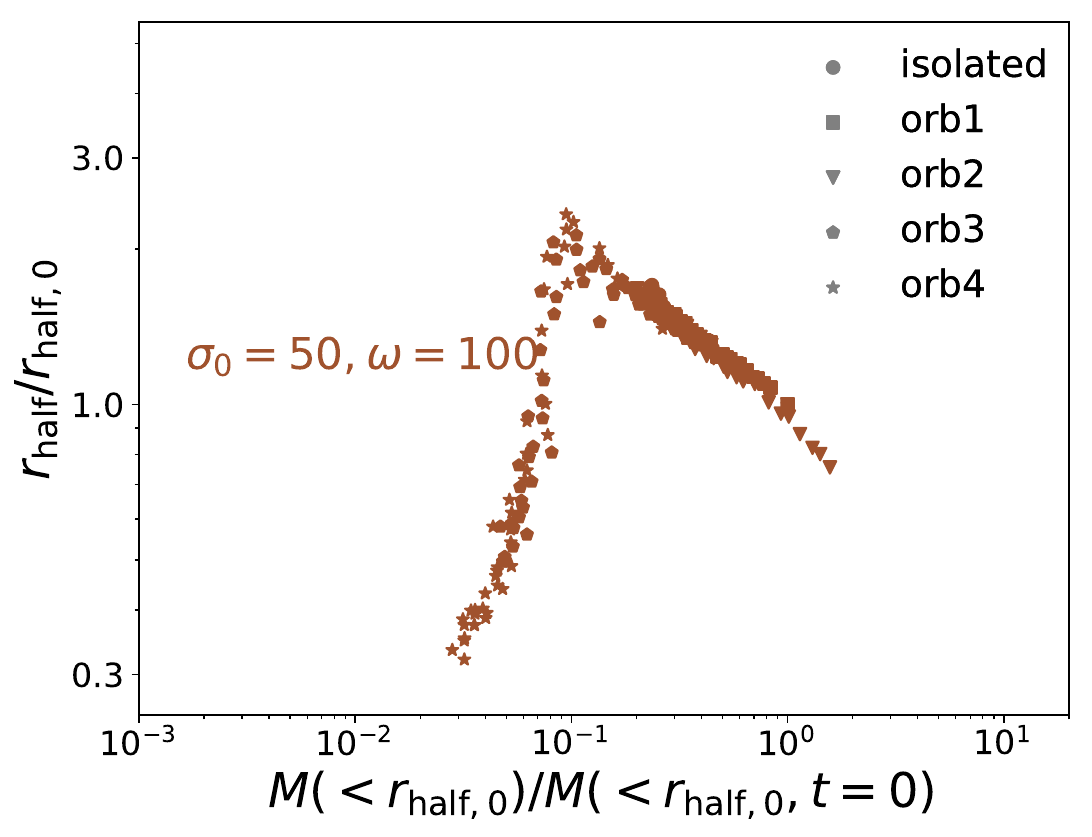} 
        \label{fig:trackall-size-e}
    \end{subfigure}
    ~
    \begin{subfigure}{0.42\textwidth}
        \centering
        \includegraphics[width=\textwidth]{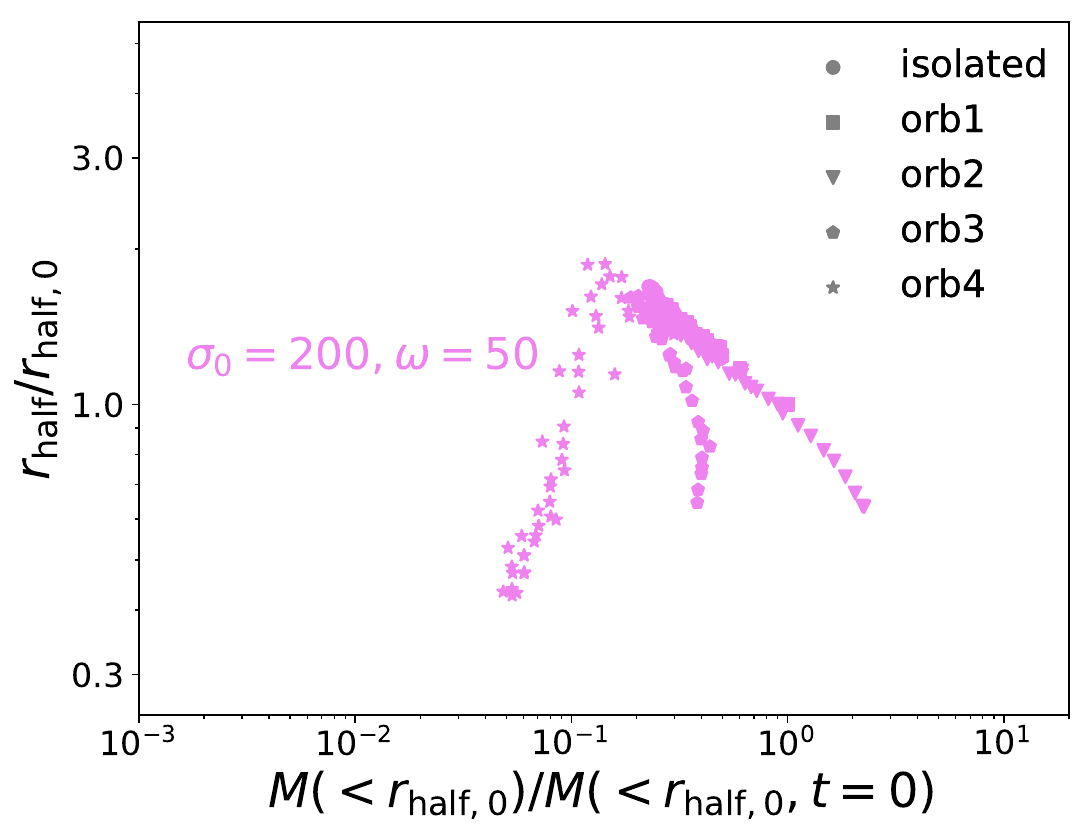} 
        \label{fig:trackall-size-f}
    \end{subfigure}
    ~
    \begin{subfigure}{0.42\textwidth}
        \centering
        \includegraphics[width=\textwidth]{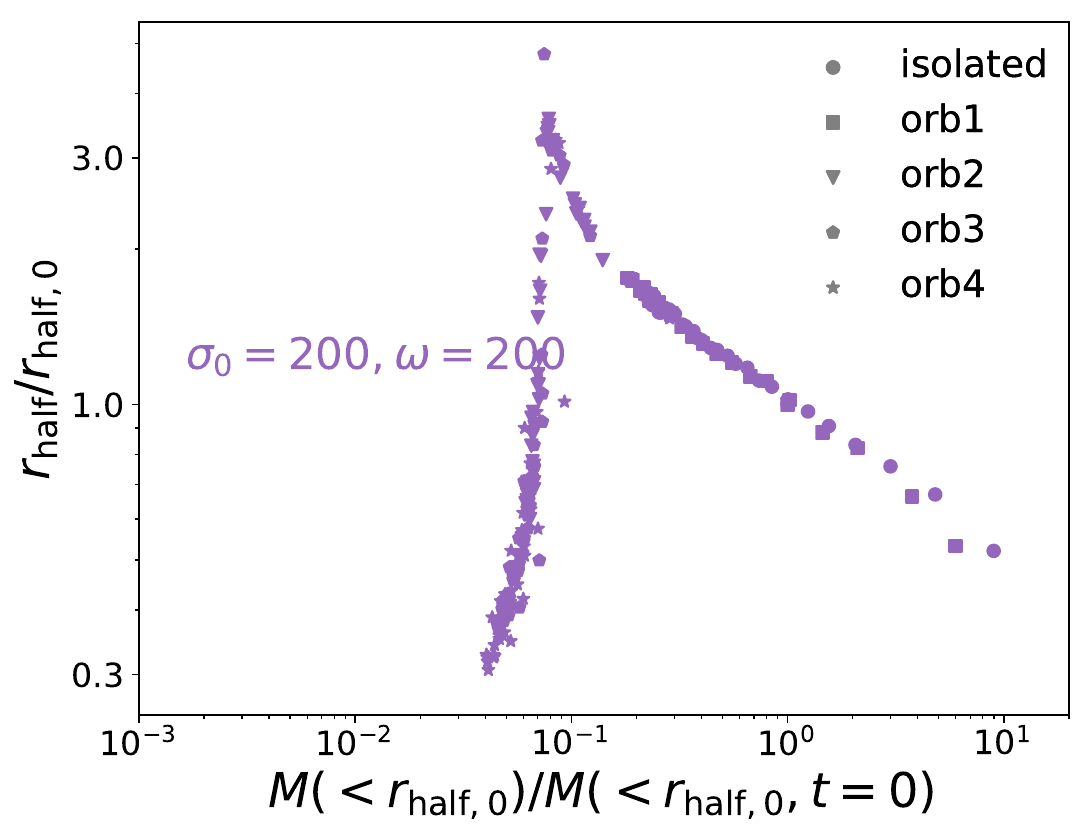} 
        \label{fig:trackall-size-g}
    \end{subfigure}
    \vspace{-0.3cm}  
    \caption{ Tidal tracks for the stellar half-light radius $r_{\rm half}$ from all simulations. }
    \label{fig:trackall-size}
\end{figure*}

\begin{figure*}
    \centering
    \begin{subfigure}{0.42\textwidth}
        \centering
        \includegraphics[width=\textwidth]{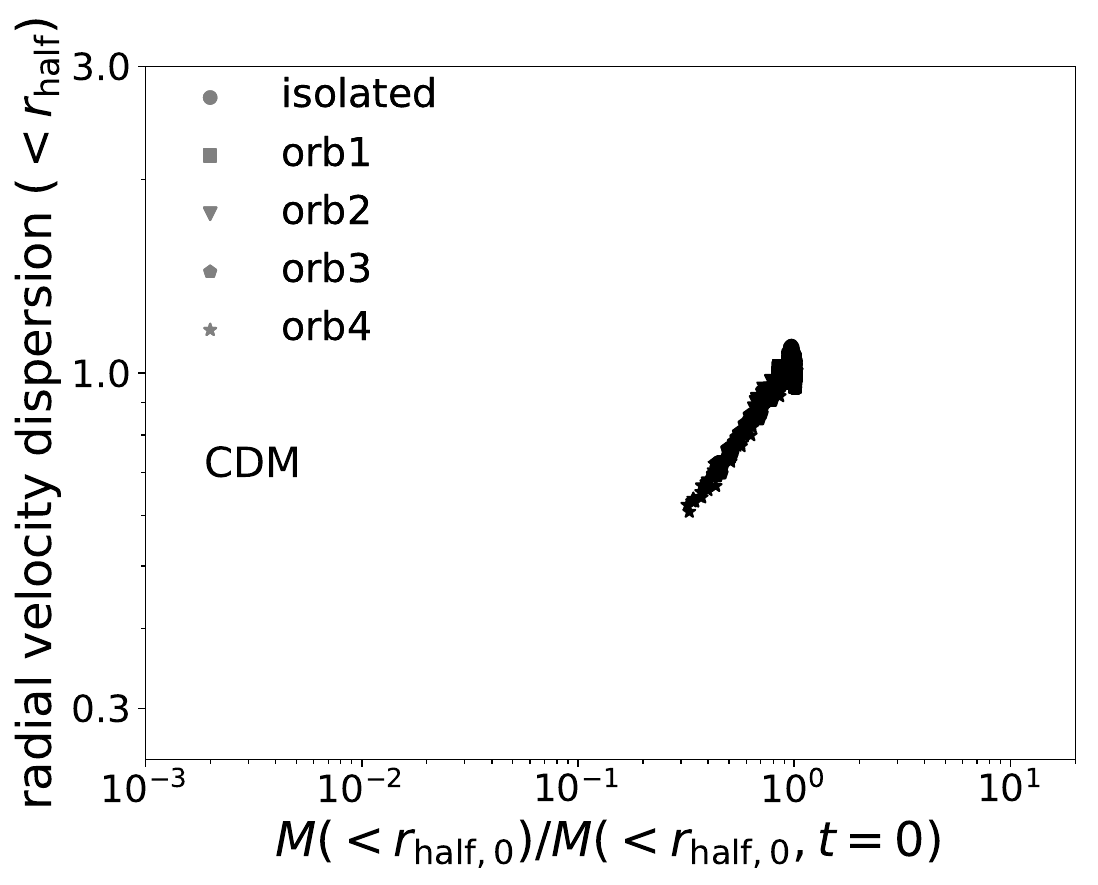} 
        \label{fig:trackall-vdisp-a}
    \end{subfigure}
    ~
    \vspace{-0.5cm}
    \begin{subfigure}{0.42\textwidth}
        \centering
        \includegraphics[width=\textwidth]{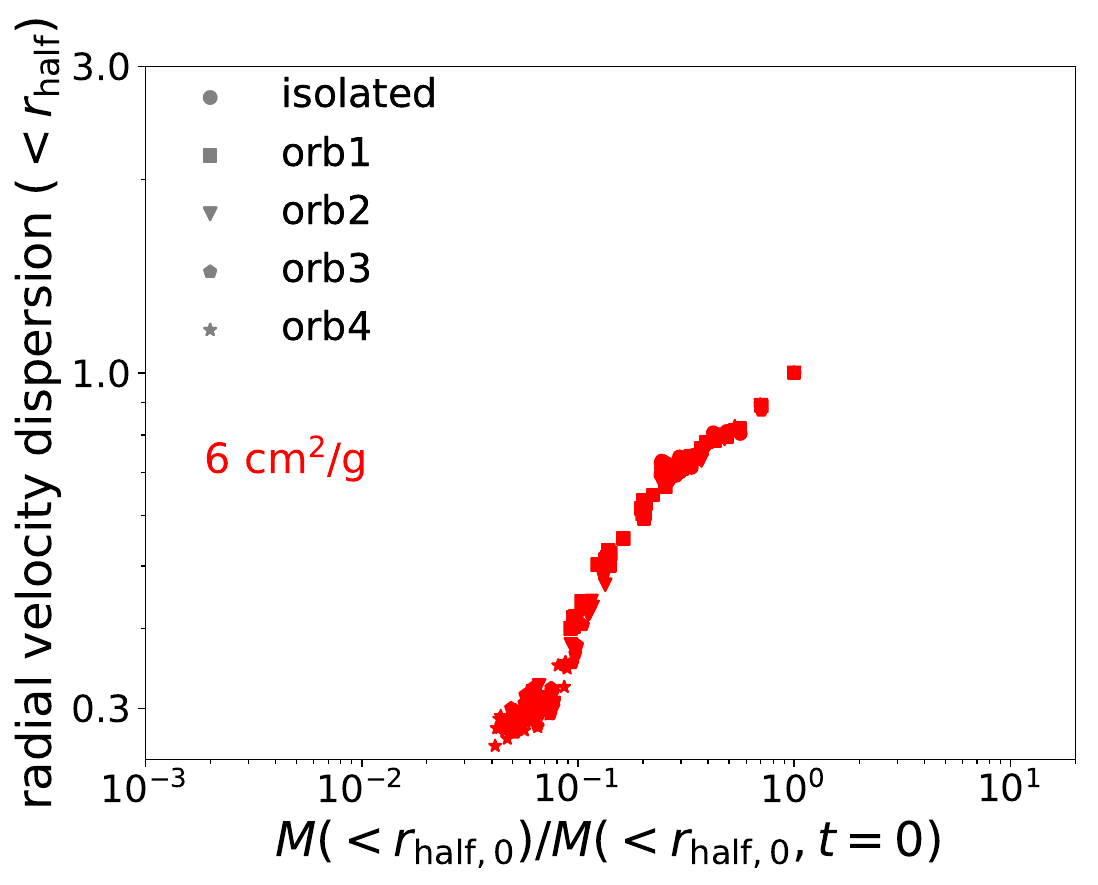} 
        \label{fig:trackall-vdisp-b}
    \end{subfigure}
    ~
    \begin{subfigure}{0.42\textwidth}
        \centering
        \includegraphics[width=\textwidth]{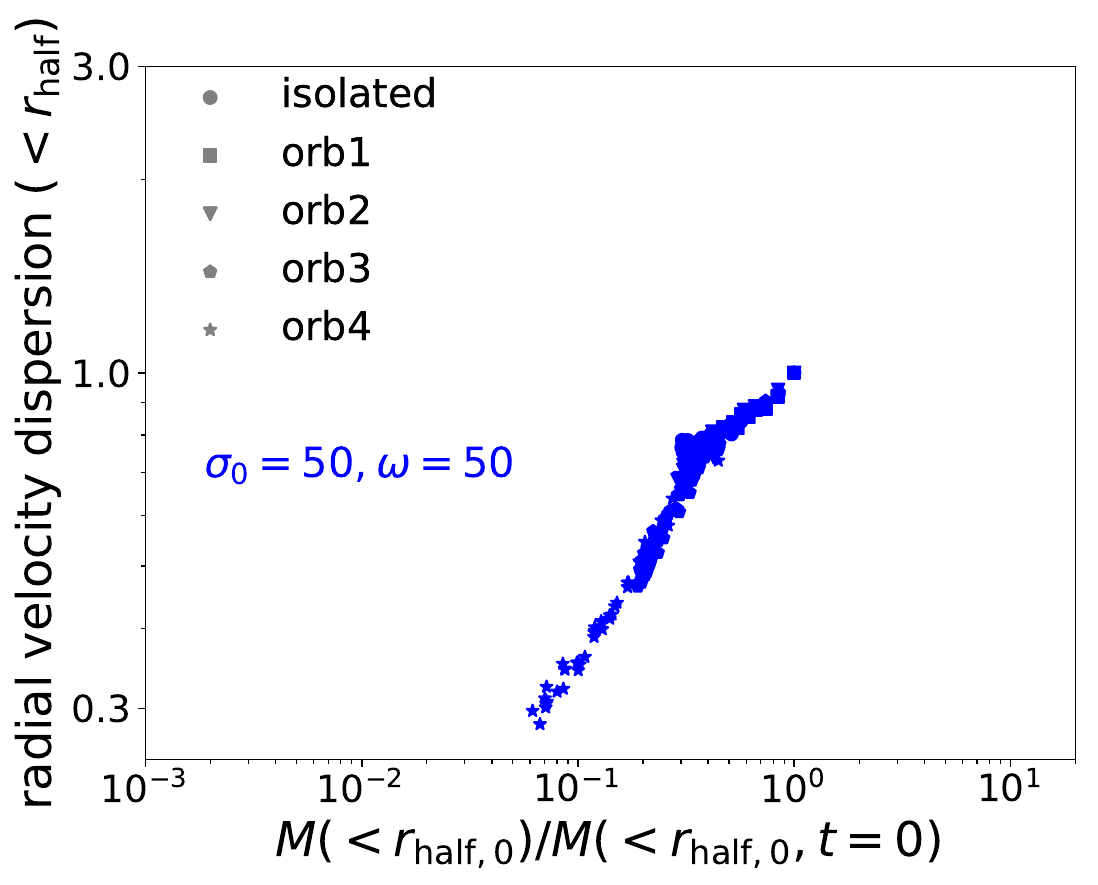} 
        \label{fig:trackall-vdisp-c}
    \end{subfigure}
    ~
    \vspace{-0.5cm}
    \begin{subfigure}{0.42\textwidth}
        \centering
        \includegraphics[width=\textwidth]{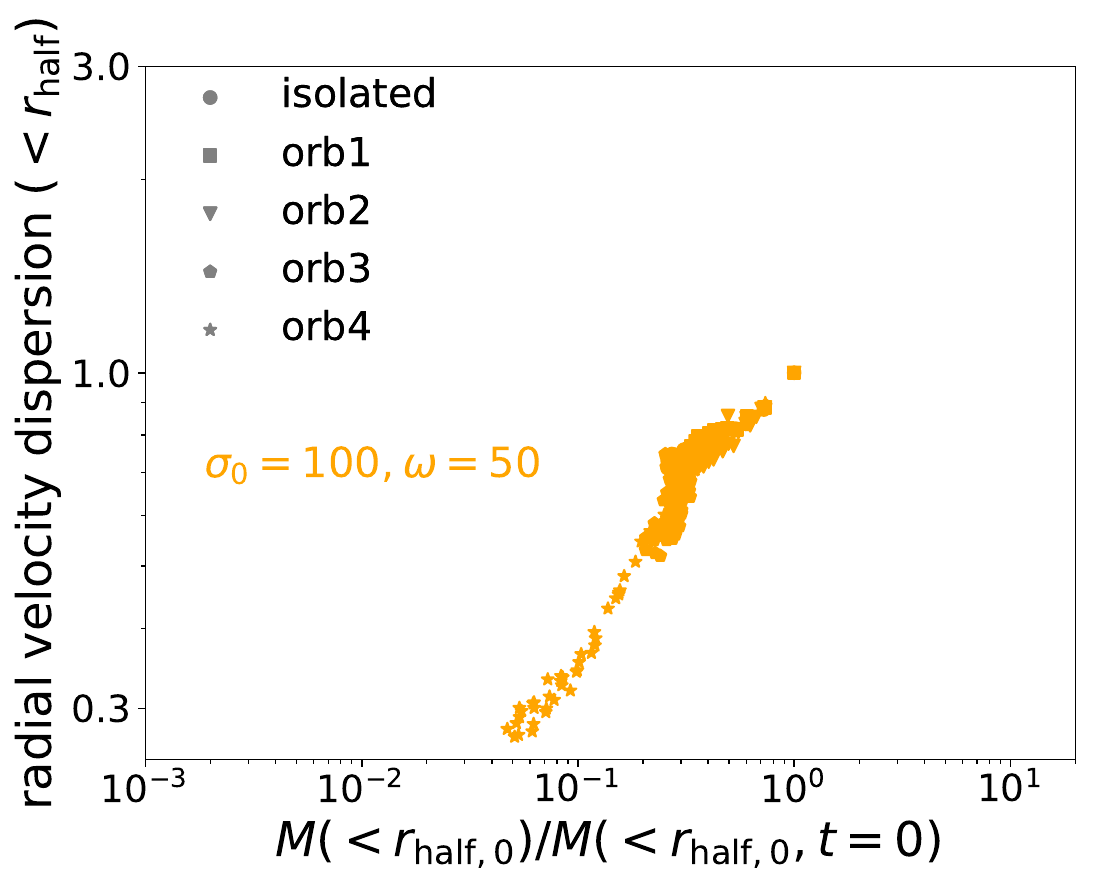} 
        \label{fig:trackall-vdisp-d}
    \end{subfigure}
    ~
    \begin{subfigure}{0.42\textwidth}
        \centering
        \includegraphics[width=\textwidth]{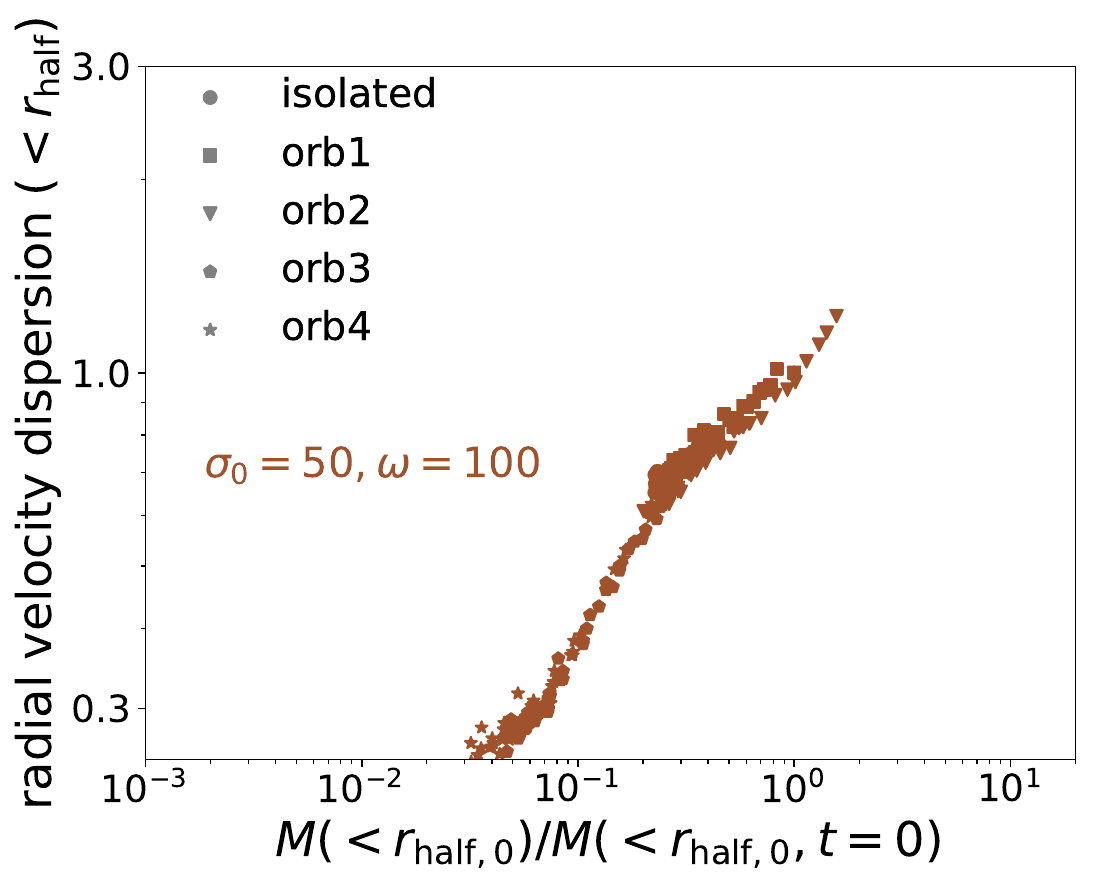} 
        \label{fig:trackall-vdisp-e}
    \end{subfigure}
    ~
    \vspace{-0.5cm}
    \begin{subfigure}{0.42\textwidth}
        \centering
        \includegraphics[width=\textwidth]{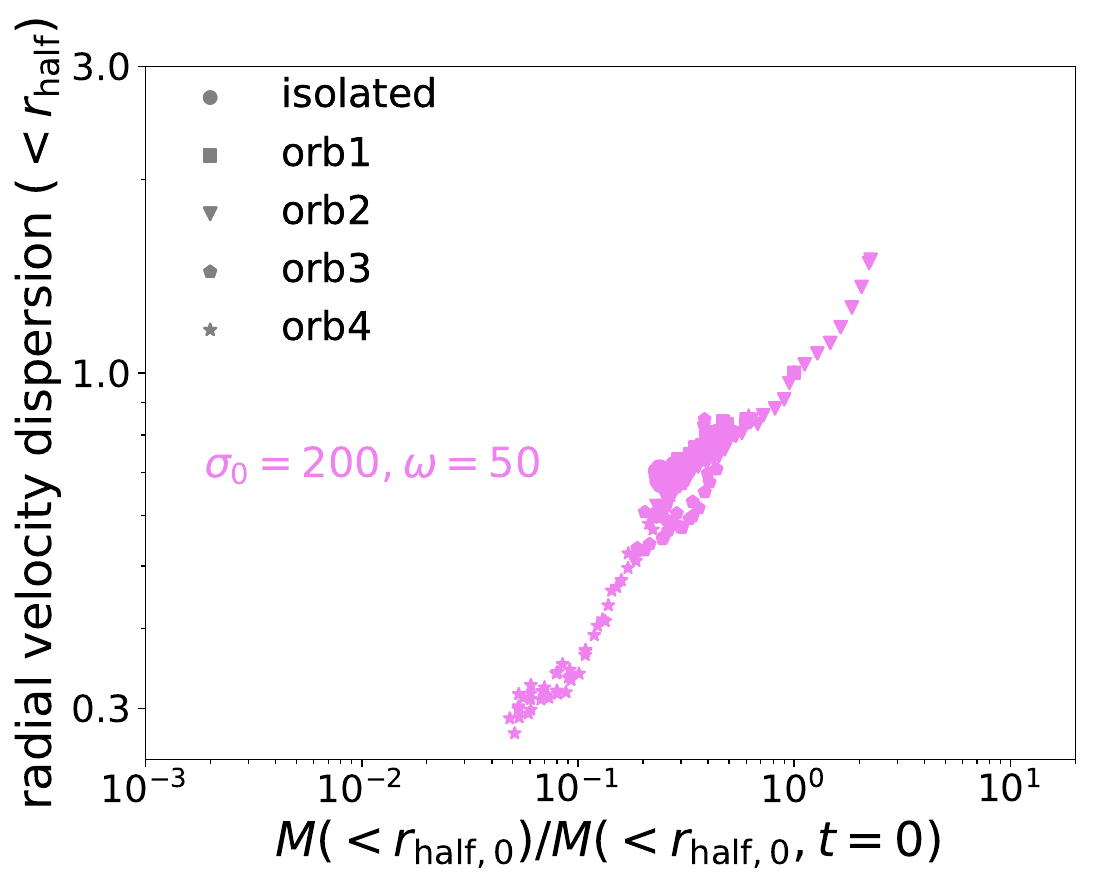} 
        \label{fig:trackall-vdisp-f}
    \end{subfigure}
    ~
    \begin{subfigure}{0.42\textwidth}
        \centering
        \includegraphics[width=\textwidth]{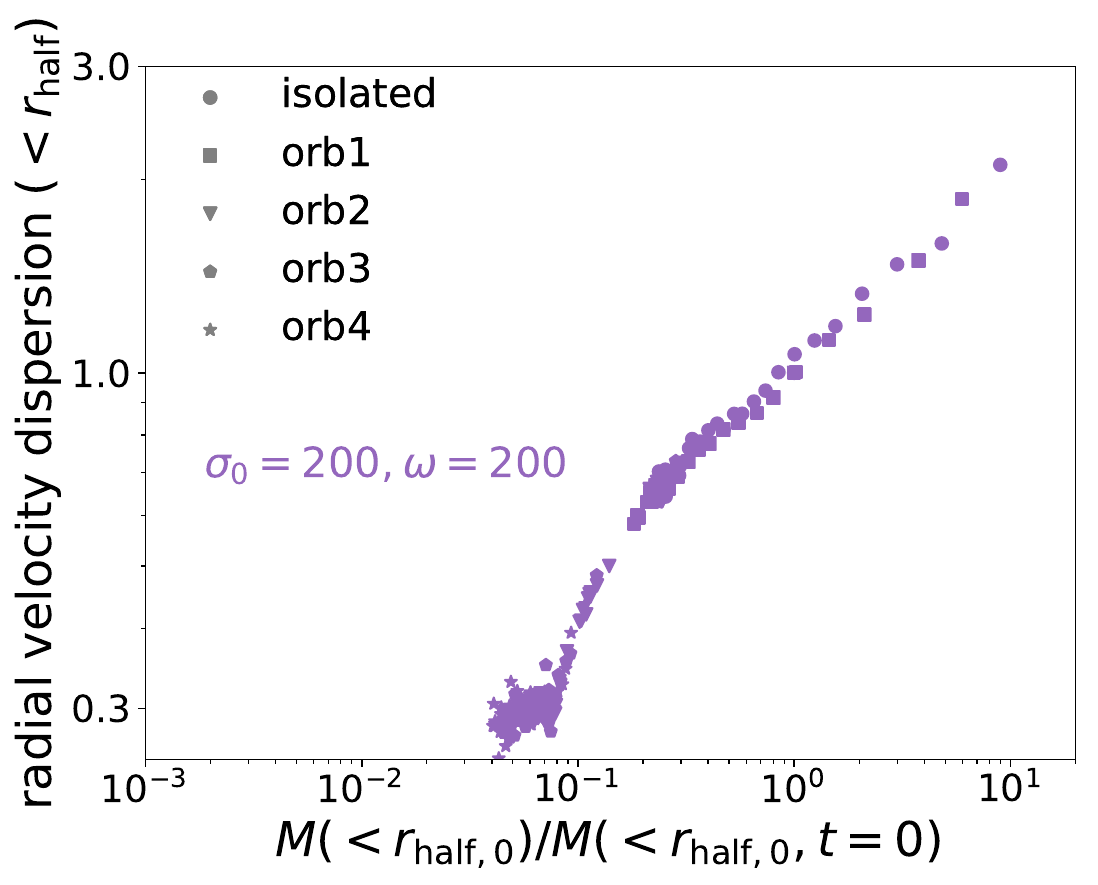} 
        \label{fig:trackall-vdisp-g}
    \end{subfigure}
    \vspace{-0.5cm}
    \caption{ Tidal tracks for the (radial) stellar velocity dispersion within $r_{\rm half}$ from all simulations. }
    \label{fig:trackall-vdisp}
\end{figure*}

\begin{figure*}
    \centering
    \begin{subfigure}{0.42\textwidth}
        \centering
        \includegraphics[width=\textwidth]{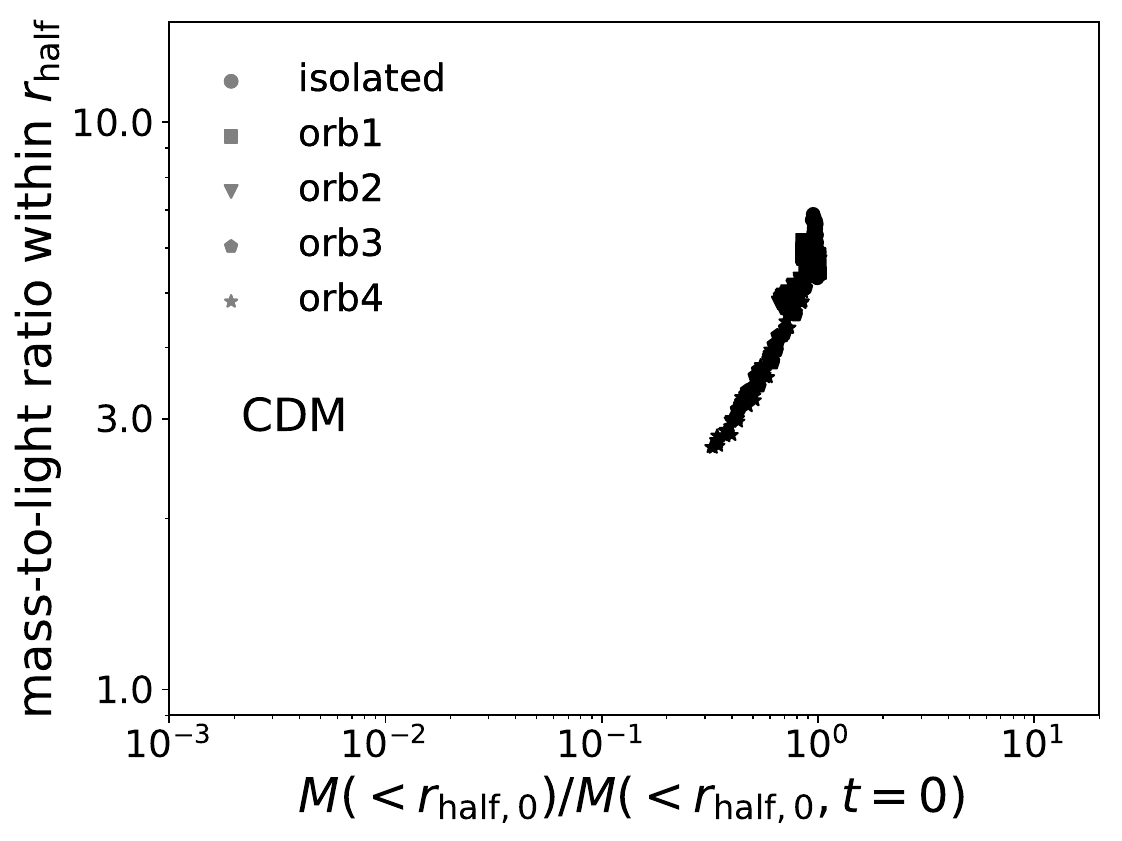} 
        \label{fig:trackall-mlratio-a}
    \end{subfigure}
    ~
    \vspace{-0.5cm}
    \begin{subfigure}{0.42\textwidth}
        \centering
        \includegraphics[width=\textwidth]{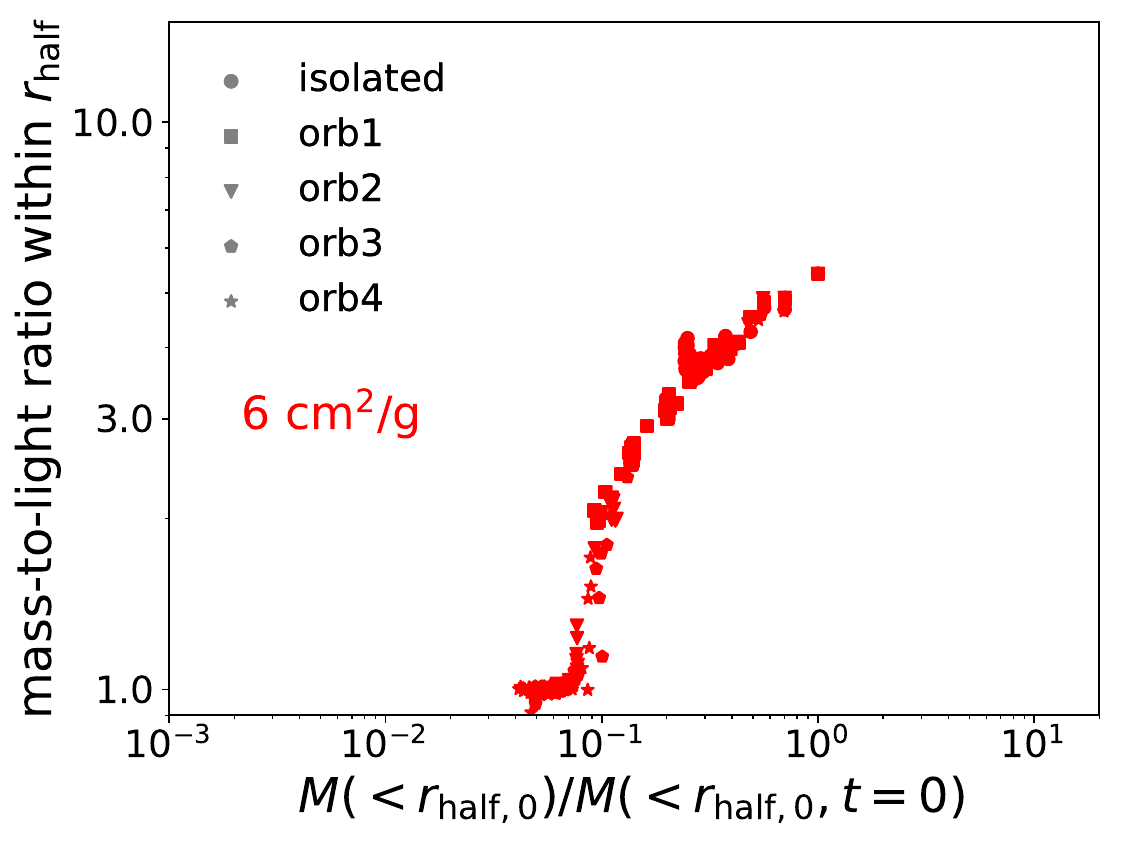} 
        \label{fig:trackall-mlratio-b}
    \end{subfigure}
    ~
    \vspace{-0.5cm}
    \begin{subfigure}{0.42\textwidth}
        \centering
        \includegraphics[width=\textwidth]{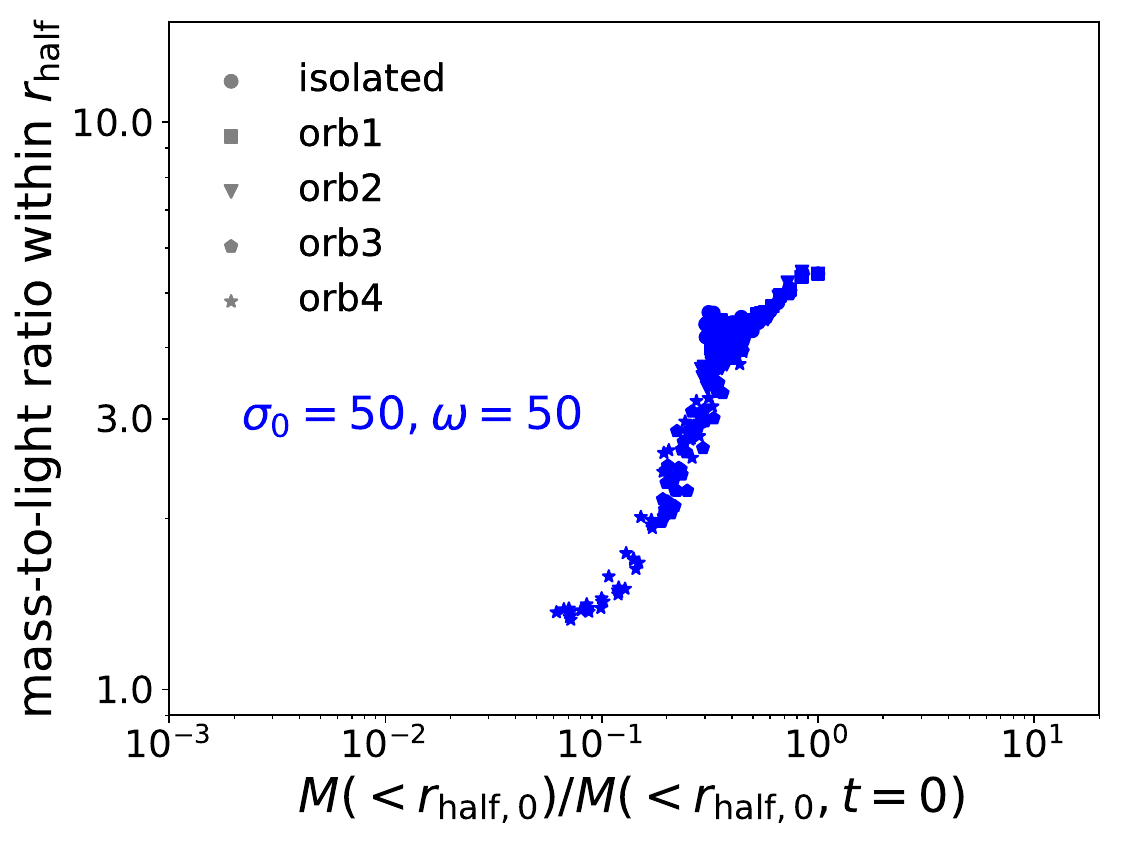} 
        \label{fig:trackall-mlratio-c}
    \end{subfigure}
    ~
    \begin{subfigure}{0.42\textwidth}
        \centering
        \includegraphics[width=\textwidth]{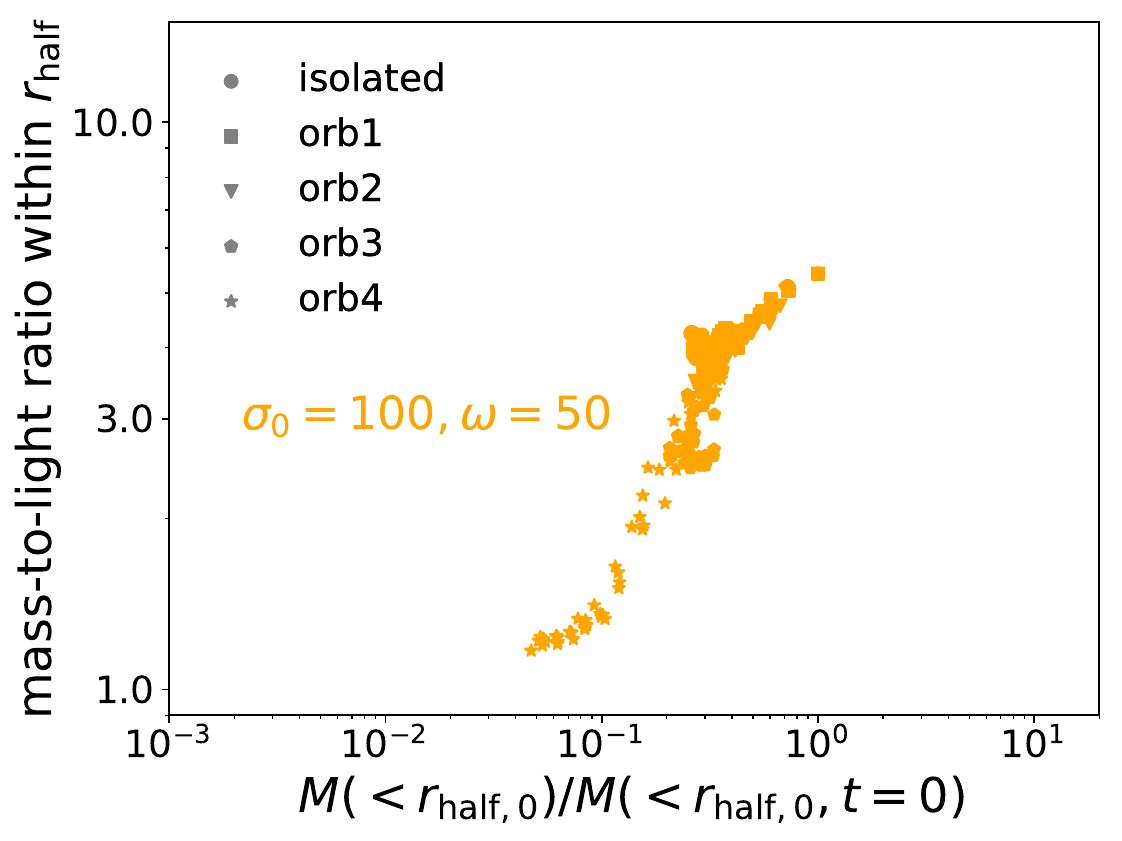} 
        \label{fig:trackall-mlratio-d}
    \end{subfigure}
    ~
    \vspace{-0.5cm}
    \begin{subfigure}{0.42\textwidth}
        \centering
        \includegraphics[width=\textwidth]{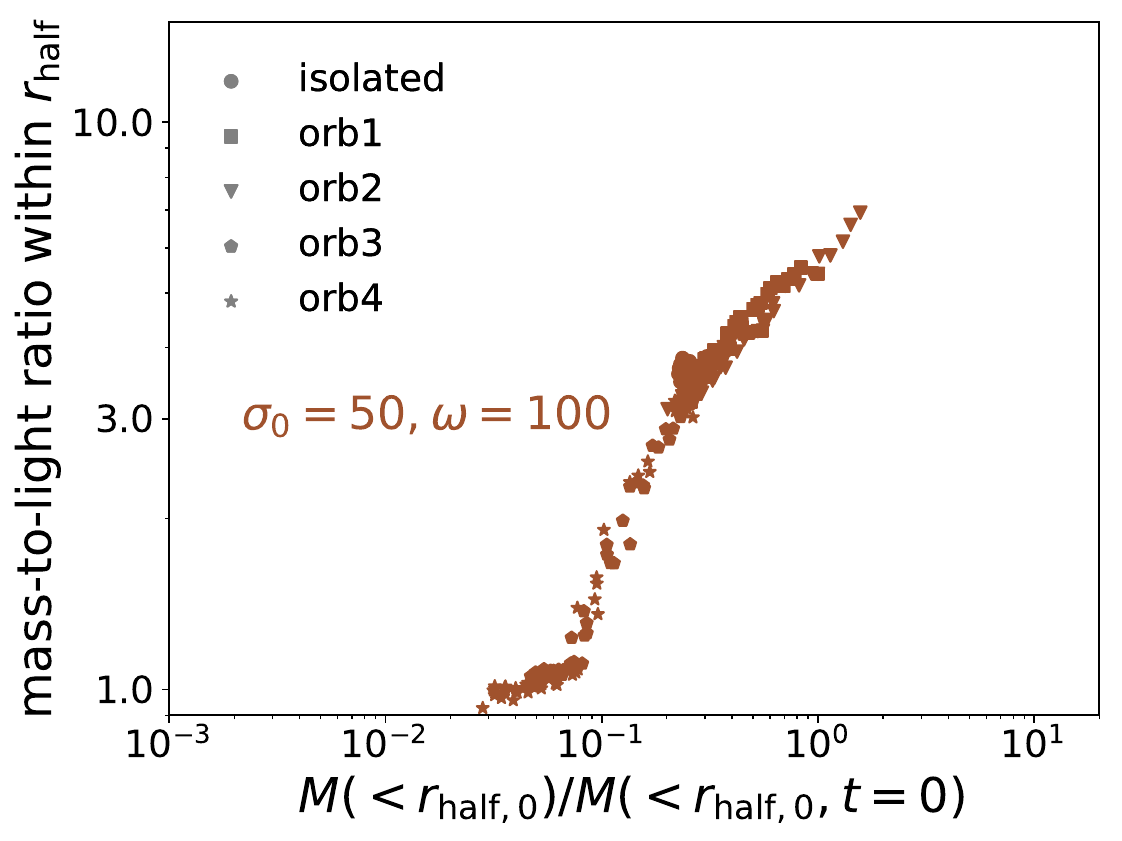} 
        \label{fig:trackall-mlratio-e}
    \end{subfigure}
    ~
    \begin{subfigure}{0.42\textwidth}
        \centering
        \includegraphics[width=\textwidth]{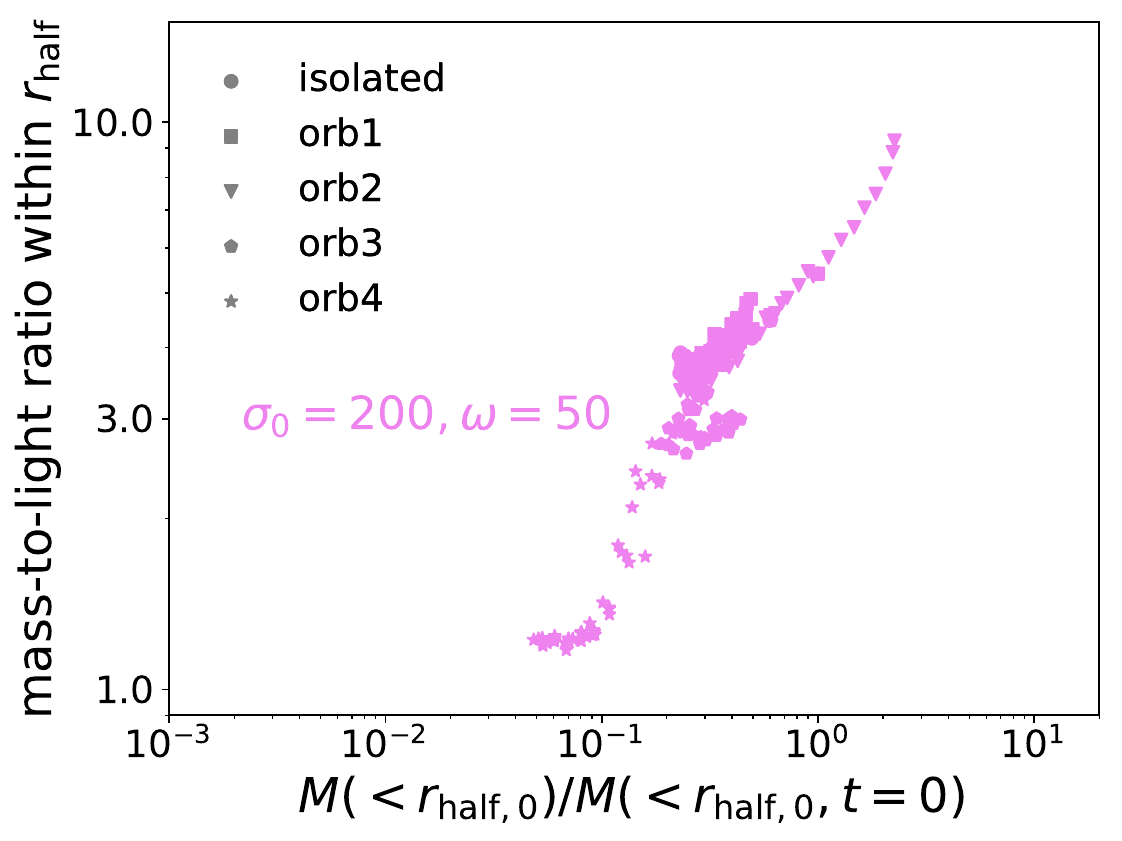} 
        \label{fig:trackall-mlratio-f}
    \end{subfigure}
    ~
    \vspace{-0.5cm}
    \begin{subfigure}{0.42\textwidth}
        \centering
        \includegraphics[width=\textwidth]{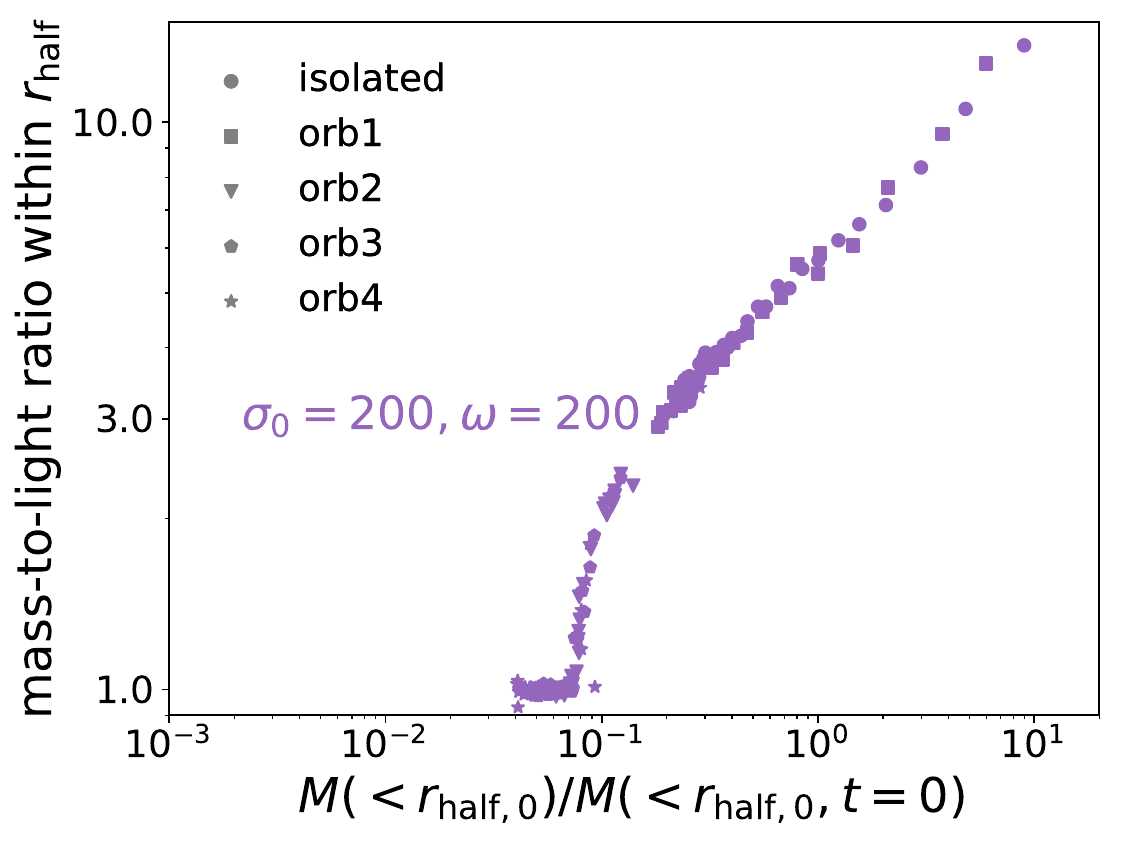} 
        \label{fig:trackall-mlratio-g}
    \end{subfigure}
    \caption{ Tidal tracks for the mass-to-light ratio within $r_{\rm half}$ from all simulations. }
    \label{fig:trackall-mlratio}
\end{figure*}

In Section \ref{sec:track}, we show only the selected, representative evolution tracks of dwarf satellites' stellar properties under SIDM halo evolution and orbital effects, with the same Hernquist initial stellar distribution as in Section \ref{sec:track}, aiming to showcase a few common and extreme cases (and thus the range of `diversity'). In this appendix, we present the tidal tracks of all individual (sub)halo-dwarf galaxy systems that we have simulated in this work, which are combinations of seven SIDM models: CDM, 6 $\rm cm^2/g$, $\{\sigma_0=50,\omega=50\}$,  $\{\sigma_0=50,\omega=100\}$, $\{\sigma_0=100,\omega=50\}$, $\{\sigma_0=200,\omega=50\}$, $\{\sigma_0=200,\omega=200\}$; and five orbits: from isolated to orb4 (see Table \ref{table:orb}). We show the tidal evolution tracks for stellar size in Fig. \ref{fig:trackall-size},  for stellar velocity dispersion in Fig. \ref{fig:trackall-vdisp}, and for mass-to-light ratio in Fig. \ref{fig:trackall-mlratio}.

\section{Supplementary figures}\label{appdx:suppfigs}

\begin{figure}
    \centering
    \begin{subfigure}{0.48\textwidth}
        \centering     \includegraphics[width=\textwidth]{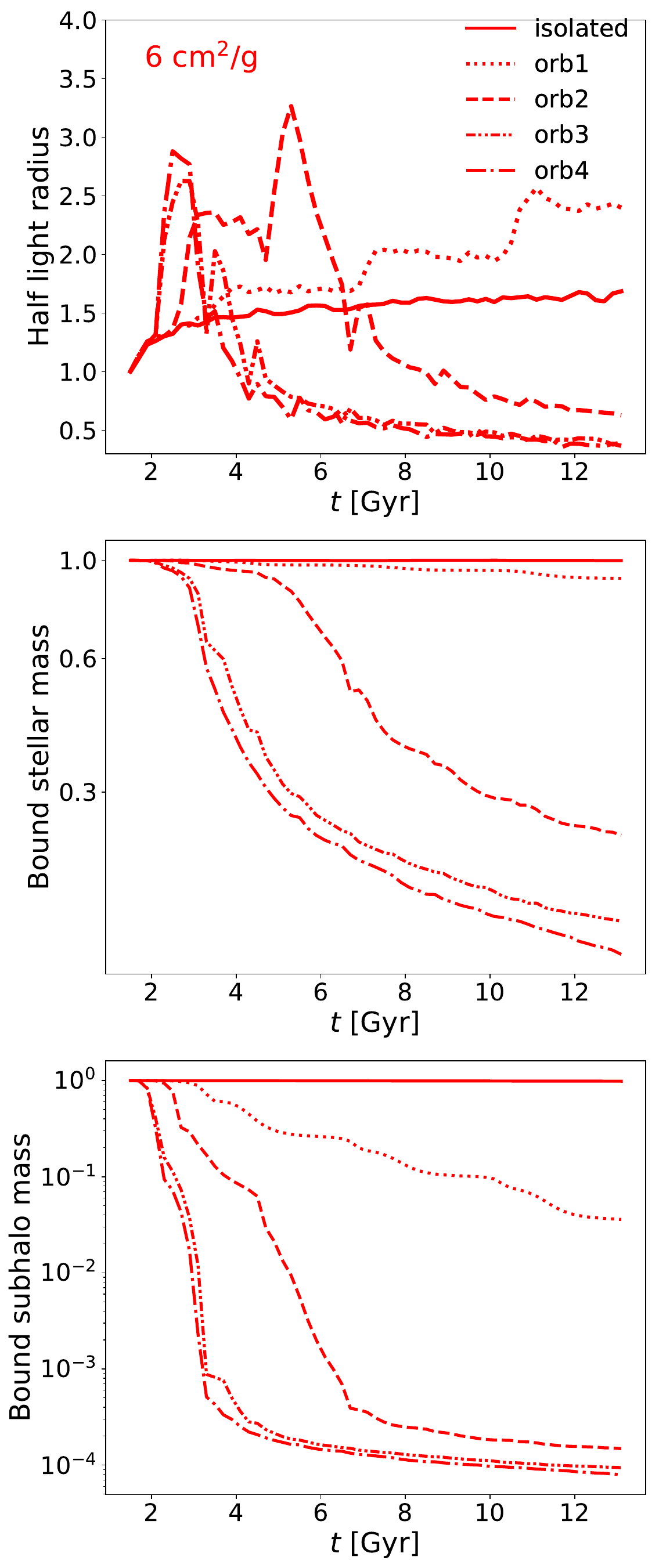} 
    \end{subfigure}
    \caption{From top to bottom: the half-light radius, bound stellar mass and bound subhalo mass (star+DM) as functions of time, for the 6 $\rm cm^2/g$ satellites on all orbits. This figure aims to show the correlation of these three panels. Quantities in each panel have been normalized to the respective initial values.  }
    \label{fig:rhalf-corr}
\end{figure}

\begin{figure}
    \centering
    \begin{subfigure}{0.48\textwidth}
        \centering
        \includegraphics[width=\textwidth]{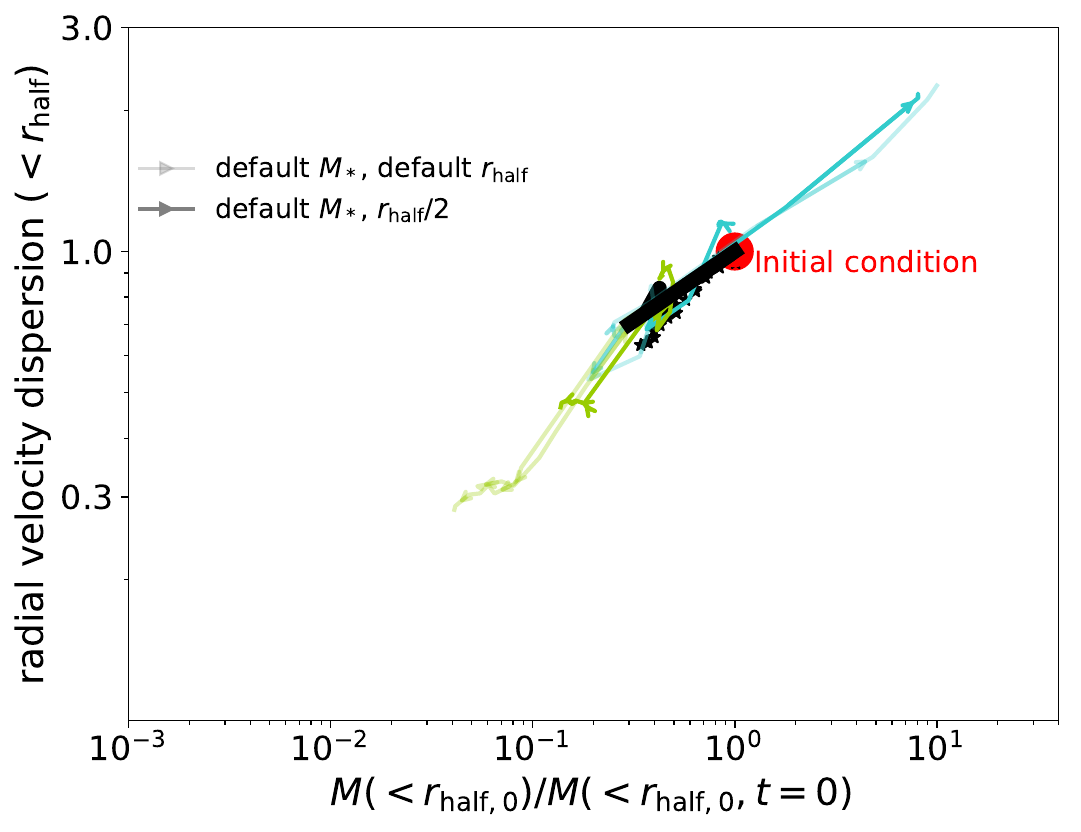} 
        \caption{}
        \label{fig:track-var-vdisp-a}
    \end{subfigure}
    ~
    \begin{subfigure}{0.48\textwidth}
        \centering
        \includegraphics[width=\textwidth]{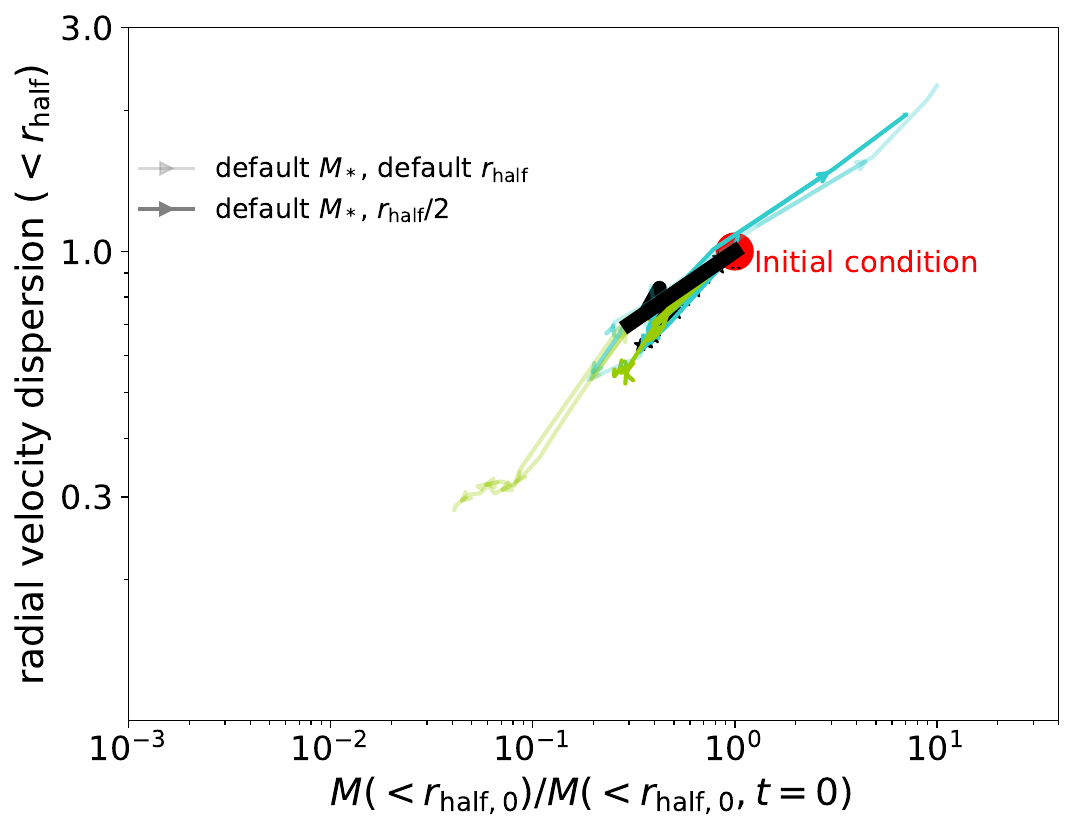} 
        \caption{}
        \label{fig:track-var-vdisp-b}
    \end{subfigure}
    ~
    \begin{subfigure}{0.48\textwidth}
        \centering
        \includegraphics[width=\textwidth]{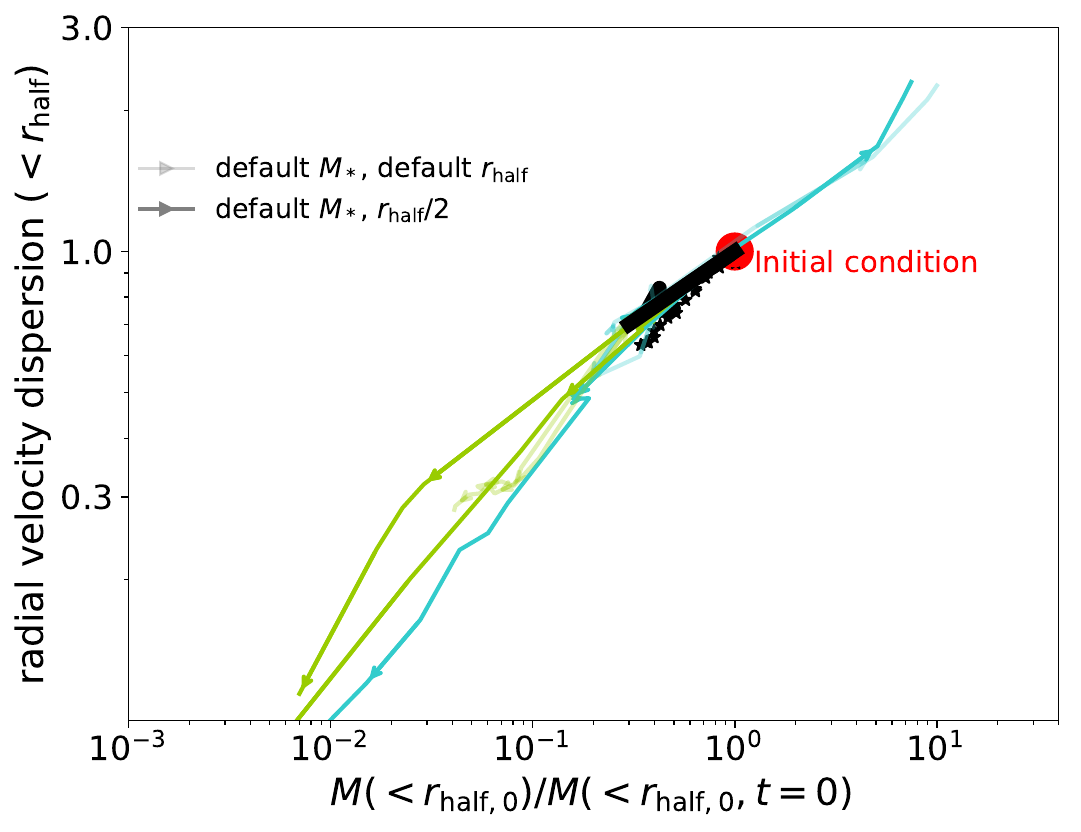} 
        \caption{}
        \label{fig:track-var-vdisp-c}
    \end{subfigure}
    \caption{ Tidal tracks for the stellar radial velocity dispersion within $r_{\rm half}$, with varied initial stellar properties, similar to Fig. \protect\ref{fig:track-var}. }
    \label{fig:track-var-vdisp}
\end{figure}

\begin{figure*}
    \centering
    \begin{subfigure}[t]{0.48\textwidth}
        \centering
        \includegraphics[width=\textwidth, clip,trim=0.2cm 0cm 0.2cm 0cm]{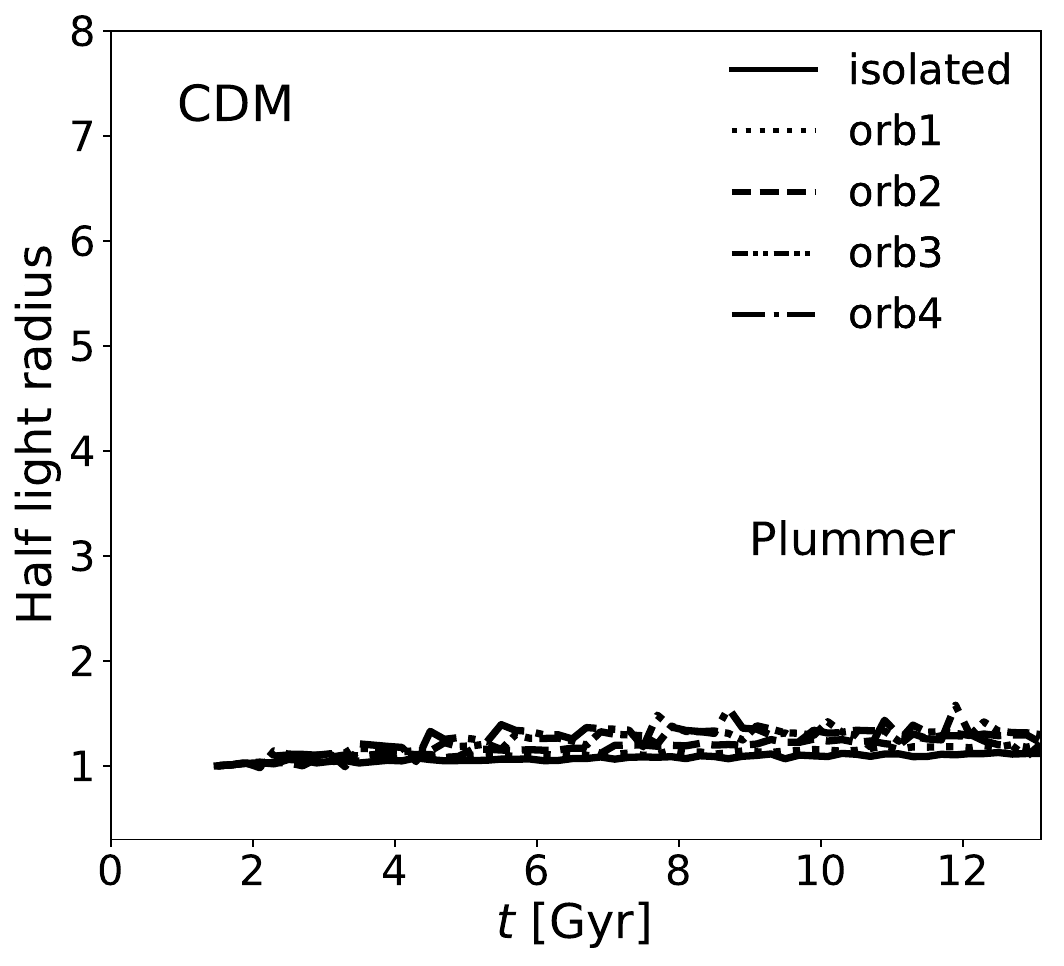}
        \caption{}
        \label{fig:rhalfp-a}
    \end{subfigure}
    ~
    \begin{subfigure}[t]{0.48\textwidth}
        \centering
        \includegraphics[width=\textwidth, clip,trim=0.2cm 0cm 0.2cm 0cm]{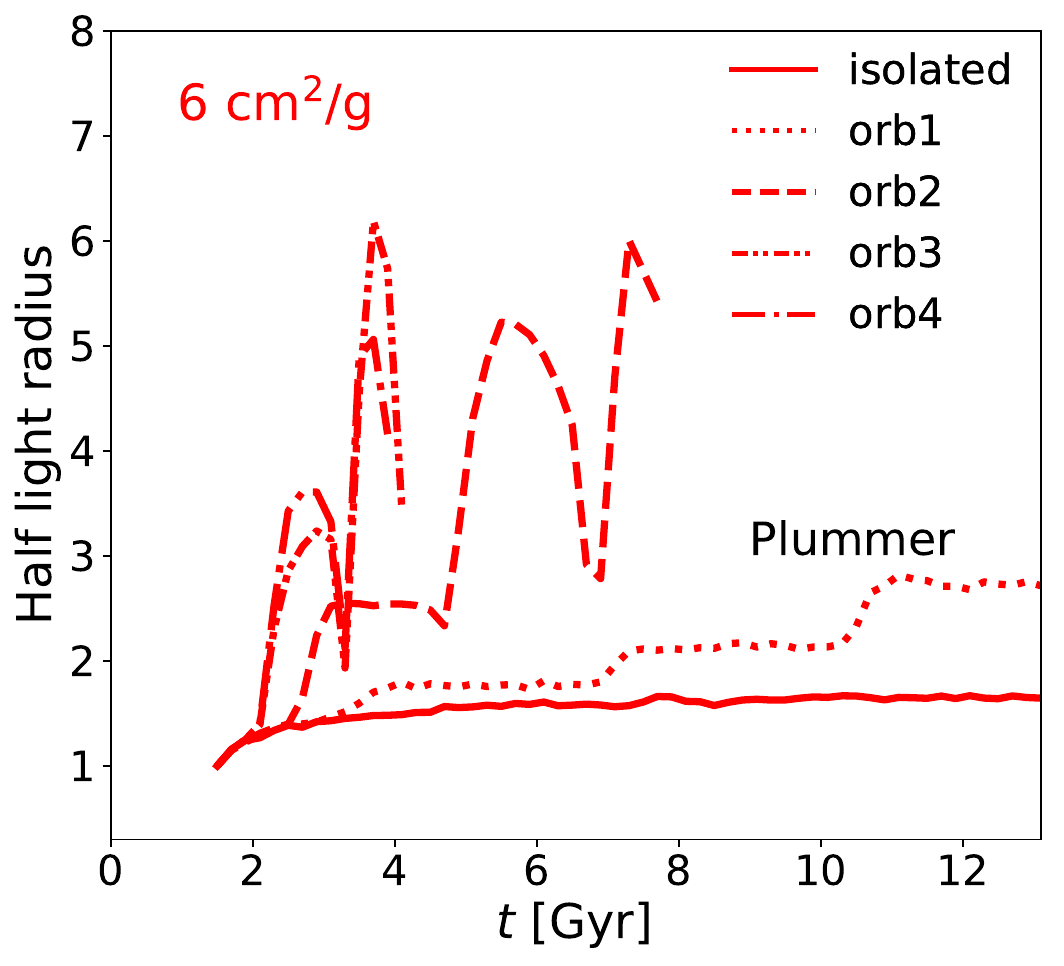}
        \caption{}
        \label{fig:rhalfp-b}
    \end{subfigure}
    ~
    \begin{subfigure}[t]{0.48\textwidth}
        \centering
        \includegraphics[width=\textwidth, clip,trim=0.2cm 0cm 0.2cm 0cm]{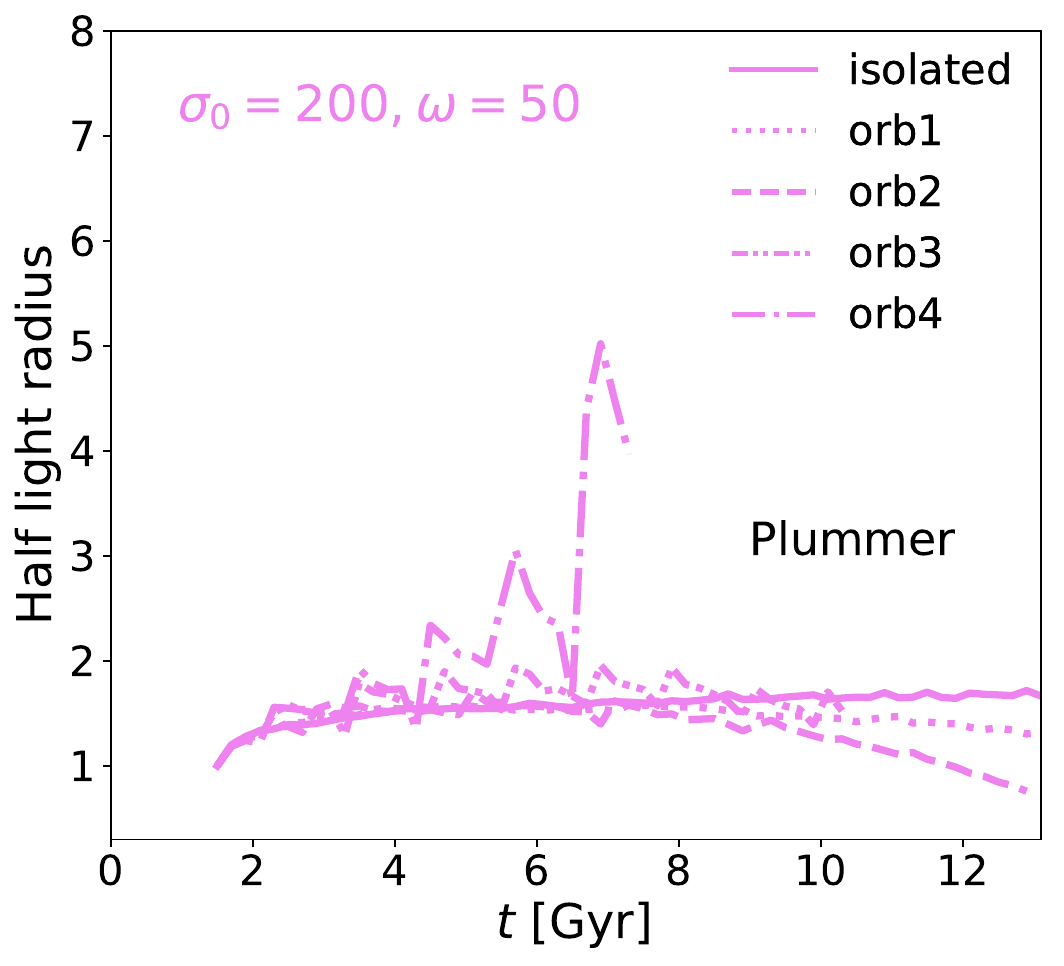}
        \caption{}
        \label{fig:rhalfp-c}
    \end{subfigure}
    ~
    \begin{subfigure}[t]{0.48\textwidth}
        \centering
        \includegraphics[width=\textwidth, clip,trim=0.2cm 0cm 0.2cm 0cm]{rhalf-DG_200_200-sat-plummer.pdf}
        \caption{}
        \label{fig:rhalfp-d}    
    \end{subfigure}
    \caption{Similar to Fig. \ref{fig:rhalf}, but showing the time evolution of the half-light radius for the satellites initialized with a Plummer profile. }
    \label{fig:rhalfp}
\end{figure*}

\begin{figure*}
    \centering
    \begin{subfigure}[t]{0.48\textwidth}
        \centering
        \includegraphics[width=\textwidth, clip,trim=0.2cm 0cm 0.2cm 0cm]{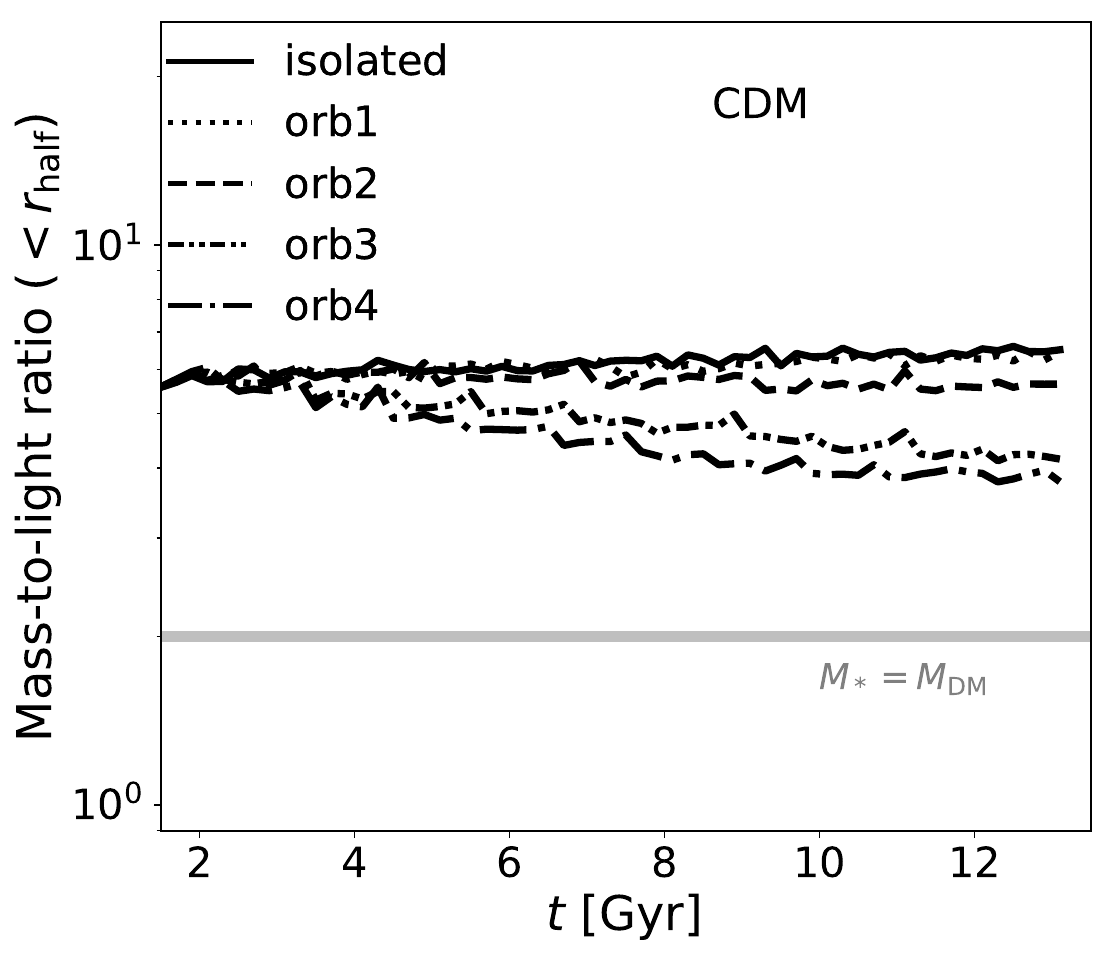}
        \caption{}
        \label{fig:mlp-a}
    \end{subfigure}
    ~
    \begin{subfigure}[t]{0.48\textwidth}
        \centering
        \includegraphics[width=\textwidth, clip,trim=0.2cm 0cm 0.2cm 0cm]{mass-light-s6-sat-plummer.pdf}
        \caption{}
        \label{fig:mlp-b}
    \end{subfigure}
    ~
    \begin{subfigure}[t]{0.48\textwidth}
        \centering
        \includegraphics[width=\textwidth, clip,trim=0.2cm 0cm 0.2cm 0cm]{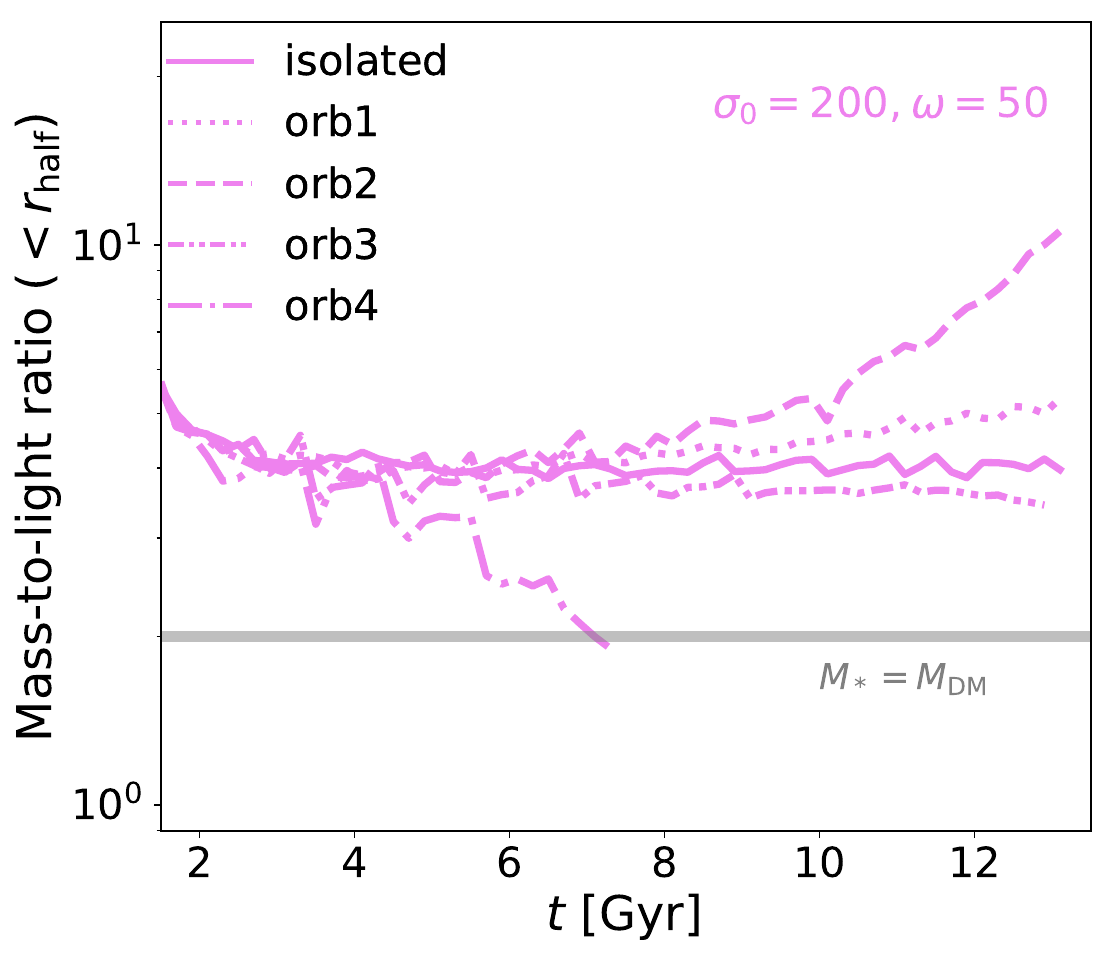}
        \caption{}
        \label{fig:mlp-c}
    \end{subfigure}
    ~
    \begin{subfigure}[t]{0.48\textwidth}
        \centering
        \includegraphics[width=\textwidth, clip,trim=0.2cm 0cm 0.2cm 0cm]{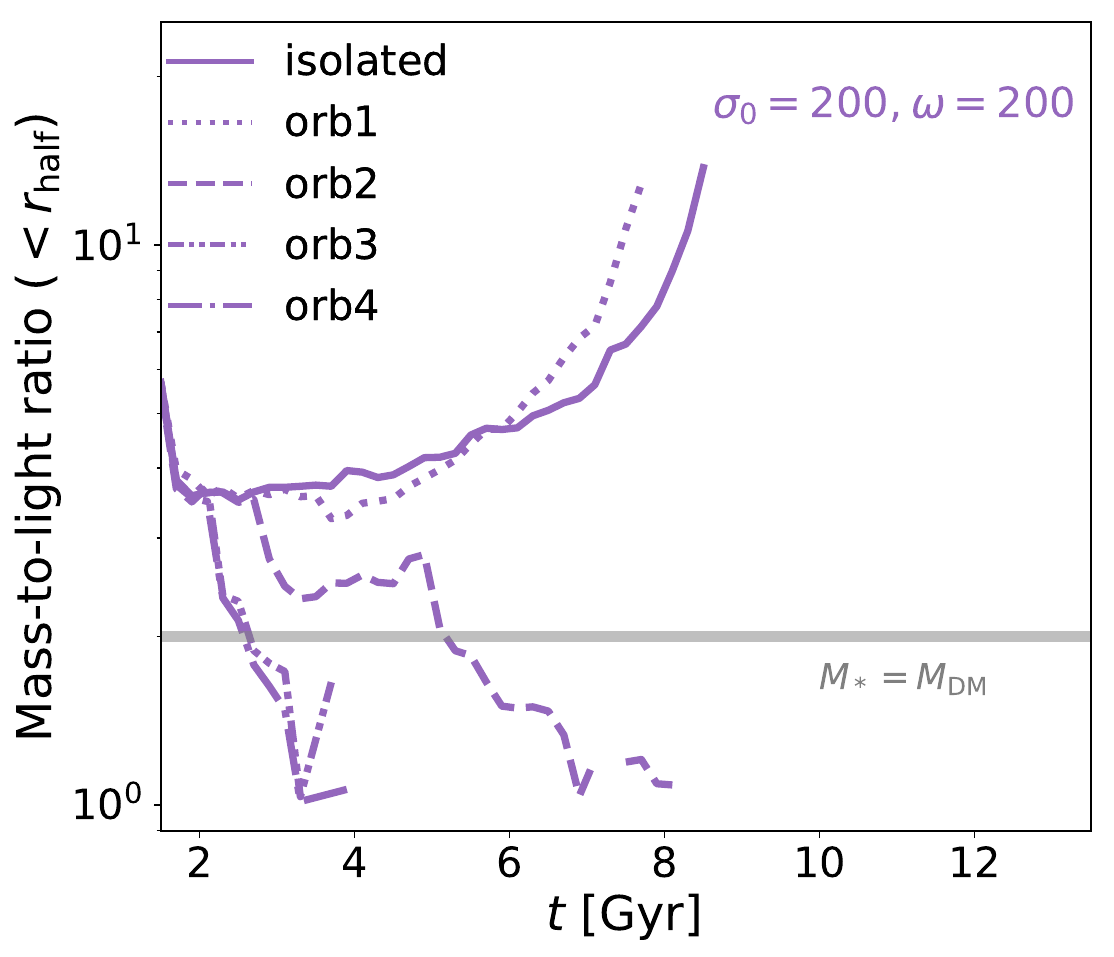}
        \caption{}
        \label{fig:mlp-d}    
    \end{subfigure}
    \caption{Similar to Fig. \ref{fig:ml}, but showing the time evolution of the mass-to-light ratio for the satellites initialized with a Plummer profile.}
    \label{fig:mlp}
\end{figure*}

\begin{figure*}
    \centering
    \begin{subfigure}[t]{0.48\textwidth}
        \centering
        \includegraphics[width=\textwidth, clip,trim=0.2cm 0cm 0.2cm 0cm]{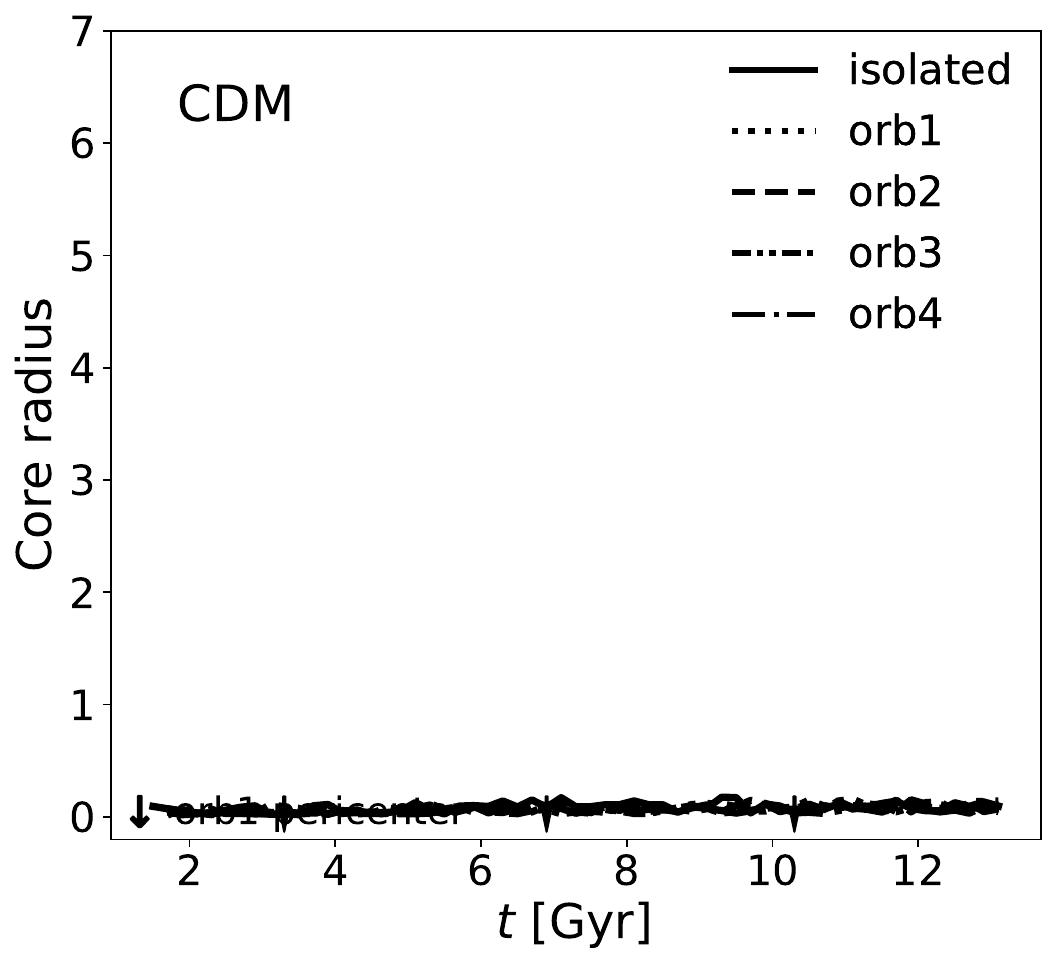}
        \caption{}
        \label{fig:rcore-a}
    \end{subfigure}
    ~
    \begin{subfigure}[t]{0.48\textwidth}
        \centering
        \includegraphics[width=\textwidth, clip,trim=0.2cm 0cm 0.2cm 0cm]{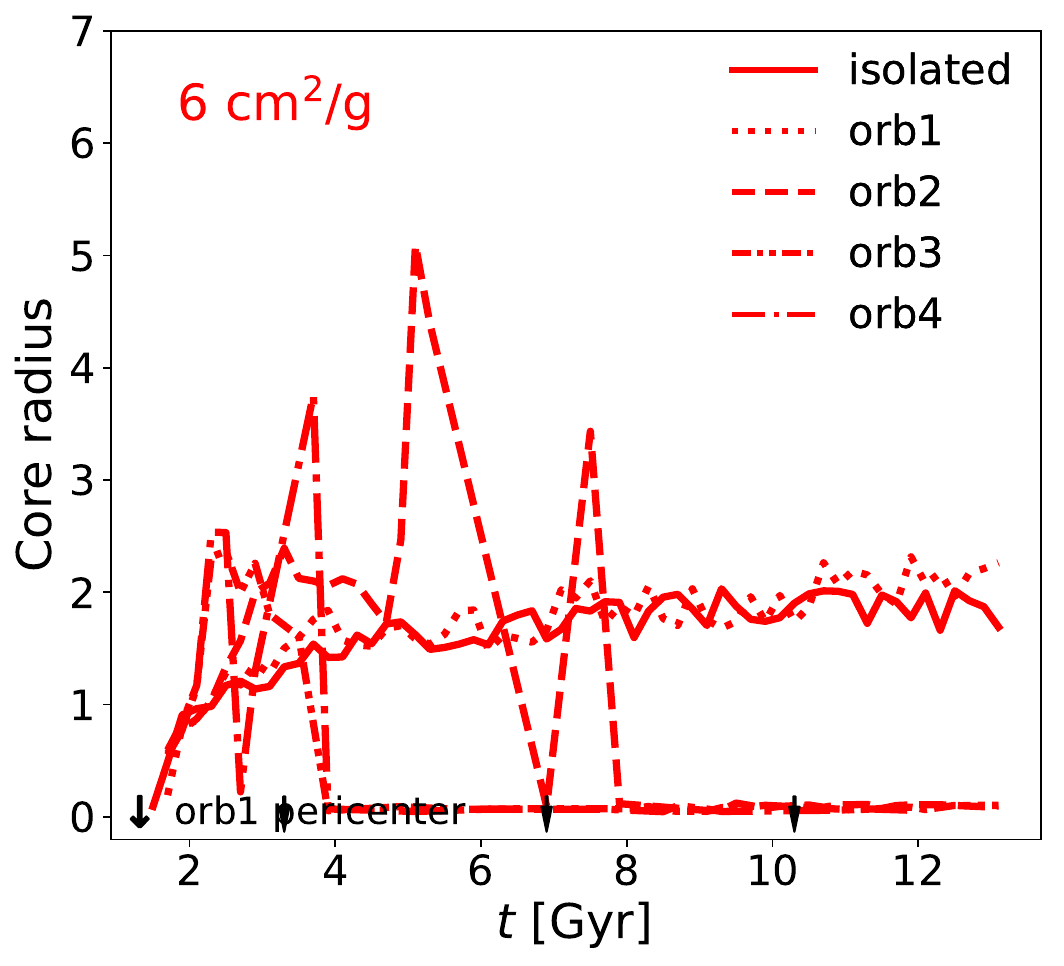}
        \caption{}
        \label{fig:rcore-b}
    \end{subfigure}
    ~
    \begin{subfigure}[t]{0.48\textwidth}
        \centering
        \includegraphics[width=\textwidth, clip,trim=0.2cm 0cm 0.2cm 0cm]{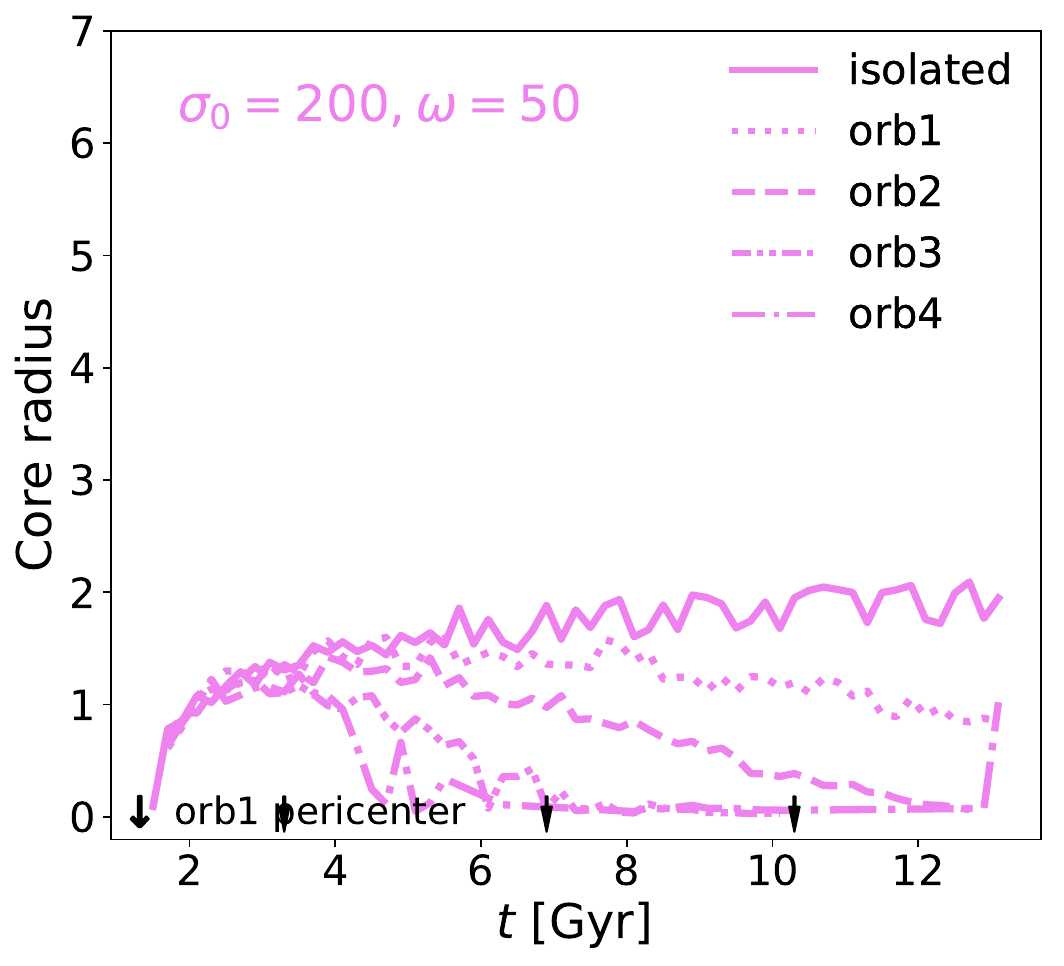}
        \caption{}
        \label{fig:rcore-c}
    \end{subfigure}
    ~
    \begin{subfigure}[t]{0.48\textwidth}
        \centering
        \includegraphics[width=\textwidth, clip,trim=0.2cm 0cm 0.2cm 0cm]{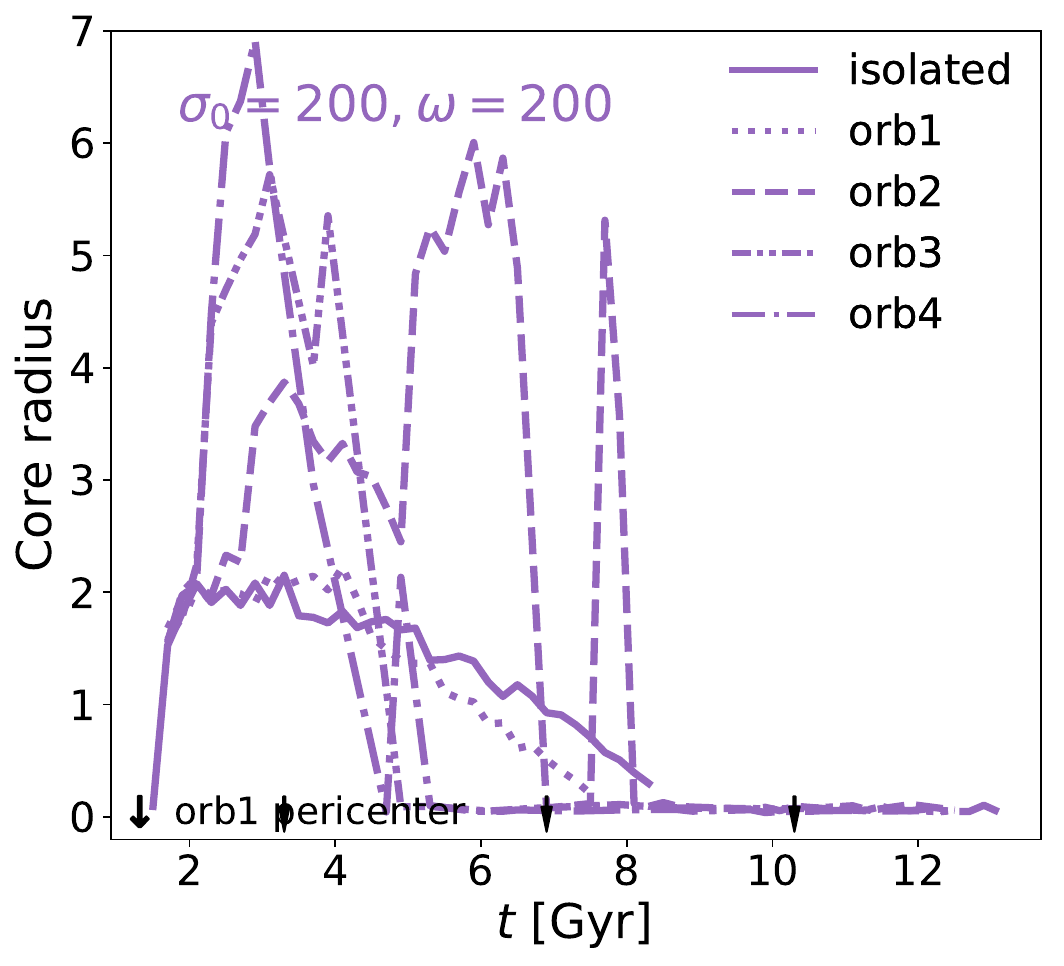}
        \caption{}
        \label{fig:rcore-d}    
    \end{subfigure}
    \caption{The time evolution of the dark matter density core size $r_{\rm core}$.}
    \label{fig:rcore}
\end{figure*}

\begin{figure*}
    \centering
    \begin{subfigure}[t]{0.48\textwidth}
        \centering
        \includegraphics[width=\textwidth, clip,trim=0.2cm 0cm 0.2cm 0cm]{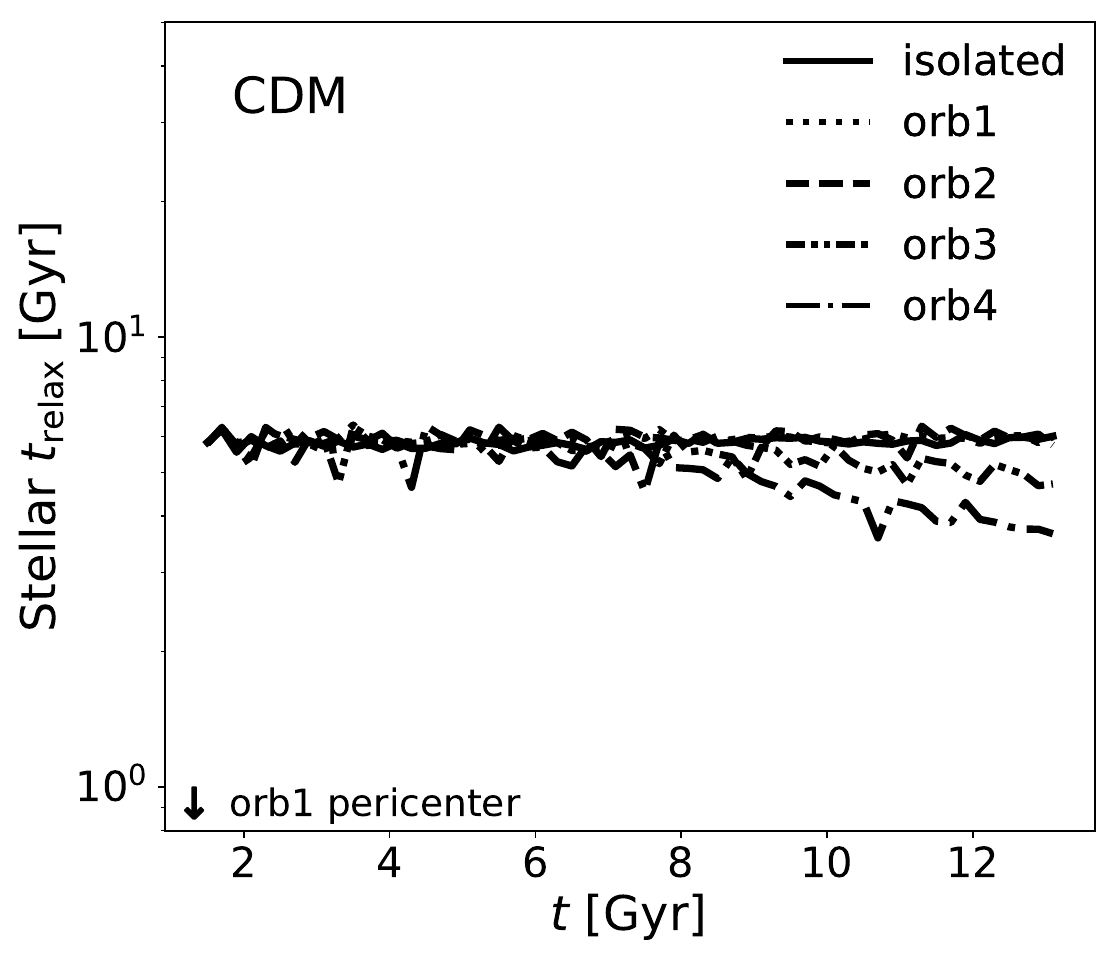}
        \caption{}
        \label{fig:stellar_trelax-a}
    \end{subfigure}
    ~
    \begin{subfigure}[t]{0.48\textwidth}
        \centering
        \includegraphics[width=\textwidth, clip,trim=0.2cm 0cm 0.2cm 0cm]{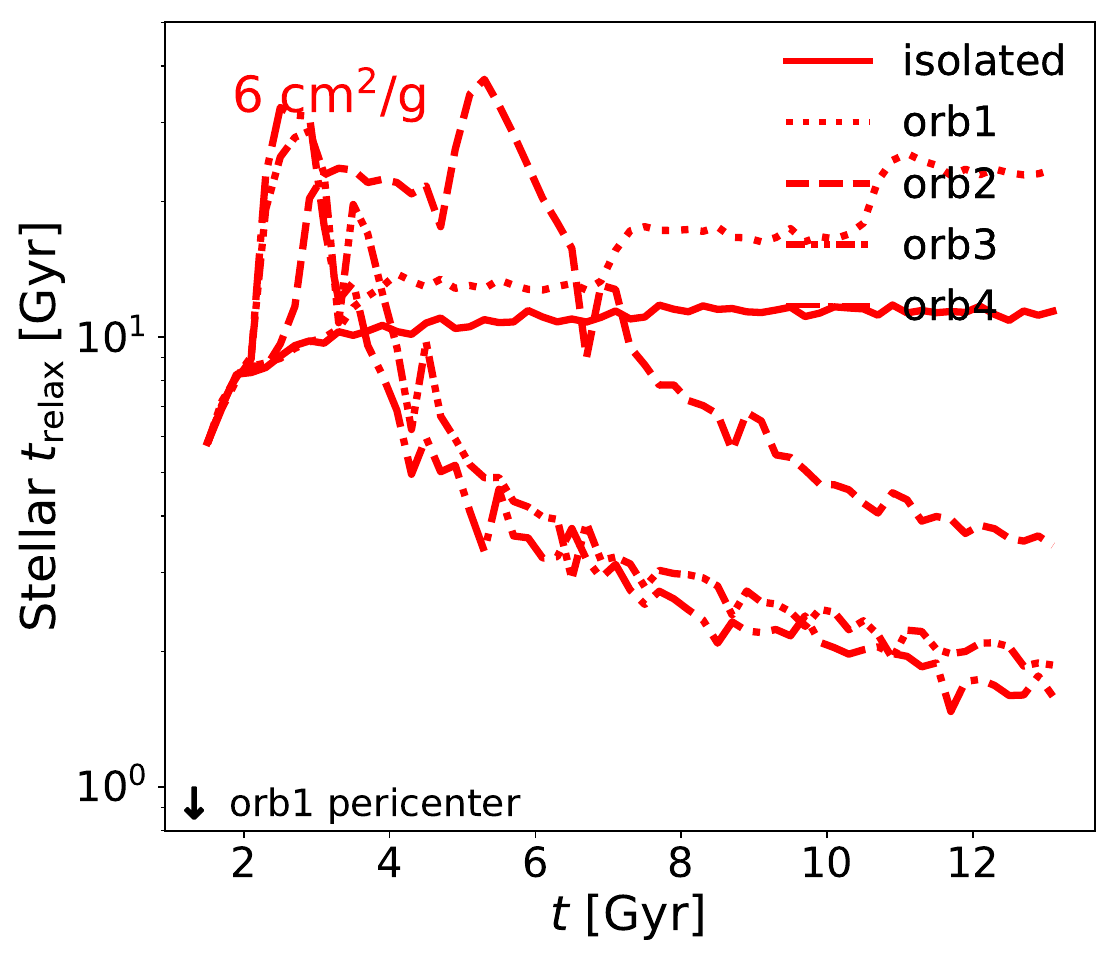}
        \caption{}
        \label{fig:stellar_trelax-b}
    \end{subfigure}
    ~
    \begin{subfigure}[t]{0.48\textwidth}
        \centering
        \includegraphics[width=\textwidth, clip,trim=0.2cm 0cm 0.2cm 0cm]{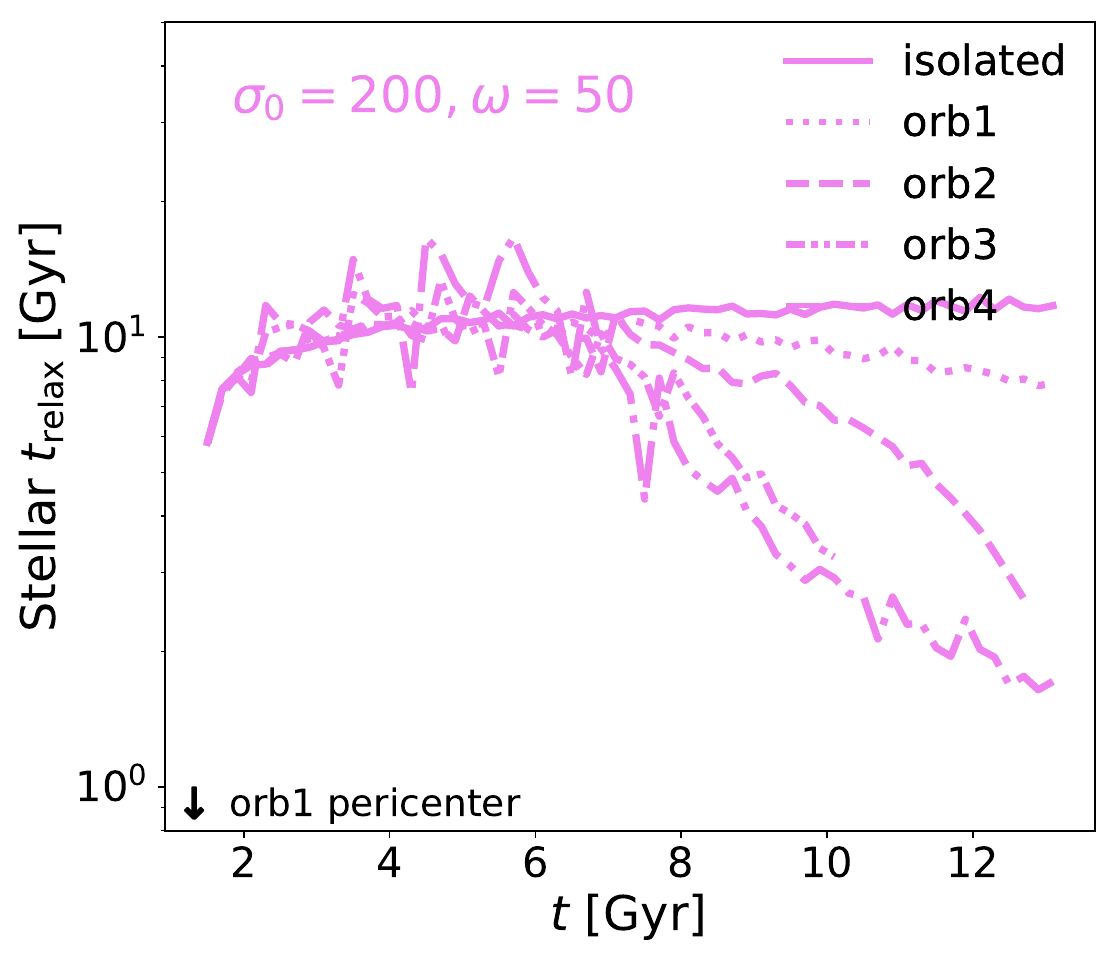}
        \caption{}
        \label{fig:stellar_trelax-c}
    \end{subfigure}
    ~
    \begin{subfigure}[t]{0.48\textwidth}
        \centering
        \includegraphics[width=\textwidth, clip,trim=0.2cm 0cm 0.2cm 0cm]{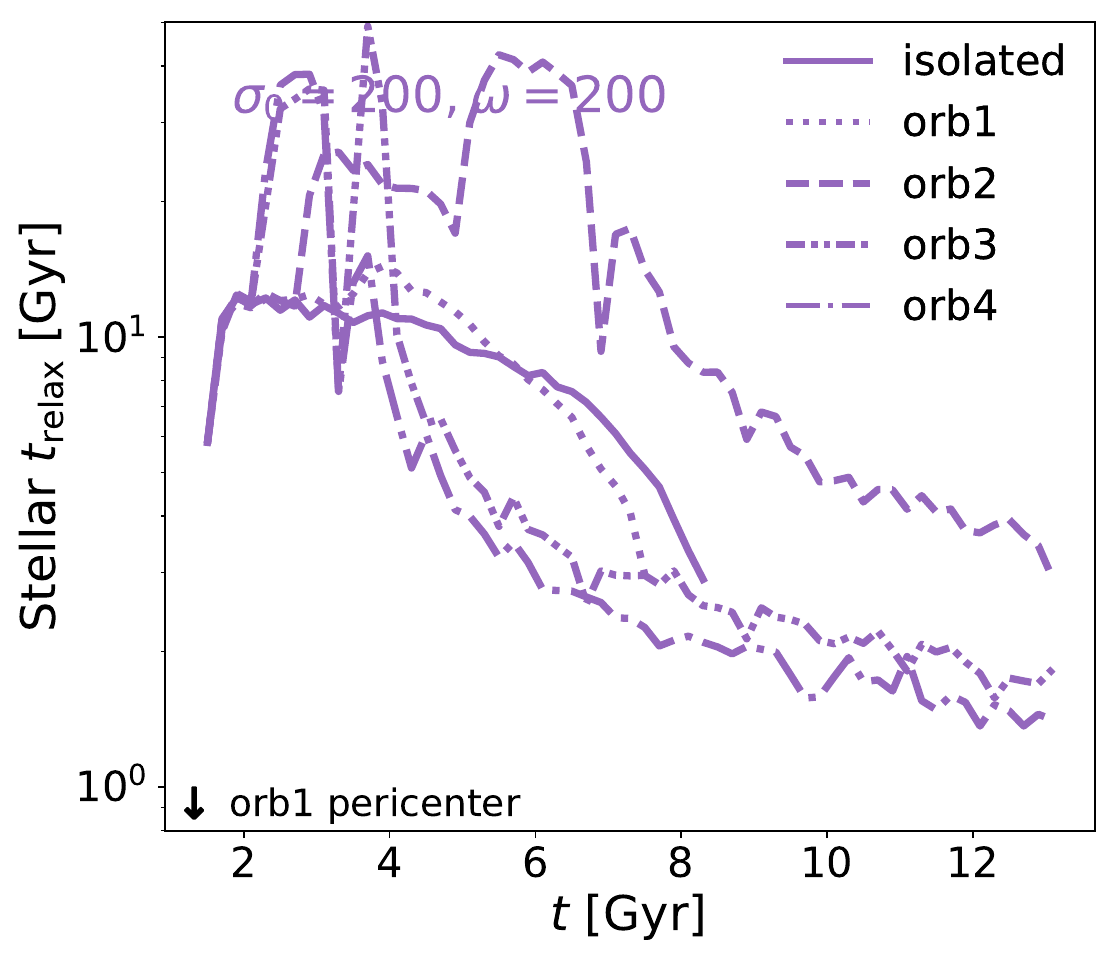}
        \caption{}
        \label{fig:stellar_trelax-d}    
    \end{subfigure}
    \caption{The time evolution of the stellar particles relaxation time.}
    \label{fig:stellar_trelax}
\end{figure*}

\begin{itemize}
    \item Fig. \ref{fig:rhalf-corr} shows the time evolution of $r_{\rm half}$, together with the bound stellar mass and bound total subhalo mass (star+DM), for satellites with 6 $\rm cm^2/g$ on all orbits. This supplementary figure aims to help explain the correlation between these three quantities.   Abrupt DM mass loss caused by evaporation causes a boost in $r_{\rm half}$.  However, when the DM mass loss becomes extreme, there will also be noticeable stellar mass loss, which causes $r_{\rm half}$ to decrease again.

    \item Fig. \ref{fig:track-var-vdisp} shows the effects on the tidal tracks of stellar velocity dispersion when varying the stellar initial conditions, presented in a style similar to Fig. \ref{fig:track-var}.

    \item Fig. \ref{fig:rhalfp} shows the time evolution of the satellites' $r_{\rm half}$, similar to Fig. \ref{fig:rhalf}, but here for the galaxies initialized with a cored, Plummer density profile. Note that these Plummer galaxies do not have a cuspy center to hold themselves together in a tidal field, hence they dissolve in the host soon after the DM mass loss gets extreme. This is reflected by the early termination of the evolution curves. For the CDM satellite galaxies with Plummer stellar distributions, we find an increase in  $r_{\rm half}$ first before decreasing at later times. This is because the tidal heating puffs up the subhalo center before tidal stripping is able to reach the inner region, as also reported in \cite{errani15}.

    \item Fig. \ref{fig:mlp} shows the time evolution of the Plummer satellites' mass-to-light ratio within $r_{\rm half}$, similar to Fig. \ref{fig:ml}.

    \item Fig. \ref{fig:rcore} shows the time evolution of the DM core radius, where the core is defined as the region in the DM density profile where its slope $\frac{d \ln\rho }{d\ln r}>-1$.

    \item Fig. \ref{fig:stellar_trelax} shows the stellar relaxation timescales for the satellites we simulate. The stellar relaxation time is estimated by $t_{\rm relax} \approx \frac{N_p}{8\ln{\Lambda}} t_{\rm cross} \approx \frac{N_p}{8\ln{\Lambda}} \frac{r_{\rm half}}{\sigma_v}$ \cite{bt08}, where $\Lambda\equiv \frac{b_{\rm max}}{b_{\rm min}}= \frac{r_{\rm half}}{\epsilon_{\rm soft}}$ is the ratio between the maximum and minimum impact parameters for the two-body gravitational scattering and $N_p$ is the number of particles within $r_{\rm half}$.
    
\end{itemize}

\bibliography{apssamp}

\end{document}